\documentclass[final,1p,times]{elsarticle}

\usepackage{lineno,hyperref}
\modulolinenumbers[5]

\usepackage{amsmath}
\usepackage{latexsym}
\usepackage{booktabs}
\usepackage{comment}
\usepackage{algorithm}
\usepackage{mdframed}
\usepackage{changepage}
\usepackage{multirow}
\usepackage{multido}
\usepackage{paralist}
\usepackage{listings}
\usepackage{color}
\usepackage{comment}
\usepackage{etex}
\usepackage{algorithmic}
\usepackage[normalem]{ulem}
\usepackage[makeroom]{cancel}
\usepackage{minibox}
\usepackage{stmaryrd}
\usepackage{amssymb}
\usepackage{wrapfig}

\definecolor{dkgreen}{rgb}{0,0.6,0}
\definecolor{gray}{rgb}{0.5,0.5,0.5}
\definecolor{mauve}{rgb}{0.58,0,0.82}

\usepackage{pstricks}
\usepackage{pst-node}
\usepackage{pst-blur}
\usepackage{pst-rel-points}
\usepackage{xspace}
\usepackage{pst-coil}

\usepackage{tikz}
\usetikzlibrary{patterns}
\usepackage{url}
\usepackage[normalem]{ulem}

\newcommand{\mips}{\text{MIPS}\xspace}
\newcommand{\mipss}{\text{MIPS}\xspace}
\newcommand{\mop}{\text{MOP}\xspace}
\newcommand{\mfp}{\text{MFP}\xspace}
\newcommand{\cfp}{\text{CFP}\xspace}
\newcommand{\cfg}{\text{CFG}\xspace}
\newcommand{\cfgs}{\text{CFGs}\xspace}
\newcommand{\cfps}{\text{{CFP\/}s}\xspace}

\newcommand{\KPadd}[1]{{\blue \protect#1}}

\newcommand{\Agreeddel}[1]{{}}
\newcommand{\mofp}{\text{MOFP}\xspace}

\newcommand{\pairc}[2]{\text{$\langle #1,#2 \rangle$}}

\newcommand{\symp}{\text{$\mathcal{P}$}\xspace}
\newcommand{\hlat}{\text{$\overline{\mathcal{L}}$}\xspace}

\newcommand{\set}[1]{\{\protect #1\}}
\newcommand{\bset}[1]{\big\{\protect #1\big\}}
\newcommand{\bbrace}[1]{\big(\protect #1\big)}

\newcommand{\foldmeet}{\text{$\emph{fold}_{\sqcap}$}}

\newcommand{\tentb}{\fourtb\fourtb\twotb}

\newcommand{\mpowerset}{\text{$\mathcal{P}(\setmips)$}\xspace}

\newcommand{\fun}{\text{\em f} }
\newcommand{\fune}{\text{\em g} }
\newcommand{\bFe}{\overline{\fune}}

\newcommand{\eequal}{\overset{\mathrm{\emph{end}}}{=}}

\newcommand{\rangeZT}{\text{$l\mapsto[0,2]$}\xspace}
\newcommand{\rangeZ}{\text{$l\mapsto[0,0]$}\xspace}
\newcommand{\rangeT}{\text{$l\mapsto[2,2]$}\xspace}

\newcommand{\arangeZF}{\text{$a\mapsto[0,5]$}\xspace}
\newcommand{\arangeZ}{\text{$a\mapsto[0,0]$}\xspace}
\newcommand{\arangeF}{\text{$a\mapsto[5,5]$}\xspace}

\newcommand{\xrangeZ}{\text{$x\mapsto[0,0]$}\xspace}
\newcommand{\xrangeO}{\text{$x\mapsto[1,1]$}\xspace}

\newcommand{\zrangeZT}{\text{$z\mapsto[0,2]$}\rule[-5pt]{0em}{15pt}\xspace}
\newcommand{\zrangeZ}{\text{$z\mapsto[0,0]$}\xspace}
\newcommand{\zrangeO}{\text{$z\mapsto[1,1]$}\rule[-5pt]{0em}{15pt}\xspace}
\newcommand{\zrangeT}{\text{$z\mapsto[2,2]$}\rule[-5pt]{0em}{15pt}\xspace}

\newcommand{\zrangePositive}{\text{$z\mapsto[1,\infty]$}\xspace}

\newcommand{\zrangeNegativeZ}{\text{$z\mapsto[-\infty,0]$}\xspace}
\newcommand{\zrangeII}{\text{$z\mapsto[-\infty,\infty]$}\xspace}

\newcommand{\setSn}{\text{$\mathcal{S}_n$}\xspace}
\newcommand{\setS}{\text{$\mathcal{S}$}\xspace}
\newcommand{\setc}{\text{$\mathcal{C}$}\xspace}
\newcommand{\setV}{\text{$\mathcal{V}$}\xspace}
\newcommand{\nomips}{\text{$\{\}$}\xspace}
\newcommand{\drangeO}{\text{$d\mapsto[1,1]$}\xspace}
\newcommand{\FI}{\text{FI}\xspace}

\newcommand{\symmips}{\text{$\mu$}\xspace}
\newcommand{\symmipsO}{\text{$\mu_1$}\xspace}
\newcommand{\symmipsT}{\text{$\mu_2$}\xspace}
\newcommand{\symmipsTh}{\text{$\mu_3$}\xspace}

\newcommand{\setmips}{\text{$\mathcal{U}$}\xspace}
\newcommand{\setm}{\text{$\mathcal{M}$}\xspace}
\newcommand{\setmD}{\text{$\mathcal{M}'$}\xspace}

\newcommand{\ips}{\text{\em ips}\xspace}

\newcommand{\onetb}{\ \ \ }
\newcommand{\twotb}{\ \ \ \ \ \ }
\newcommand{\threetb}{\ \ \ \ \ \ \ \ \ }
\newcommand{\fourtb}{\ \ \ \ \ \ \ \ \ \ \ \ }

\newcommand{\true}{\text{\em true}\xspace}
\newcommand{\false}{\text{\em false}\xspace}

\newcommand{\states}[2]{\text{\emph{states}$(\protect #1, \protect#2)$}\xspace}

\newcommand{\meet}{\text{$\;\sqcap\;$}}

\newcommand{\mipsstart}{\emph{start}\xspace}
\newcommand{\mipsend}{\emph{end}\xspace}
\newcommand{\mipsinner}{\emph{inner}\xspace}

\newcommand{\Boundary}{\emph{\sffamily BI}\xspace}

\newcommand{\Out}{\emph{\sffamily Out}\xspace}
\newcommand{\In}{\emph{\sffamily In}\xspace}
\newcommand{\bOut}{\text{$\overline{\emph{\sffamily Out}}$}\xspace}
\newcommand{\bIn}{\text{$\overline{\emph{\sffamily In}}$}\xspace}

\newcommand{\bbOut}{\text{$\overline{\overline{\emph{\sffamily Out}}}$}\xspace}
\newcommand{\bbIn}{\text{$\overline{\overline{\emph{\sffamily In}}}$}\xspace}

\newcommand{\lat}{\text{$\mathcal{L}$}\xspace}

\newcommand{\pIn}{\text{$\overline{\overline{\emph{\sffamily In}}}$}\xspace}

\newcommand{\zerofive}{\text{$\{a\!\mapsto\! [0,5]\}$}}

\newcommand{\bBoundary}{\text{$\overline{\emph{\sffamily BI}}$}\xspace}

\newcommand{\bF}{\text{$\overline{f}$}\xspace}
\newcommand{\bMeet}{\text{\;$\overline{\text{$\sqcap$}}$\;}\xspace}

\newcommand{\mn}{\text{$m \rightarrow n$}\xspace}

\newcommand{\fpmfp}{\text{FPMFP}\xspace}

\newcommand{\czero}{\text{$\mathbb{C}$}\xspace}

\newcommand{\oursol}{\fpmfp solution\xspace}

\newcommand{\setmu}{\text{$\set{\symmips}$}\xspace}
\newcommand{\setmuO}{\text{$\set{\symmipsO}$}\xspace}
\newcommand{\setmuT}{\text{$\set{\symmipsT}$}\xspace}

\newcommand{\setmuOT}{\text{$\set{\symmipsO,\symmipsT}$}\xspace}

\newcommand{\CPO}{\text{CPO}\xspace}
\newcommand{\prefix}[2]{\text{\emph{prefix}$(#1,#2)$}\xspace}
\newcommand{\suffix}[2]{\text{\emph{suffix}$(#1,#2)$}\xspace}

\newcommand{\finto}{contains-prefix-of\xspace}
\newcommand{\cpo}[2]{\text{$\text{\emph{cpo}}_{#1}(#2)$}\xspace}

\newcommand{\cso}[2]{\text{$\text{\emph{cso}}_{#1}(#2)$}\xspace}
\newcommand{\cpon}{\text{$\text{CPO}$}\xspace}

\newcommand{\cpob}[2]{\text{$\text{\emph{ext}}(#2,#1)$}\xspace}

\newcommand{\cpobn}{\text{\emph{ext}}\xspace}
\newcommand{\edges}[1]{\text{\emph{edges}(#1)}\xspace}

\newcommand{\setmO}{\text{$\mathcal{M}_1$}\xspace}

\psset{unit=1mm}

\def\BibTeX{{\rm B\kern-.05em{\sc i\kern-.025em b}\kern-.08emT\kern-.1667em\lower.7ex\hbox{E}\kern-.125emX}}

\newcommand{\meanDefUse}{1.75\%}
\newcommand{\avgDefUse}{2.87\%}
\newcommand{\uptoDefUse}{13.6\%}

\newcommand{\avgUninit}{18.5\%}
\newcommand{\meanUninit}{3\%}
\newcommand{\uptoUninit}{100\%}
\newcommand{\avgTime}{2.9 }

\newcommand{\meanTime}{1.6}

\newcommand{\avgMemory}{0.5 }

\begin{document}
\begin{frontmatter}

\title{Computing Maximum Fixed Point Solutions over Feasible Paths in Data Flow Analyses}

\author{Komal Pathade}
\ead{komal.pathade@tcs.com}
\address{TCS Research, India}

\author{Uday Khedker}
\ead{uday@cse.iitb.ac.in}
\address{Indian Institute of Technology Bombay, India}


\begin{abstract}
The \emph{control flow graph} (CFG) representation of a procedure used by
virtually all flow-sensitive program analyses, admits a large number of 
\emph{infeasible} control flow paths i.e., these paths do not occur in any execution of the program. 
Hence the information reaching along infeasible paths in an analysis is spurious. This affects the 
precision of the conventional \emph{maximum fixed point} (\mfp) solution of the data flow analysis, because it 
includes the information reaching along all control flow paths. The existing approaches for removing this imprecision
are either specific to a data flow problem with no straightforward
generalization or involve control flow graph restructuring which may
exponentially blow up the size of the CFG.

We lift the notion of  MFP solution to define 
the notion of \emph{feasible path MFP} (FPMFP) solutions that exclude the
data flowing along known infeasible paths. The notion of FPMFP is
generic and does not involve CFG restructuring. Instead, it takes
externally supplied information about infeasible paths and lifts any
data flow analysis to an analysis that maintains the distinctions
between different paths where these distinctions are beneficial,
and ignores them where they are not. Thus it gets the benefit of
a path-sensitive analysis where it is useful without performing a
conventional path-sensitive analysis. Hence, an FPMFP solution is
more precise than the corresponding MFP solution in most cases; it is
guaranteed to be sound in each case.
 
We implemented the proposed computation of feasible path MFP solutions for reaching definitions analysis and potentially uninitialized 
variable analysis. We evaluated the precision improvement in these two analyses by analyzing 30 benchmark applications selected from 
open source, industry, and SPEC CPU 2006.
The evaluation results indicate that the precision improvement in these two analyses respectively reduce the number def-use pairs by up to 
\uptoDefUse\ (average \avgDefUse, geometric mean \meanDefUse), and reduce the potentially uninitialized variable alarms by up to 
\uptoUninit\ (average \avgUninit, geo. mean \meanUninit). We found that the \fpmfp computation time was $\avgTime\times$ of 
the \mfp computation time on average.
\end{abstract}


\end{frontmatter}

\newdefinition{definition}{Definition}
\newdefinition{mytheorem}{Theorem}[section]
\newdefinition{mylemma}{Lemma}[section]
\newdefinition{observation}{Observation}[section]
\newcounter{example}[paragraph]
\newenvironment{example}[1][]{\vspace{3pt}\begin{mdframed} \refstepcounter{example}\par
  \noindent \textbf{Example~\theexample. #1} \rmfamily }{\end{mdframed}\vspace{3pt}}


\section{Introduction}
The effectiveness of techniques that optimize, verify, and debug computer programs depends on the availability of accurate information 
describing the program behavior during various executions. To this end, data flow analysis gathers code level information that could 
answer either a subset of queries over a program  (demand-driven analyses) or all possible queries of a particular type over a program  
(exhaustive analyses) without executing the program. It is well known that computing highly precise exhaustive data flow analysis solutions 
(like \emph{meet over paths} (\mop)~\cite{khedker2009data}) is undecidable; even in the instances where it is computable, it does not scale well 
to practical programs. On the other hand, scalable and widely used solutions  (like \emph{maximum fixed-point} (\mfp)~\cite{khedker2009data}) 
are not as precise.

\subsection{The Context of This Work}
In this work, we focus on exhaustive data flow analyses because they compute information that can be used to answer all queries of a 
particular type over all program executions. In particular, we compute \mfp solutions (details in Section~\ref{sec:Computation}) and 
try to make them more precise. 


A \mfp solution includes the data flow values that reach a program point along all \emph{control 
flow paths} (\cfps)~\cite{khedker2009data} which are paths in the \emph{control flow graph} (\cfg)~\cite{khedker2009data} representation of 
the program. 
Experimental evidence~\cite{bodik1997refining} on Linux kernel code suggests that 9-40\% of conditional statements in a program lead to 
at least one infeasible 
\cfp i.e., these \cfps do not occur in any execution of the program. Additionally, we found in our benchmarks on average 60\% 
(Geometric mean 29\%) functions had at least one infeasible \cfp in their \cfg. Inclusion 
of data flow values reaching along such infeasible \cfps makes the computed data flow information an over-approximation of the 
actual data flow information. This dilutes the usefulness of the information for client analyses.

	Various approaches reported in the literature detect infeasible \cfps~\cite{bodik1997refining,bodik1997interprocedural,
	ngo2007detecting,rui2006infeasible,delahaye2010explanation,bueno2000identification,zhuang2006using,yang2011improve,
	bertolino1994automatic,chen2007exploiting,malevris1990predictive}. Most of these approaches cater to the needs of program 
	testing, in which the knowledge of infeasible \cfps is used to refine various test coverage criteria (like branch, path coverage) 
	leading to less testing overhead. However, not 
	much attention has been given towards improving the precision of data flow analysis by avoiding the data flow values reaching along known 
	infeasible \cfps. A potential reason being that \cfp based data flow approaches like \mop computation (that considers \cfps independently, 
	and hence can inherently exclude infeasible \cfps) are not scalable, because the number of \cfps as well as the number of data flow
	values may be very large or even unbounded.  Hence, scalable approaches like \mfp computation do not maintain 
	a mapping between individual \cfps and the corresponding data flow values, thus making it non-trivial to identify (and discard) the data 
	flow values reaching along infeasible \cfps.

We name the \mfp solutions computed by restricting the data flow values at each program point to those values that reach along 
feasible \cfps as \emph{feasible path \mfp} (\fpmfp) solutions.  Existing approaches that attempt to compute such solutions are either 
unsuitable because they involve \cfg restructuring which can exponentially blow up the size of the \cfg or are analysis specific 
with no straightforward generalization.
	
\subsection{Our Contributions}

We observe that infeasible \cfps are a property of programs and not of any particular data flow analysis over programs. We use this 
observation to separate the identification of infeasible \cfps in the \cfg of a program, from computation of \fpmfp solutions that exclude 
data flow values reaching along known infeasible \cfps. These two phases are described below.

\begin{enumerate}
\item In the first phase, we use the work done by Bodik et.al~\cite{bodik1997refining} to detect infeasible \cfps in the \cfg of a program. 
In their work, they 
identify \emph{minimal infeasible path segments} (\mips) which are minimal length sub-segments of infeasible \cfps such that the following 
holds: if a \cfp $\sigma$ contains a \mips as a sub-segment then $\sigma$ is infeasible. \mips\footnote{We use the  acronym \mips as an 
irregular noun, whose plural form is same as its singular form.} captures the infeasibility property of \cfps in a concise form (illustrated 
in Section~\ref{sec:Background}).

Given a set of \mips as input, we do the following processing over them.
\begin{enumerate}
\item[(i)] We propose a novel criteria (called \emph{contains-prefix-of}) to cluster \mips. This clustering 
is necessary because treating each \mips in isolation may lead to an imprecise solution, if the \mips are intersecting 
with each other (illustrated in Section~\ref{sec:fpmfp.overview}). 
\item[(ii)] We establish an equivalence relation over the clusters of \mipss, so that the clusters that belong to the same 
equivalence class can be treated as a single unit during \fpmfp computation (illustrated in Section ~\ref{sec:optimizations}). This 
increases the scalability of \fpmfp computation without reducing the precision of the computed solution. 
\end{enumerate}
\item In the second phase, we automatically lift the existing \mfp specifications of an analysis to its \fpmfp specifications using the 
processed \mipss from the first phase. These specifications are then used to compute the \fpmfp solution. The computation of \fpmfp solution 
satisfies the following properties that distinguish \fpmfp from existing approaches: 
\begin{enumerate}
\item[(i)] Computation restricts the data flow values at each program point to those values that reach along feasible \cfps (data flow values 
reaching along known infeasible \cfps are identified and discarded).
\item[(ii)] Computation does not involve \cfg restructuring and is generically applicable for all \mfp based data flow analyses. 
\end{enumerate}
\end{enumerate}
Items (i) and (ii) in each of the above phases describe the distinguishing features of \fpmfp over existing approaches. 
Moreover, separating the above two phases is an 
useful insight that avoids the costly repetition involved in identifying infeasible \cfps for each different data flow analysis over the same 
underlying program, which happens in analysis specific approaches
~\cite{das2002esp,dhurjati2006path,hampapuram2005symbolic,xie2003archer,dillig2008sound,
dor2004software}. Instead, we perform the first phase that involves identification of infeasible \cfps once for each program, and 
the second phase that involves computation of \fpmfp is performed once for each different data flow analysis.

The core ideas of \fpmfp have been presented before~\cite{pathade2018computing,pathade2019path}. This paper enhances the core ideas 
significantly and provides a complete treatment. In particular, the following contributions are exclusive to this paper.  
\begin{enumerate}
\item We unify our earlier approaches~\cite{pathade2018computing,pathade2019path} by proposing a novel formalization 
for handling infeasible paths in the \mfp computations. This formalization is based on the contains-prefix-of relation 
(explained in Section~\ref{sec:fpmfp.overview}) that precisely distills the useful interactions between \mips from all possible 
interactions. 
The formalization is more intuitive to understand because of its declarative nature as opposed to the procedural 
formalization earlier, and its correctness is easier to prove and verify. Moreover, the new formalization of \fpmfp is close to
the standard \mfp formalization. We believe 
this will ease the adoption of the \fpmfp technique in data flow analysis.
\item We formalize the concept of \fpmfp for inter-procedural setting. In particular, we extend it to the \emph{functional approach}
~\cite{sharir1978two} of inter-procedural data flow analysis. We evaluate our inter-procedural technique of \fpmfp computation on an industry 
strength static program analysis tool called TCS ECA~\cite{TECA}. Inter-procedural static analyses are known to be desirable to 
 software developer community~\cite{blackshear2018racerd}.
\item We improve the practical applicability of the \fpmfp by adding novel optimizations that improve the scalability of the 
\fpmfp computation by the factor of 2 without affecting soundness or precision of the computed solution. The approach has now
scaled on the codesets of size 150 KLOC.
\item We prove that the \fpmfp computes an over-approximation of the most precise  
 solution of a data flow problem. The most precise solution is called the \emph{meet over feasible paths} (\mofp) 
solution and is explained in Section~\ref{sec:existingApproaches}. 
\end{enumerate}

\subsection{Organization of the Paper}
Section ~\ref{sec:existingApproaches} describes the limitations of existing approaches of data flow analysis. Section ~\ref{sec:Background} 
describes the concept of Minimal Infeasible Path Segments that facilitate \fpmfp solutions, and illustrates 
the \fpmfp solution using an example. Section ~\ref{sec:Computation} explains the computation of \fpmfp. Section ~\ref{sec:interprocedural} 
explains \fpmfp computation in inter-procedural setting. Section ~\ref{sec:optimizations} presents the optimizations that improve the 
scalability of \fpmfp computation and explains the worst-case complexity of the \fpmfp computation. 
Section ~\ref{sec:Experiments} presents experimental evaluation and Section ~\ref{sec:Relatedwork} describes 
related work. Section ~\ref{sec:Conclusion} concludes the paper. 
The appendix (Section ~\ref{sec:pos}) gives the  proof of soundness of \fpmfp solution. 

\section{Existing Approaches and Their Limitations}\label{sec:existingApproaches}
\begin{table}[t]
\setlength{\tabcolsep}{2pt}%
\centering
\begin{tabular}{@{}|c|c|c|c|c|c|c|c|@{}}
\hline
\rule[-.5em]{0em}{1.5em}%
\multirow{6}{*}{Solution} 
	& \multicolumn{3}{c|}{Traverses \cfps}
	& \multicolumn{2}{c|}{Merges information}
	& \multirow{6}{*}{Decidable}
	& \multirow{6}{*}{Precision}
	\\ \cline{2-6}	
\rule[-.35em]{0mm}{1.3em}
	& \multirow{2}{*}{Infeasible}
	& \multirow{2}{*}{Feasible}
	& 
	  \multirow{2}{*}{Spurious}
	& 
		\begin{tabular}{@{}l@{}}
		Across \\ \cfps 
		\end{tabular}
	& 
		\begin{tabular}{@{}l@{}}
		Across \\
		program \\
		points
		\end{tabular}
	&
	& 
	\\ \hline\hline
\rule{0em}{1em}%
\mofp 
	& No 
	& Yes
	& No 
	& No 
	& No 
	& No 
	& 
		\rnode{most}{\psframebox[framesep=.5mm,linestyle=none]{Most}}
	\\ \cline{1-7}
\rule{0em}{1em}%
\mop 
	& Yes
	& Yes
	& No 
	& No 
	& No
	& No 
	&
	\\ \cline{1-7}
\rule{0em}{1em}%
\mfp 
	& Yes 
	& Yes 
	& No 
	& Yes 
	& No
	& Yes 
	&
	\\ \cline{1-7}
\rule{0em}{1em}%
\FI 
	& Yes 
	& Yes 
	& Yes 
	& Yes 
	& Yes
	& Yes 
	&
		\rnode{least}{\psframebox[framesep=.5mm,linestyle=none]{Least}}%
		\ncline[doubleline=true,nodesepB=-.5mm,nodesepA=.15mm]{->}{most}{least}
	\\ \hline
\end{tabular}
\vspace{5pt}
\caption{Solutions of a data flow analysis. As we go down the rows, more
entries become ``Yes'' leading to increased efficiency and decreased precision.}
\label{fig:dfa.solutions}
\end{table}

\begin{figure}[t]
\begin{pspicture}(0,20)(25,68)
\small
\putnode{n00}{origin}{70}{68}{All Paths}
\putnode{n0}{n00}{15}{-20}{$\begin{array}{c}\text{Spurious} \\ \text{Paths}\end{array}$}
\putnode{n1}{n00}{-15}{-20}{$\begin{array}{c}\text{\cfg} \\ \text{Paths}\end{array}$}
\putnode{n1l}{n1}{-15}{-23}{$\begin{array}{c}\text{Infeasible} \\ \text{Paths}\end{array}$}
\putnode{n1r}{n1}{15}{-23}{$\begin{array}{c}\text{Feasible} \\ \text{Paths}\end{array}$}
\ncline[nodesep=0.5]{->}{n00}{n0}
\ncline[nodesep=0.5]{->}{n00}{n1}
\ncline{->}{n1}{n1l}
\ncline{->}{n1}{n1r}
\end{pspicture}
\caption{Paths Traversed During Data Flow Analyses}\label{figallpaths}
\end{figure}
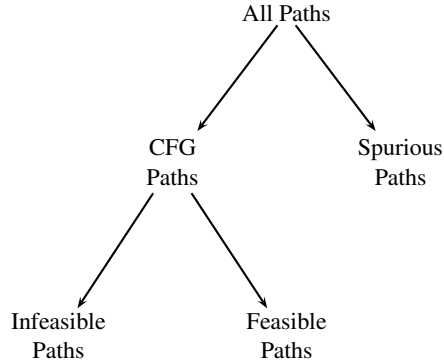

Table~\ref{fig:dfa.solutions} lists the conventional solutions of a data flow analysis
and mentions their computability and (relative) precision by describing the
nature of paths traversed in computing the solutions. These paths are illustrated in Figure~\ref{figallpaths}. In particular, 
the paths that are traversed by a data flow analysis consist of \cfg paths or spurious paths 
(described below) or both. \cfg paths (\cfps) that do not represent any possible execution of the program are \emph{infeasible}; others 
are \emph{feasible}. Apart from feasible and infeasible \cfps, a data flow
analysis could also traverse \emph{spurious} \cfps which are paths
that result from an over-approximation of \cfps---they do not appear in the
\cfg but are effectively traversed by a specific method of data flow analysis 
as shown in Table~\ref{fig:dfa.solutions} and as explained below.

\begin{itemize}
\item Meet Over Feasible Paths (\mofp) solution captures the
      information reaching a program point along \emph{feasible}
      \cfps only and does not merge
      information across \cfps. Since 
      a program could have infinitely many \cfps, 
	computing this solution is \emph{undecidable}. 
       The information defined by \mofp
      precisely represents the possible runtime information at each
      program point because it does not include information reaching along infeasible and spurious \cfps.

\item Meet Over Paths (\mop)~\cite{khedker2009data} solution captures the information reaching  a program point
	along \emph{feasible} and \emph{infeasible} \cfps. Since the information reaching along infeasible \cfps is also considered, \mop is 
	less precise compared to \mofp. Computing \mop is also
      \emph{undecidable}.

\item Maximum Fixed Point (\mfp)~\cite{khedker2009data} solution differs from \mop in that the
      information is effectively merged in the shared segments
      of \cfps because the solution is inductively computed. Since only one piece
      of information is stored at a program point regardless of the
      number of paths passing through it, computing \mfp is
      \emph{decidable}. \mfp is less precise than \mop
      because information across paths is
      merged and this over-approximation cannot distinguish between 
      feasible and infeasible paths. Spurious \cfps are not traversed in \mfp computation.

\item Flow Insensitive (\FI)~\cite{khedker2009data} solution merges
	information across all program points by
      ignoring the control flow. Effectively, it over-approximates the
      set of \cfps and hence includes information along feasible, infeasible, and spurious \cfps.
\end{itemize}

We propose an approach to compute \fpmfp solutions in which the computation at each program point is restricted to those data flow values 
that reach along feasible \cfps (data flow values reaching along known infeasible \cfps are excluded and spurious \cfps are not 
traversed). 

Our work can be seen in the mould of trace partitioning~\cite{mauborgne2005trace} which tries to 
maintain some disjunction over data flow values based on path conditions that represent control flow of values instead of merging the values 
indiscriminately across all paths to compute an \mfp solution. However, in trace partitioning, the desirable disjunctions (over data flow 
values) are not generic, instead, they need to be specified for each data flow analysis separately which limits the usefulness of trace 
partitioning. 

Towards this end, we use the infeasibility of control flow paths\textemdash a property of programs and not of any particular 
data flow analysis over the 
program\textemdash as the criteria for trace partitioning. This allows us to achieve a generic and practical trace partitioning that 
can be automatically incorporated in any data flow analysis to obtain a more precise solution. 

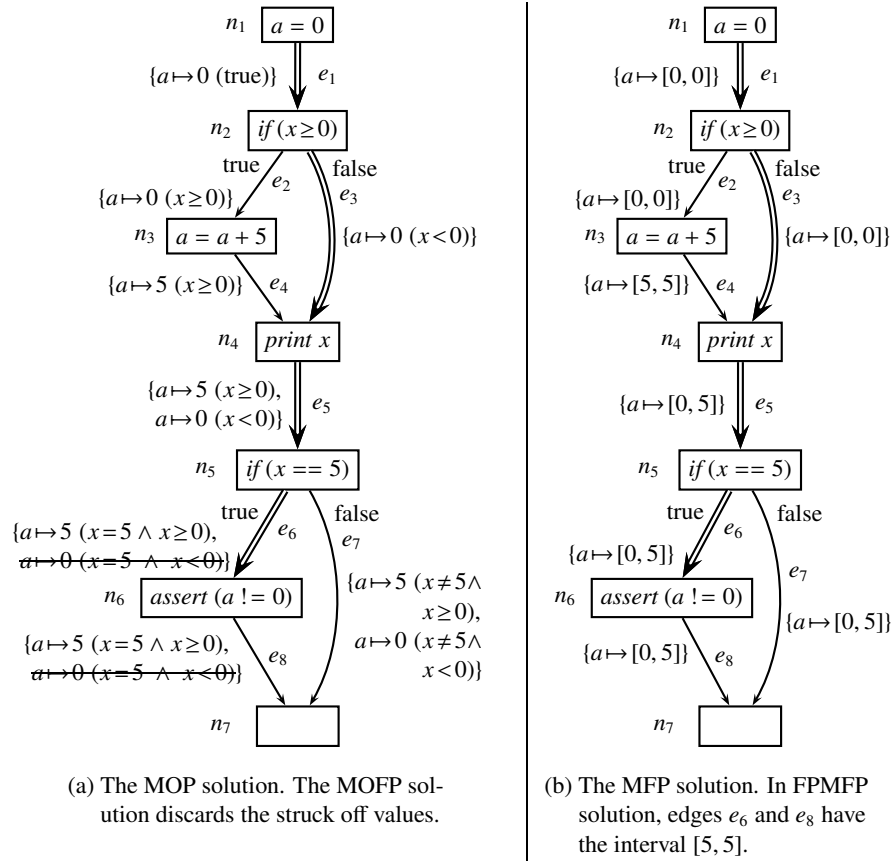
\begin{figure*}[t!]
\psset{unit=1mm}
\centering
\small
\begin{tabular}{c|c}
\begin{tabular}{@{}c@{}}
\begin{pspicture}(-135,-93)(-68,8)

\putnode{entry1}{origin}{-96}{5}{\psframebox{$a=0$}}
	\putnode{w}{entry1}{-8}{0}{$n_1$}
\putnode{c11}{entry1}{0}{-14}{\psframebox{\em if $(x\!\geq\!0)$}}
	\putnode{w}{c11}{-10}{0}{$n_2$}
\putnode{n11}{c11}{-10}{-14}{\psframebox{$a=a+5$}}
	\putnode{w}{n11}{-10}{0}{$n_3$}
\putnode{n31}{n11}{10}{-14}{\psframebox{\em print $x$}}
	\putnode{w}{n31}{-9}{0}{$n_4$}
\putnode{c21}{n31}{0}{-17}{\psframebox{\em if $(x==5)$}}
	\putnode{w}{c21}{-12}{0}{$n_5$}
\putnode{n51}{c21}{-10}{-17}{\psframebox{\em assert $(a\ !\!=0)$}}
	\putnode{w}{n51}{-14}{0}{$n_6$}
\putnode{exit1}{n51}{10}{-17}{\psframebox{\white print x}}
	\putnode{w}{exit1}{-10}{0}{$n_7$}
%

\ncline[doubleline=true]{->}{entry1}{c11}
	\naput{$e_1$}
	\nbput{$\{a\!\mapsto\! 0 \; \text{(true)}\}$}
\ncline{->}{c11}{n11}
	\naput[npos=.35,labelsep=.5]{$e_2$}
	\nbput[npos=.35,labelsep=.5]{true}
	\nbput[npos=.9,labelsep=.5]{$\{a \!\mapsto\! 0 \; (x\!\geq\!0)\}$}
\nccurve[doubleline=true,angleA=300,angleB=60]{->}{c11}{n31}
	\naput[npos=.3,labelsep=.5]{$e_3$}
	\naput[npos=.15,labelsep=.5]{false}
	\naput[npos=.5,labelsep=.5]{$\{a\!\mapsto\! 0 \;(x\!<\!0)\}$}
\ncline{->}{n11}{n31}
	\naput[npos=.6,labelsep=.5]{$e_4$}
	\nbput[npos=.3,labelsep=.5]{$\{a\!\mapsto\! 5 \;(x\!\geq\!0)\}$}

\ncline[doubleline=true]{->}{n31}{c21}
	\naput[labelsep=1]{$e_5$}
	\nbput[npos=.5,labelsep=1]{$\begin{array}{@{}l@{}}
			\{a \!\mapsto\! 5 \;(x\!\geq\!0), 
				\\
			\phantom{\{} a\!\mapsto\! 0 \;(x\!<\!0)\}
				\end{array}$}
\ncline[doubleline=true]{->}{c21}{n51}
	\naput[npos=.35,labelsep=.5]{$e_6$}
	\nbput[npos=.35,labelsep=.2]{true}
	\nbput[npos=0.9,labelsep=-.5]{$\begin{array}{@{}l@{}}
			\{a \!\mapsto\! 5 \;(x\!=\!5 \land x\!\geq\!0),
				\\
			\phantom{\{} \text{\sout{$a \!\mapsto\! 0 \;(x\!=\!5\ \land\ x\!<\!0)$}}\}
			\end{array}$}
\nccurve[angleA=300,angleB=60]{->}{c21}{exit1}
	\naput[npos=.27,labelsep=1]{$e_7$}
	\naput[npos=.15,labelsep=1]{false}
	\naput[npos=.6,labelsep=.5]{$\begin{array}{@{}l@{}}
				\{a \!\mapsto\! 5 \;(\begin{array}[t]{@{}l@{}}
							x\!\neq\!5 \land 
							\\
							x\!\geq\!0),
							\end{array}
					\\
				\phantom{\{} a\!\mapsto\! 0 \;(\begin{array}[t]{@{}l@{}}
								x\!\neq\!5 \land 
							\\ 
							x\!<\!0)\}
						\end{array}
				\end{array}$}
\ncline{->}{n51}{exit1}
	\naput[npos=.55,labelsep=.5]{$e_8$}
	\nbput[npos=.25,labelsep=-.5]{$\begin{array}{@{}l@{}}
			\{a \!\mapsto\! 5 \;(x\!=\!5 \land x\!\geq\!0),
				\\
			\phantom{\{} \text{\sout{$a \!\mapsto\! 0 \;(x\!=\!5\ \land\ x\!<\!0)$}}\}
			\end{array}$}
	
\end{pspicture}
\end{tabular}
&
\begin{tabular}{@{}c@{}}
\begin{pspicture}(-108,-93)(-65,8)

\putnode{entry1}{origin}{-82}{5}{\psframebox{$a=0$}}
	\putnode{w}{entry1}{-8}{0}{$n_1$}
\putnode{c11}{entry1}{0}{-14}{\psframebox{\em if $(x\!\geq\!0)$}}
	\putnode{w}{c11}{-10}{0}{$n_2$}
\putnode{n11}{c11}{-9}{-14}{\psframebox{$a=a+5$}}
	\putnode{w}{n11}{-10}{0}{$n_3$}
\putnode{n31}{n11}{9}{-14}{\psframebox{\em print $x$}}
	\putnode{w}{n31}{-9}{0}{$n_4$}
\putnode{c21}{n31}{0}{-17}{\psframebox{\em if $(x==5)$}}
	\putnode{w}{c21}{-12}{0}{$n_5$}
\putnode{n51}{c21}{-9}{-17}{\psframebox{\em assert $(a\ !\!=0)$}}
	\putnode{w}{n51}{-14}{0}{$n_6$}
\putnode{exit1}{n51}{9}{-17}{\psframebox{\white print x}}
	\putnode{w}{exit1}{-10}{0}{$n_7$}
%

\ncline[doubleline=true]{->}{entry1}{c11}
	\naput{$e_1$}
	\nbput{$\{a\!\mapsto\! [0,0]\}$}
\ncline{->}{c11}{n11}
	\naput[npos=.35,labelsep=.5]{$e_2$}
		\nbput[npos=.35,labelsep=.5]{true}
	\nbput[npos=.9,labelsep=.5]{$\{a \!\mapsto\! [0,0]\}$}
\nccurve[doubleline=true,angleA=300,angleB=60]{->}{c11}{n31}
	\naput[npos=.3,labelsep=.5]{$e_3$}
		\naput[npos=.15,labelsep=.5]{false}
	\naput[npos=.5,labelsep=.5]{$\{a\!\mapsto\! [0,0] \}$}
\ncline{->}{n11}{n31}
	\naput[npos=.6,labelsep=.5]{$e_4$}
	\nbput[npos=.3,labelsep=.5]{$\{a\!\mapsto\! [5,5] \}$}

\ncline[doubleline=true]{->}{n31}{c21}
	\naput[labelsep=1]{$e_5$}
	\nbput[npos=.5,labelsep=1]{$\begin{array}{@{}l@{}}
			\{a \!\mapsto\! [0,5] \}
				\end{array}$}
\ncline[doubleline=true]{->}{c21}{n51}
	\naput[npos=.35,labelsep=.5]{$e_6$}
		\nbput[npos=.35,labelsep=.2]{true}
	\nbput[npos=.9,labelsep=.5]{$\begin{array}{@{}l@{}}
			\{a \!\mapsto\! [0,5] \}
			\end{array}$}
\nccurve[angleA=300,angleB=60]{->}{c21}{exit1}
	\naput[npos=.4,labelsep=1]{$e_7$}
		\naput[npos=.15,labelsep=1]{false}
	\naput[npos=.6,labelsep=.5]{$\begin{array}{@{}l@{}}
				\{a \!\mapsto\! [0,5] \} 
				\end{array}$}
\ncline{->}{n51}{exit1}
	\naput[npos=.55,labelsep=.5]{$e_8$}
	\nbput[npos=.3,labelsep=.5]{$\{a\!\mapsto\! [0,5]\}$}
	
\end{pspicture}
\end{tabular}
\\
\small (a) 
	\begin{tabular}[t]{@{}l@{}}
	The \mop solution.  The \mofp sol- \\ution discards the struck off values.
	\end{tabular}
	&\small  (b) 
	\begin{tabular}[t]{@{}l@{}}
	The \mfp solution.  In \fpmfp \\ solution, edges $e_6$ and $e_8$ have \\ the interval $[5,5]$.
	\end{tabular}
\end{tabular}
\caption{Different solutions of value analysis for an example program. The nodes in the \cfg are numbered $n_i$, and the edges 
	are numbered $e_i$. A value $a\!\mapsto\!k\;(p)$ at edge $e$ indicates that the variable $a$ has value $k$ at edge $e$ along \cfp 
	represented by condition $p$. The path reaching $n_6$ shown by double lines is infeasible. 
	For simplicity we have shown the values for $a$ alone. The \FI solution merges values across all program
	points resulting in a single interval $[0,5]$.}
\label{motivating_example}
\end{figure*}

The main challenge in achieving this automatic approach
is that it is difficult to eliminate the effect of an infeasible path when data flow values are merged which happens in \mfp computation. 
We meet this challenge by identifying
\mips (computed using existing approaches~\cite{bodik1997refining}) and
creating equivalence classes of \cfps; such that all \cfps in one equivalence class overlap with the same set of \mipss. These classes 
allow us to lift any data flow analysis to a data flow analysis that computes one data flow value for each equivalence class. The 
resulting solution ignores data flow values corresponding to equivalence classes containing infeasible \cfps and hence is more precise 
than \mfp (and sometimes more precise than \mop and as precise as \mofp like in Figure \ref{motivating_example}).

\begin{example}
Figure \ref{motivating_example} shows an example of different solutions 
for an analysis that determines the values of variables. 
For \mop and \mofp, each data flow value has a constraint that represents the \cfps
along which the value reaches the program point. Since \mfp, \fpmfp, and \FI solutions merge the values, they
compute a range of values in terms of lower and upper bounds. 
\fpmfp solution is more precise than \mop and \mfp because it excludes
the data flow value $a \!\mapsto\! [0,0]$ reaching along infeasible path at node $n_6$ (the infeasible path is shown by double lines).
\end{example}

\section{Feasible Path \mfp Solutions}\label{sec:Background}

In this section, we define minimal infeasible path segments and show how they allow us to define the \fpmfp solution. 
Section~\ref{sec:Computation} describes how \fpmfp solution is computed. 

\subsection{Background}

\subsubsection*{Control Flow Paths} A control flow path is a path in the control flow graph representation of a 
procedure. The start node of a \cfp is always the start node of the \cfg, however, the end node of a \cfp can be any node in the \cfg. 
We use the term ``\cfps that reach node n'' to refer to all \cfps that have $n$ as the end node. All \cfps referred henceforth are 
intra-procedural, unless explicitly stated otherwise.

\subsubsection*{Infeasible Control Flow Paths}
We use the following concepts related to infeasible paths~\cite{bodik1997refining}. 

We denote a \cfp by \text{$\rho: n_1\xrightarrow{e_1} n_2\xrightarrow{e_2} n_3\xrightarrow{e_3} \ldots \rightarrow 
n_p\xrightarrow{e_p} n_{p+1}$}, \text{$p\geq 2$}, where $n_i$ denotes the node at the $i^{th}$ position in the path, and $e_i$ denotes its 
out edge. \cfp $\rho$ is an \emph{infeasible control flow path} if
there is some conditional node $n_k,\,k\le p$ such that the subpath from $n_1$ to $n_k$ of $\rho$ is a prefix of some execution path
but the subpath from $n_1$ to $n_{k+1}$ is not a prefix of any execution path.\footnote{Note that
the subscripts used in the nodes and edges in a path segment signify positions of the nodes and edges in the path segment and
should not be taken as labels. By abuse of notation, we also use $n_1$, $e_2$ etc. as labels of nodes and edge in a 
control flow graph and then juxtapose them when we write path segments.}
A path segment \text{$\symmips: n_i\xrightarrow{e_i} \ldots \rightarrow n_k\xrightarrow{e_k} n_{k+1},\ e_i\not=e_k$} of \cfp $\rho$ above
is a \emph{minimal infeasible path segment} (\mips) if \symmips is not a subpath of any execution path but every subpath of
\symmips is a subpath of some execution path. Infeasibility occurs when inconsistent conditions are expected to hold at two different edges 
across a path segment. Hence every infeasible path must have at least two edges.\footnote{We assume that the infeasible paths arising from
the conditional statements of the form \emph{if}(\emph{cond}) where \emph{cond} is known to be either definitely zero or definitely non-zero, 
have been eliminated by the compiler in the \cfg.} 


 For \mips \symmips of the previous paragraph, we call all nodes $n_j,\ i<j<k+1$ as the \emph{intermediate} nodes of \symmips.
Additionally, we call the edges $e_i,\ e_k$ as the \mipsstart and the \mipsend edge of \symmips respectively, 
while all edges $e_j,\ i<j<k$ are called as the \mipsinner edges of \mips \symmips. As a rule, the end edge of a \mips is always a 
conditional edge, and inner edges exist only if a \mips has at least 3 edges. For the \fpmfp computation, we only admit \mips that do not 
contain cycles. All \mips referred henceforth are \mips without cycles.
  
\begin{figure}[t]
\setlength{\tabcolsep}{1pt}
\begin{tabular}{@{}c|c@{}}
\begin{tabular}{c}
\begin{minipage}{84mm}
Let $a$ be an integer variable, then the \emph{states} representing possible value range for variable $a$ at each edge over all executions are:
\begin{itemize}
\item $\states{e_0}{\{a\}}=\left\{\{a \mapsto  i\}\mid -\infty \leq i \leq \infty\right\}$
\item $\states{e_1}{\{a\}}=\left\{\{a \mapsto  0\}\right\}$
\item $\states{e_2}{\{a\}}=\left\{\{a \mapsto  i\}\mid -\infty \leq i \leq \infty\right\}$
\item $\states{e_3}{\{a\}}=\left\{\{a \mapsto  i\}\mid -\infty \leq i \leq \infty\right\}$
\item $\states{e_4}{\{a\}}=\left\{\{a \mapsto  i\}\mid 5 < i \leq \infty\right\}$
\item $\states{e_5}{\{a\}}=\left\{\{a \mapsto  i\}\mid -\infty \leq i \leq 5\right\}$
\end{itemize}

Path segment \text{$\symmips_1: n_1\xrightarrow{e_1} n_3\xrightarrow{e_3} n_4\xrightarrow{e_4} n_5$} is a \mips because
$e_4$ cannot be reached from $e_1$ in any execution although it can be reached from $e_3$ in some execution i.e., 
$\states{e_1}{\{a\}}\cap \states{e_4}{\{a\}}=\Phi$, 

$\states{e_3}{\{a\}}\cap\states{e_4}{\{a\}}=\set{\{a \mapsto  i\}\mid 5 < i \leq \infty}$

The \mipsstart, \mipsinner, and \mipsend edges of $\symmips_1$ are given by,
$\mipsstart(\symmips_1)=\set{e_1},\ \mipsinner(\symmips_1)=\set{e_3},\ \mipsend(\symmips_1)=\set{e_4}$
\end{minipage}
\end{tabular}
\;\;
&
\begin{tabular}{c}
\begin{pspicture}(-1,0)(35,48)
\small
\putnode{n0}{origin}{17}{50}{\framebox{$z<1$}}
	\putnode{w}{n0}{-7}{0}{$n_0$}

\putnode{n1}{n0}{-8}{-14}{\framebox{$a\!=\!0$}}
	\putnode{w}{n1}{-7}{0}{$n_1$}

\putnode{n3}{n1}{8}{-14}{\framebox{\em use a}}
	\putnode{w}{n3}{-7}{0}{$n_3$}

\putnode{n4}{n3}{0}{-14}{\framebox{$a\!>\!5$}}
	\putnode{w}{n4}{-7}{0}{$n_4$}

\putnode{n5}{n4}{-8}{-14}{\framebox{\em use a}}
	\putnode{w}{n5}{-7}{0}{$n_5$}

\putnode{n6}{n5}{15}{0}{\framebox{\white\em use a}}

\ncline{->}{n0}{n1}
\nbput[npos=0.4,labelsep=.5]{$e_0$}
\nbput[npos=0.7,labelsep=0.2]{\true}
\ncline[doubleline=true,doublesep=.2]{->}{n1}{n3}
\nbput[npos=0.35,labelsep=0]{$e_1$}
\nccurve[angleA=300,angleB=60]{->}{n0}{n3}
\naput[npos=0.5,labelsep=.5]{$e_2$}
\naput[npos=0.65,labelsep=0.2]{\false}
\ncline[doubleline=true,doublesep=.2]{->}{n3}{n4}
\nbput[npos=0.5,labelsep=.5]{$e_3$}
\ncline[doubleline=true,doublesep=.2]{->}{n4}{n5}
\naput[npos=0.5,labelsep=0.2]{$e_4$}
\nbput[npos=0.7,labelsep=0.2]{\true}
\ncline{->}{n4}{n6}
\naput[npos=0.33,labelsep=.2]{$e_5$}
\naput[npos=0.8,labelsep=.2]{\false}

\end{pspicture}
\end{tabular}
\end{tabular}
\caption{Illustrating minimal infeasible path segment (\mips).}

\label{fig:mips}
\end{figure}
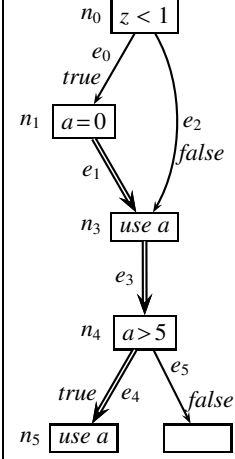

\begin{example}\label{example:mips.illustration}
Figure~\ref{fig:mips} illustrates infeasible paths and \mips. Observe that the \cfp 
$\displaystyle\rho: n_0 \xrightarrow{e_0} n_1\xrightarrow{e_1} n_3\xrightarrow{e_3} n_4\xrightarrow{e_4} n_5$ is infeasible but 
not minimal
. However, its suffix $\symmips_1$ is a \mips, \text{$\symmips_1: n_1\xrightarrow{e_1} n_3\xrightarrow{e_3} n_4\xrightarrow{e_4} n_5$}, 
because $\symmips_1$ is infeasible but no sub-segment of $\symmips_1$ is infeasible. 
\end{example}


\newcommand{\fold}{\text{\sf\em fold$_k$}\xspace}
\newcommand{\foldR}{\text{\sf\em fold$_k\mid s$}\xspace}

\newcommand{\pairname}{\text{pair}\xspace}
\newcommand{\pairnames}{\text{pairs}\xspace}
\subsection{Our Key Idea}

We define the following notations that allow us to describe the key ideas. For a given \mips $\symmips$, let $\mipsstart(\symmips)$, 
$\mipsinner(\symmips)$, and $\mipsend(\symmips)$ denote the set of start, inner, 
and end edges of $\symmips$. Sets $\mipsstart(\symmips)$ and $\mipsend(\symmips)$ are singleton because each \mips has exactly one 
start and one end edge. \mipss exhibit the following interesting property: 

\begin{observation}\label{observation:mips.block}
\cfps that contain a \mips \symmips as a sub-segment are infeasible. 
Consequently,
the data flow values computed along \cfps that contain \symmips are unreachable at the end edge of
\symmips, so these values should be blocked at the end edge. 
Thus, if we define a data flow analysis that separates the data flow values that 
are to be blocked, we can eliminate them thereby avoiding the effect of infeasible paths. 
\end{observation}

\begin{example}
In Figure~\ref{fig:mips}, a \cfp $\sigma: n_0\xrightarrow{e_0}n_1\xrightarrow{e_1}n_3\xrightarrow{e_3}n_4\xrightarrow{e_4}n_5$ is 
infeasible because it contains \mips $\symmips_1: n_1\xrightarrow{e_1} n_3\xrightarrow{e_3} n_4\xrightarrow{e_4} n_5$. Hence, the 
data flow value that reaches along $\sigma$ (e.g., \arangeZ, \zrangeNegativeZ) is blocked at $e_4$ (which is end edge of 
\mips $\symmips_1$).
\end{example}

Observation \ref{observation:mips.block} leads us to the following key idea to handle infeasible paths: at each program point, we 
separate the data flow values reaching along \cfps that contain a \mips, from the data flow values reaching along \cfps that do not 
contain any \mips. 
This allows us to discard these values at the end of the \mips (because these values are unreachable as per Observation 
\ref{observation:mips.block}).

\begin{table}
\center
\begin{tabular}{|c|c|r|r|r|}
\cline{3-5}

	\multicolumn{1}{c}{ }
  &\multicolumn{1}{c|}{ }
	& \multicolumn{2}{|c|}{Data Flow Values}
	& \multicolumn{1}{c|}{Meet} 
	\\ \cline{3-4}
	\multicolumn{1}{c}{ }
	&\multicolumn{1}{c|}{ }
	& \multicolumn{1}{|c|}{flowing through $\symmips_1$}
	& \multicolumn{1}{|c|}{not flowing through $\symmips_1$}
	& \multicolumn{1}{|c|}{$\sqcap$}
	\\ \cline{1-5}
	\multirow{4}{*}{\rotatebox{90}{\hspace*{-.3cm}Edges}}
	&$e_0$ 
	& $\top$
	& \zrangeNegativeZ
	& \zrangeNegativeZ
	\\ \cline{2-5}
	&$e_2$
	& $\top$
	& \zrangePositive
	& \zrangePositive
	\\ \cline{2-5}
	&$e_1$
	& \zrangeNegativeZ 
	& $\top$
	& \zrangeNegativeZ
	\\ \cline{2-5}
	&$e_3$
	& \zrangeNegativeZ
	& \zrangePositive
	& \zrangeII
	\\ \cline{2-5}
	&$e_4$
	& $\top$ \sout{\zrangeNegativeZ}
	& \zrangePositive
	& \zrangePositive
	\\ \cline{2-5}
	&$e_5$
	& \zrangeNegativeZ
	& \zrangePositive
	& \zrangeII
	\\ \cline{1-5}
\end{tabular}
\caption{Separating the Values Flowing through \mips $\symmips_1$ for example in Figure~\ref{fig:mips}: $\top$ represents top value of 
the lattice, the last column contains the result of meet of all data flow values present in a row at each edge, for simplicity values 
of only $z$ are shown.}
\label{tab:mips.separation}
\end{table}

For a \mips \symmips, following is a sufficient condition to block the data flow values that reach along \cfps that contain \symmips.
 \begin{quote}
	\czero: At the end edge of \symmips, block the data flow values that \emph{flow through} \symmips (i.e., values that flow along 
	\symmips\textemdash from start edge of \symmips till the end edge of \symmips).
 \end{quote}

\begin{example}
For example in Figure~\ref{fig:mips},  the values (of variable $z$) that \emph{flow through} \mips \symmipsO are shown in 
Table~\ref{tab:mips.separation}. The value \zrangeNegativeZ reaches the start edge ($e_1$) of \symmipsO, and flows 
through \symmipsO. This value is separated from the values that do not flow through \symmipsO e.g., \zrangePositive. This 
separation allows us to identify and block the value $\zrangeNegativeZ$ at the end edge ($e_4$) of \symmipsO. 

The final value at each edge, except the end edge, is computed by taking a meet of values that \emph{flow through} 
\symmipsO, with values that do not flow through \symmipsO. However, the final value at end edge contains only the 
values that do not flow through \symmipsO. Thus, we get a precise value range for $z$ at edge $e_4$ (i.e., \zrangePositive) 
because value \zrangeNegativeZ is blocked (a \mfp solution that includes data flow values along infeasible \cfps gives value 
range \zrangeII at edge $e_4$).
\end{example}

Note that the data flow values that reach the start edge of a \mips \symmips may change or get killed at intermediate nodes or 
edges when they flow through \symmips, depending on the type of data flow analysis. In this case, we block the updated data flow 
value accordingly, as illustrated in example \ref{ex:data.flowthrough} below.

\begin{example}\label{ex:data.flowthrough}
Consider a \mips \symmips: $n_1\xrightarrow{e_1} n_2\xrightarrow{e_2} n_3\xrightarrow{e_3} n_4$ with $e_1,\ e_2,\ e_3$ as the 
start, the inner, and the end edge respectively. Let $x$ be an integer variable such that $n_1: x=0$, $n_2: x=x+1$ 
be two assignment nodes, and nodes $n_3, n_4$ do not modify the value of the variable $x$. 

The data flow value \xrangeZ at the start edge ($e_1$) of \symmips will be changed to \xrangeO as it flows through \symmips. 
Hence \xrangeO is blocked at $e_3$. 
\end{example}

\subsection{An Overview of \fpmfp computation}\label{sec:fpmfp.overview}
We now briefly explain the steps of \fpmfp computation that ensure that condition \czero (from the previous section) is satisfied 
for all \mips in the program. The formal details of \fpmfp computation are presented in sections \ref{sec:psmfp.definition} and 
\ref{sec:Computation}. 

We first explain the \fpmfp computation in presence of single \mips in the program. Later, we introduce the necessary changes
 to handle multiple \mipss in the program. We use the example in Figure~\ref{fig:intersecting.mipsone}a as a running example 
for explaining \fpmfp computation. Two \mips in Figure~\ref{fig:intersecting.mipsone}a are 
$\symmipsO:e_3\rightarrow e_4\rightarrow e_5$ and $\symmipsT:e_3\rightarrow e_4\rightarrow e_8$ (They are marked with double 
arrows). For simplicity of exposition, we restrict the discussion to computation of value range of variable $l$ only.

In \fpmfp computation for the example in Figure~\ref{fig:intersecting.mipsone}a, we separate the value \rangeZ that flows 
through \symmipsO from the value \rangeT that does not flow through \symmipsO as shown in Figure~\ref{fig:intersecting.mipsone}b. 
This allows us to block \rangeZ at the end edge ($e_5$) of \symmipsO. Thus, the \fpmfp solution \textemdash computed by meet of 
data flow values in each row at each edge \textemdash contains \rangeT (a precise value) at edge $e_5$. 

The \fpmfp computation explained in previous paragraph becomes imprecise if we additionally separate the data flow values that 
flow through \mips \symmipsT, as explained below. For \mips \symmipsT: $e_3\rightarrow e_4\rightarrow e_8$, we separate the 
values that flow through \symmipsT (i.e., \rangeZ) as shown in Figure~\ref{fig:intersecting.mipsone}c. This allows us to block 
\rangeZ at end edge ($e_8$) of \symmipsT. However, the \fpmfp solution (computed in Figure~\ref{fig:intersecting.mipsone}c) 
is imprecise because it includes the value \rangeZ at edge $e_5$ and $e_8$. In particular, at edge $e_5$, \rangeZ is blocked 
within \symmipsO but it is retained in \symmipsT because $e_5$ is not end edge of \symmipsT. Similarly, at edge $e_8$, 
\rangeZ is blocked within \symmipsT but is retained in \symmipsO.

\begin{figure*}[t!]
\psset{unit=1mm}
\centering
\small
\begin{tabular}{l|l}
\begin{tabular}{@{}c@{}}
\begin{pspicture}(-126,-83)(-90,8)
\putnode{n01}{origin}{-107}{7}{\psframebox{$\begin{array}{@{}l@{}}
				 \emph{l=2} \\
				 \emph{input c} \\
				 \emph{if$(p1>0)$}
				\end{array}$}}

\putnode{n02}{n01}{-15}{-20}{\psframebox{\!\!\!$\begin{array}{c}l=0\\c=0\end{array}$\!\!\!}}
	
\putnode{n03}{n01}{0}{-30}{\psframebox{\emph{print l}}}

\putnode{n04}{n03}{0}{-15}{\psframebox{$\emph{switch$(c)$}$}}
	
\putnode{n05}{n04}{-14}{-22}{\psframebox{\!\!\!$\begin{array}{c}assert\\(l!\!=0)\end{array}$\!\!\!}}

\putnode{n06}{n04}{11}{-22}{\psframebox{\!\!\!$\begin{array}{c}assert\\(l!\!=0)\end{array}$\!\!\!}}
				
\putnode{exit01}{n04}{0}{-38}{\psframebox{\emph{exit}}}
	

\ncline{->}{n01}{n02}
	\nbput[npos=.2,labelsep=.2]{$e_1$}
	\nbput[npos=.6,labelsep=.5]{true}
	
\ncline{->}{n01}{n03}
	\naput[npos=.2,labelsep=.2]{false}
	\nbput[npos=.35,labelsep=.5]{$e_2$}
	
\ncline[doubleline=true]{->}{n02}{n03}
	\nbput[npos=.6,labelsep=.5]{$e_3$}
	
\ncline[doubleline=true]{->}{n03}{n04}
	\nbput[npos=.6,labelsep=.5]{$e_4$}
	
\ncline[doubleline=true]{->}{n04}{n05}
	\nbput[npos=.9,labelsep=.1]{$e_5$}
	\nbput[npos=.45,labelsep=.1,nrot=0]{case 1}
	                            
\ncline[doubleline=true]{->}{n04}{n06}
	\naput[npos=.9,labelsep=.1]{$e_8$}
	\naput[npos=.45,labelsep=.01,nrot=0]{case 2}
	                            
\ncline{->}{n04}{exit01}

	\nbput[npos=.2,labelsep=.5]{$e_7$}
	\nbput[npos=.45,labelsep=.5]{\rotatebox{90}{default}}
	                              
\ncline{->}{n05}{exit01}
	\nbput[npos=.6,labelsep=.5]{$e_6$}
	
\ncline{->}{n06}{exit01}
	\naput[npos=.6,labelsep=.5]{$e_9$}

\end{pspicture}
\end{tabular}
&
\hspace*{-10pt}
\begin{tabular}{l}
\setlength{\tabcolsep}{0.1pt}
\begin{tabular}{@{}l@{}}  
$\begin{array}{l}
  \text{Paths and \mipss:}\\
 \bullet\ \mips\ \symmips_1:e_3\rightarrow e_4\rightarrow e_5\\
 \bullet\ \mips\ \symmips_2:e_3\rightarrow e_4\rightarrow e_8\\
 \bullet\ \mipsstart(\symmips_1)=\mipsstart(\symmips_2)=e_3\\
 \bullet\ \mipsinner(\symmips_1)=\mipsinner(\symmips_2)=e_4\\
 \bullet\ \mipsend(\symmips_1)=e_5, \mipsend(\symmips_2)=e_8\\
\end{array}$
\end{tabular}
\\
\\
\begin{tabular}{|c|r|r|r|}
\cline{1-4}
	\multicolumn{1}{|c}{\multirow{2}{*}{\rotatebox{90}{\hspace*{-0.5cm}Edges}}}
	& \multicolumn{2}{|c|}{Data Flow Values}
	& \multicolumn{1}{c|}{\fpmfp} 
	\\ \cline{2-3}
	\multicolumn{1}{|c}{ }
	& \multicolumn{1}{|c|}{flowing}
	& \multicolumn{1}{|c|}{not flowing}
	& \multicolumn{1}{|c|}{$\sqcap$}
	\\ 
	\multicolumn{1}{|c}{ }
	
	& \multicolumn{1}{|c|}{through $\symmips_1$}
	& \multicolumn{1}{|c|}{through $\symmips_1$}
	& 
	\\ \cline{1-4}
	$e_3$ 
	& \rangeZ
	& $\top$
	& \rangeZ
	\\ \cline{1-4}
	$e_4$
	& \rangeZ
	& \rangeT
	& \rangeZT
	\\ \cline{1-4}
	$e_5$
	& $\top$ 
	& \rangeT
	& \rangeT
	\\ \cline{1-4}
	$e_8$
	& \rangeZ
	& \rangeT
	& \rangeZT
	\\ \cline{1-4}
\end{tabular}
\\\multicolumn{1}{c}{(b)}
\\
\\
\begin{tabular}{|c|r|r|r|r|}
\cline{1-5}
	\multicolumn{1}{|c}{\multirow{2}{*}{\rotatebox{90}{\hspace*{-1cm}Edges}}}
	& \multicolumn{3}{|c|}{Data Flow Values}
	& \multicolumn{1}{|c|}{\fpmfp} 
	\\ \cline{2-4}
	\multicolumn{1}{|c|}{ }
	& \multicolumn{2}{|c|}{flowing through}
	& \multicolumn{1}{|c|}{not flowing}
	& \multicolumn{1}{|c|}{$\sqcap$}
	\\ 
	\multicolumn{1}{|c|}{ }
	& \multicolumn{2}{|c|}{\mips}
	& \multicolumn{1}{|c|}{through}
	& 
	\\ \cline{2-3}
	  \multicolumn{1}{|c}{ }
	& \multicolumn{1}{|c|}{$\symmips_1$}
	& \multicolumn{1}{|c|}{$\symmips_2$}
	& \multicolumn{1}{|c|}{\mips}
	&
	\\ \cline{1-5}
	$e_3$ 
	& \rangeZ
	& \rangeZ
	& $\top$
	& \rangeZ 
	\\ \cline{1-5}
	$e_4$
	& \rangeZ 
	& \rangeZ
	& \rangeT
	& \rangeZT
	\\ \cline{1-5}
	$e_5$
	& $\top$
	& \rangeZ
	& \rangeT
	& \rangeZT
	\\ \cline{1-5}
	$e_8$
	& \rangeZ
	& $\top$
	& \rangeT
	& \rangeZT
	\\ \cline{1-5}
\end{tabular}
\end{tabular}
\\
\multicolumn{1}{c}{(a)}&\multicolumn{1}{c}{(c)}
\end{tabular}
\caption{Computing the \fpmfp solution for an example having overlapping \mipss.}
\label{fig:intersecting.mipsone}
\end{figure*}

We explicate the reasons for this imprecision below: we separate the data flow values that reach along \cfps that contain a \mips, 
however, a \mips can overlap with other \mips in a way that both \mips follow the same \cfp resulting in the same data flow values 
flowing through them. In such a case, at the end edge of a \mips, the data flow along that \mips will be blocked but the flow 
along the other overlapping \mips will be allowed unless the edge is end edge of both the \mips. This leads to imprecision. 
Hence we need to handle the overlapping \mips case explicitly. For this purpose, we define a \emph{\finto} relation between \mips below.

\begin{definition}(Contains-prefix-of (\cpon)) \label{definition.cpo}
For a \mips \symmips and an edge $e$ in \symmips, let \prefix{\symmips}{e} be the subsegment of \symmips from $\mipsstart(\symmips)$ 
to $e$.  Then, 
we say that  \mips \symmipsO \finto  \mips \symmipsT at $e$, iff \symmipsO contains \prefix{\symmipsT}{e}.
\end{definition}

For example in Figure~\ref{fig:intersecting.mipsone}, the \mips \symmipsO \finto \symmipsT\footnote{In this case \symmipsT also 
\finto \symmipsO at $e_3$ and $e_4$, however in general \finto may not be a symmetric relation.} at edges $e_3$ and $e_4$. For a 
\mips \symmipsO and an edge $e$, all \mips \symmipsT that satisfy \symmipsO \CPO \symmipsT are obtained as follows: let \setmips 
be the set of all \mips in the \cfg of the program.
  \begin{align}\label{eq:cpo.one}
  \cpo{e}{\symmipsO}&=\set{\symmipsT\mid \symmipsT\in \setmips,\ \symmipsO\ \text{contains}\ \prefix{\symmipsT}{e}}
  \end{align}

We observe the following property of \cpo{e}{\symmipsO}.
\begin{observation}\label{observation:cpo}
Because of the prefix relation, all \mips in \cpo{e}{\symmipsO} essentially follow the same control flow path till edge $e$ 
although their start edges may appear at different positions in the path. Next, the values along this path flow through each 
\mips in \cpo{e}{\symmipsO}. Hence, these values should be blocked at the end edge of every \mips in \cpo{e}{\symmipsO}. 
\end{observation}

For an edge sequence $e\rightarrow e'$, \cpo{e}{\symmips} may be different from \cpo{e'}{\symmips} because all \mips that reach $e$ may 
not reach $e'$, and some new \mips may start at $e'$. Hence, we define \cpobn function below that computes \cpo{e'}{\symmips}
 from \cpo{e}{\symmips}.
 
Let \edges{\symmips} be the set of all edges in \symmips, and $\setS=\cpo{e}{\symmips}$ then  
  \begin{align}\label{eq:cpo.set}
  \cpob{e'}{\setS}&= \set{\symmips\mid \symmips\in \setS,\,e'\in \edges{\symmips}}\ \cup\  \set{\symmips\mid \symmips\in\setmips,\ 
	e'\in\mipsstart(\symmips)}
  \end{align}
	
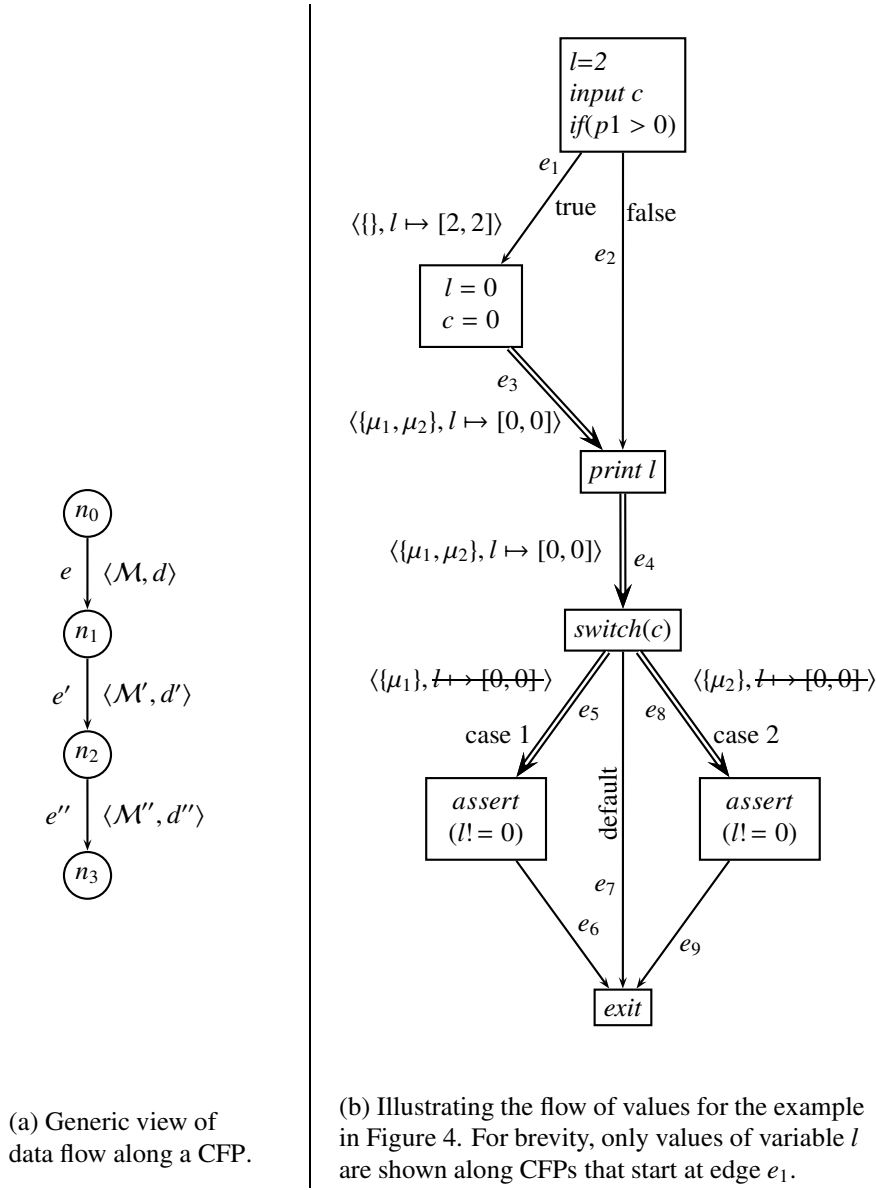
\begin{figure}[t!]
\begin{tabular}{l|l}
\begin{tabular}{c}
\begin{pspicture}(0,0)(35,52)
\putnode{n0}{origin}{10}{30}{\pscirclebox{$n_0$}}
\putnode{n1}{n0}{0}{-16}{\pscirclebox{$n_1$}}
\putnode{n2}{n1}{0}{-16}{\pscirclebox{$n_2$}}
\putnode{n3}{n2}{0}{-16}{\pscirclebox{$n_3$}}
\ncline{->}{n0}{n1}
\nbput{$e$}
\naput{$\pairc{\setm}{d}$}
\ncline{->}{n1}{n2}
\nbput{$e'$}
\naput{$\pairc{\setm'}{d'}$}
\ncline{->}{n2}{n3}
\nbput{$e''$}
\naput{$\pairc{\setm''}{d''}$}
\end{pspicture}
\end{tabular}
&
\begin{tabular}{c}
\begin{pspicture}(0,-83)(-72,60)
\putnode{n01}{origin}{-35}{48}{\psframebox{$\begin{array}{@{}l@{}}
				 \emph{l=2} \\
				 \emph{input c} \\
				 \emph{if$(p1>0)$}
				\end{array}$}}

\putnode{n02}{n01}{-20}{-28}{\psframebox{$\begin{array}{c}l=0\\ c=0\end{array}$}}
	
\putnode{n03}{n01}{0}{-50}{\psframebox{\emph{print l}}}

\putnode{n04}{n03}{0}{-21}{\psframebox{$\emph{switch$(c)$}$}}
	
\putnode{n05}{n04}{-18}{-25}{\psframebox{$\begin{array}{c}assert\\ (l!\!=0)\end{array}$}}

\putnode{n06}{n04}{18}{-25}{\psframebox{$\begin{array}{c}assert\\ (l!\!=0)\end{array}$}}
				
\putnode{exit01}{n04}{0}{-50}{\psframebox{\emph{exit}}}
	

\ncline{->}{n01}{n02}
	\nbput[npos=.2,labelsep=.2]{$e_1$}
	\naput[npos=.4,labelsep=.5]{true}
	\nbput[npos=0.8]{\pairc{\set{}}{\rangeT}}
	
\ncline{->}{n01}{n03}
	\naput[npos=.2,labelsep=.2]{false}
	\nbput[npos=.35,labelsep=.5]{$e_2$}
	
\ncline[doubleline=true]{->}{n02}{n03}
	\nbput[npos=.2,labelsep=.5]{$e_3$}
	\nbput[npos=.6,labelsep=-0.4]{\pairc{\setmuOT}{\rangeZ}}
	
\ncline[doubleline=true]{->}{n03}{n04}
	\naput[npos=.6,labelsep=.5]{$e_4$}
	\nbput{\pairc{\setmuOT}{\rangeZ}}

\ncline[doubleline=true]{->}{n04}{n05}
	\naput[npos=.4,labelsep=.1]{$e_5$}
	\nbput[npos=.75,labelsep=.1,nrot=0]{case 1}
	\nbput[npos=0.4]{\pairc{\setmuO}{\text{\sout{\rangeZ}}}}
                            
\ncline[doubleline=true]{->}{n04}{n06}
	\nbput[npos=.4,labelsep=.1]{$e_8$}
	\naput[npos=.75,labelsep=.01,nrot=0]{case 2}
	\naput[npos=0.4]{\pairc{\setmuT}{\text{\sout{\rangeZ}}}}
                            
\ncline{->}{n04}{exit01}

	\nbput[npos=.7,labelsep=.5]{$e_7$}
	\nbput[npos=.45,labelsep=.5]{\rotatebox{90}{default}}
	                              
\ncline{->}{n05}{exit01}
	\naput[npos=.6,labelsep=.5]{$e_6$}
	
\ncline{->}{n06}{exit01}
	\naput[npos=.6,labelsep=.5]{$e_9$}

\end{pspicture}
\end{tabular}
\\
$\begin{array}{l}
\text{(a) Generic view of }\\
\text{data flow along a \cfp.}
\end{array}$ 
& 
$\begin{array}{l} 
\text{(b) Illustrating the flow of values for the example}\\ 
\text{in Figure~\ref{fig:intersecting.mipsone}. For brevity, only values of variable $l$ }\\
\text{are shown along \cfps that start at edge $e_1$.}
\end{array}$
\end{tabular}
\caption{Computing the \fpmfp solution}
\label{fig:abstract.eff}
\end{figure}

At an edge $e$ of \mips \symmips, we associate the data flow values that flow through \symmips with \cpo{e}{\symmips} (instead of \symmips).
These associations are abstractly illustrated in  Figure~\ref{fig:abstract.eff}a and instantiated to an example in 
Figure~\ref{fig:abstract.eff}b. We explain each of these below. 

In Figure~\ref{fig:abstract.eff}a, the data flow values $d,\, d'\!,\, d''$ reach edges $e,\, e'\!,\, e''$ respectively, along the 
following \cfp: $\sigma: n_0\xrightarrow{e}n_1\xrightarrow{e'} n_2\xrightarrow{e''}n_3$. The data flow value $d'=f_{n1}(d)$ and 
$d''=f_{n2}(d')$ where $f_{ni}(x)$ computes the effect of node $n_i$ on input data flow value $x$. The corresponding associations 
with sets of \mips $\setm,\, \setm',\,\setm''$ are computed as follows: 
\begin{align*}
\setm &=  \cpob{e}{\set{}}\\
\setm' &= \cpob{e'}{\setm}\\
\setm'' &= \cpob{e''}{\setm'}
\end{align*}  
The idea is that a \pairname $\pairc{\setm}{d}$ at edge $e$ transforms into another \pairname $\pairc{\setm'}{d'}$ at a successor 
edge $e'$ because
\begin{itemize}
  \item the \CPO relation between \mips at $e'$ may be different from that at $e$, and
 \item $d'$ is computed from $d$ as $d'=f_{n1}(d)$.
\end{itemize}

In Figure~\ref{fig:abstract.eff}b, the data flow values \rangeT at edge $e_1$ is associated with \set{} because no \mips 
contains $e$. However, at edges $e_3,\, e_4$, the data flow value \rangeZ is associated with \setmuOT because \symmipsO \CPO \symmipsT 
at these edges i.e., $\cpob{e_3}{\setmuO}=\setmuOT$, $\cpob{e_4}{\setmuOT}=\setmuOT$. At $e_5$, \rangeZ is associated with \setmuO because 
$\cpob{e_5}{\setmuOT}=\setmuO$. Since $e_5$ is end edge of \symmipsO, \rangeZ is blocked at $e_5$. Similarly, \rangeZ is blocked at 
$e_8$. Rest of the associations at each edge are presented in Table~\ref{tab:mips.intersecting}. 

\begin{table}[t]
\center
\begin{tabular}{|c|r|r|r|}
\cline{1-4}
Edges & Associations ($\pairc{\setm}{d}$) & \fpmfp & \mfp\\ \cline{1-4}
$e_1$ & $\pairc{\set{}}{\rangeT}$&\rangeT&\rangeT\\\cline{1-4}
$e_2$ & $\pairc{\set{}}{\rangeT}$&\rangeT&\rangeT\\\cline{1-4}
$e_3$ & $\pairc{\set{\symmipsO,\symmipsT}}{\rangeZ}$&\rangeZ&\rangeZ\\\cline{1-4}
$e_4$ & $\pairc{\set{\symmipsO,\symmipsT}}{\rangeZ},\ \pairc{\set{}}{\rangeT}$&\rangeZT&\rangeZT\\\cline{1-4}
$e_5$ & $\pairc{\set{\symmipsO}}{\text{\sout{\rangeZ}}}, \pairc{\set{}}{\rangeT}$&\rangeT&\rangeZT\\\cline{1-4}
$e_7$ & $\pairc{\set{}}{\rangeZT}$&\rangeZT&\rangeZT\\\cline{1-4}
$e_8$ & $\pairc{\set{\symmipsT}}{\text{\sout{\rangeZ}}}, \pairc{\set{}}{\rangeT}$&\rangeT&\rangeZT\\\cline{1-4}
\end{tabular}
\caption{The \fpmfp computation for example in Figure~\ref{fig:intersecting.mipsone}. A \pairname $\pairc{\setm}{d}$ indicates that 
the data flow value $d$ flows through all \mips in set of \mips \setm. The final \fpmfp solution is computed by taking meet of values 
over all \pairnames at each edge.}
\label{tab:mips.intersecting}
\end{table}

We now introduce the  notations used in the \fpmfp computation. 
We compute a set of \pairnames of the form $\pairc{\setm}{d}$ at each edge $e$ where \setm is a \emph{set of \mips}, and $d$ is 
the data flow value that flows through all \mips in \setm. For completeness, we have a \pairname $\pairc{\setm'}{d}$ corresponding to 
each possible subset $\setm'$ of \setmips at each edge $e$, where $d=\top$ if the \mips in $\setm'$ are not related with each other 
by \cpon relation i.e., there is no \mips \symmips in $\setm'$ such that 
$\cpo{e}{\symmips}=\setm'$.

At an edge $e$, presence of a pair $\pairc{\setm}{d},\ d\not=\top$ indicates the data flow value $d$ flows through each \mips 
contained in \setm, and for some $\symmips\in \setm, \setm=\cpo{e}{\symmips}$. Similarly, a pair $\pairc{\nomips}{d'}$ indicates 
that the data flow value $d'$ does not flow through any \mips at edge $e$. 

\begin{example}
Figure \ref{fig:motivating_example2} illustrates the \fpmfp computation for the example in Figure \ref{motivating_example}. Since 
\symmips is the only \mips in the example, each data flow value $d$ is mapped with either $\set{\symmips}$ or \nomips. The data flow 
value \arangeZ reaches the start edge ($e_3$) of \symmips, hence, a \pairname \pairc{\set{\symmips}}{\arangeZ} is created at 
$e_3$. Similarly, at edge $e_4$, data flow value \arangeF is mapped with \nomips i.e., a \pairname \pairc{\nomips}{\arangeF} is created. 

Edge $e_6$ is end edge of \symmips, hence, the data flow value \arangeZ that is mapped with \symmips in \pairname 
\pairc{\set{\symmips}}{\arangeZ} is discarded i.e., replaced with $\top$ (the resulting \pairname \pairc{\set{\symmips}}{\top} 
at $e_6$ is not shown in figure for brevity).
\end{example}

For a program containing $k$ \mips, theoretically we need $2^k$ \pairnames\footnote{In Section~\ref{sec:complexity.analysis}, we prove that 
at any program point at most $k$ \pairnames  can contain the data flow values other than $\top$, 
where $k$ is the total number of input \mips. Further, our empirical evaluation shows that the number of such \pairnames to be much 
smaller than $k$. Moreover, 
ignoring  the \pairnames that contain $\top$ value does not affect the \fpmfp solution (details are given in Section~\ref{sec:remove.top}).} 
at each program point (one pair corresponding to 
each possible subset of \emph{set of all \mips} present in the program). 
The pairs computed in a \fpmfp solution satisfy the following property:
\begin{quote}
At each program point, the meet of data flow values over all \pairnames gives at least as precise information as in \mfp solution, 
and may give more precise information if some \pairnames contain information reaching from infeasible \cfps{} because such information 
is discarded. The precision is achieved only when the discarded information has weaker value or incomparable value compared to rest of 
the information reaching a node. Let $D_n$ be the data flow value in \mfp solution at node $n$, and \setSn be the set of \pairnames 
computed in \fpmfp analysis at node $n$ then
\begin{align*}
D_n \sqsubseteq \bigsqcap_{\pairc{\setm}{d}\in \setSn} d
\end{align*}
\end{quote}

\begin{example}
In Figure \ref{fig:motivating_example2}, at the edge $e_5$ the meet of data flow values over all \pairnames i.e.,

\text{$\pairc{\set{\symmips}}{\arangeZ}$} and $\pairc{\nomips}{\arangeF}$ is \arangeZF which is same as given by \mfp solution in 
Figure \ref{motivating_example}; however, at the edge $e_6$ the meet of data flow values over all \pairnames i.e., 
$\pairc{\set{\symmips}}{\top}$ and $\pairc{\nomips}{\arangeF}$ is \arangeF which is more precise compared to \arangeZF given by 
\mfp solution in Figure \ref{motivating_example}. 
\end{example}

\subsection{Defining \fpmfp solution}
\label{sec:psmfp.definition}
We now explain how these ideas are incorporated in a data flow analysis. We compute the \fpmfp solution in the following two steps. 
\begin{enumerate}
\item \emph{Step 1}: we lift a data flow analysis to an analysis that computes separate data flow values for different \mipss 
(Equation \ref{eq:dfe21} and \ref{eq:dfe22} define the lifted analysis). The lifted analysis blocks the data flow values associated 
with each \mips \symmips at the end edge of \symmips.
\item \emph{Step 2}: we merge the data flow values that are computed by the lifted analysis (over all pairs at each program point) 
to obtain the \fpmfp solution (Equation \ref{eq:dfe01} and \ref{eq:dfe02} define the merging of data flow values).
\end{enumerate}

We describe each of these steps in detail now. Let \lat denote the lattice of data flow values of 
the underlying data flow analysis. Then the \mfp solution of the data flow analysis for a program is a set of values in \lat, 
represented by data flow variables $\In_n/\Out_n$ (Equation \ref{eq:dfe11} and \ref{eq:dfe12}) for every node $n$ of the \cfg of 
the program, when the equations are solved using $\top$ as the initial value of $\In_n/\Out_n$. Assume we are given a set \setmips 
that contains all \mipss in the \cfg of the program, then each $\setm\subseteq\setmips$ could be associated with any data flow 
value $d$ in \lat, if $d$ flows through each \mips contained in \setm. These associations are represented by the elements of the 
lattice $\hlat: \mpowerset\rightarrow \lat $, where \mpowerset is the powerset of \setmips.

We lift the underlying analysis that computes values $\In_n/\Out_n\in \lat$ to an analysis that computes $\bIn_n/\bOut_n\in \hlat$. 
Thus $\bIn_n$, $\bOut_n$ are sets of \pairnames
$\pairc{\setm}{d}$ such that $\setm \subseteq \setmips$, $d \in \lat$. Each of these sets represent a \emph{function} that maps each 
value in \mpowerset to a value in \lat i.e., $\In_n/\Out_n$ cannot contain two pairs $\pairc{\setm}{d}$ and $\pairc{\setm'}{d'}$ where 
$\setm=\setm'$ but $d\not=d'$.
This collection of \pairnames gives us the distinctions required to achieve precision by separating the data flow values along 
infeasible paths from those along feasible paths. 

For a data flow value $\bIn_n$, the meaning of association of a data flow value $d$ with a set of \mips \setm is as follows:
\begin{itemize}
 \item Pair $\pairc{\setm}{d} \in \bIn_n$, indicates that the data flow value $d$ flows through all \mips in \setm. If no such 
value exists at $n$, then $d=\top$.
 \item Pair $\pairc{\nomips}{d} \in \bIn_n$, indicates that the data flow value $d$ does not flow through any \mips at node $n$. 
If no such value exists (i.e., every data flow value flows through some or the other \mips) then $d=\top$.
\end{itemize}
 
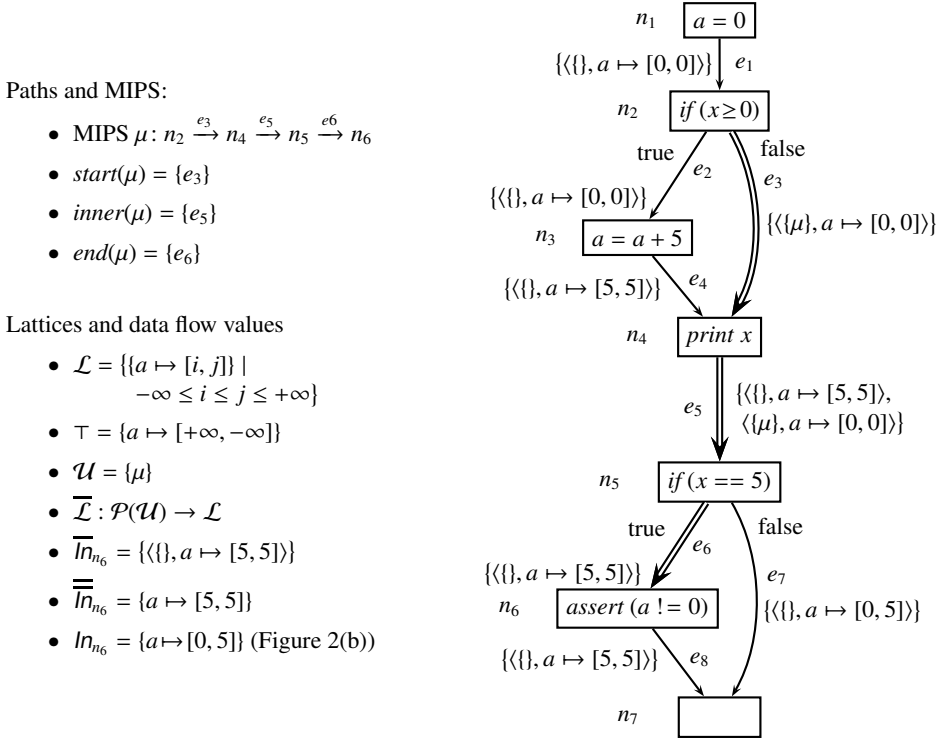
\begin{figure*}[t!]
\centering
\psset{unit=1.2mm}
\small
\begin{tabular}{@{}cc@{}}
\begin{tabular}{@{}c@{}}
\begin{minipage}{62mm}
Paths and \mipss:
\begin{itemize}
\item \mips $\symmips\!: n_2 \xrightarrow{e_3}  n_4 \xrightarrow{e_5}   n_5 \xrightarrow{e6} n_6$
\item $\mipsstart(\symmips)  =  \{ e_3 \}$
\item $\mipsinner(\symmips)  =  \{ e_5 \}$
\item $\mipsend(\symmips)    =  \{ e_6 \}$
\end{itemize}

\bigskip

Lattices and data flow values
\begin{itemize}
\item $\lat=\begin{array}[t]{@{}l@{}l} \big\{\{ & a\mapsto [i,j]\}\mid 
		\\
		& -\infty\leq i\leq j\leq +\infty\big\}
		\end{array}$
\item $\top=\{a\mapsto [+\infty,-\infty]\}$
\item $\setmips=\set{\symmips}$
\item $\hlat:\mpowerset \rightarrow \lat$
\item $\bIn_{n_6} = \bset{\pairc{\nomips}{\arangeF}}$ 
\item $\pIn_{n_6} = \set{\arangeF}$ 
\item $\In_{n_6} = \zerofive$ (Figure~\ref{motivating_example}(b))
\end{itemize}
\end{minipage}
\end{tabular}
&
\begin{tabular}{@{}c@{}}

\begin{pspicture}(-116,-70)(-56,8)

\putnode{entry1}{origin}{-93}{7}{\psframebox{$a=0$}}
	\putnode{w}{entry1}{-8}{0}{$n_1$}
\putnode{c11}{entry1}{0}{-10}{\psframebox{\em if $(x\!\geq\!0)$}}
	\putnode{w}{c11}{-10}{0}{$n_2$}
\putnode{n11}{c11}{-9}{-14}{\psframebox{$a=a+5$}}
	\putnode{w}{n11}{-10}{0}{$n_3$}
\putnode{n31}{n11}{9}{-11}{\psframebox{\em print $x$}}
	\putnode{w}{n31}{-9}{0}{$n_4$}
\putnode{c21}{n31}{0}{-16}{\psframebox{\em if $(x==5)$}}
	\putnode{w}{c21}{-12}{0}{$n_5$}
\putnode{n51}{c21}{-9}{-14}{\psframebox{\em assert $(a\ !\!=0)$}}
	\putnode{w}{n51}{-14}{0}{$n_6$}
\putnode{exit1}{n51}{9}{-12}{\psframebox{\white print x}}
	\putnode{w}{exit1}{-10}{0}{$n_7$}
%
\ncline{->}{entry1}{c11}
	\naput{$e_1$}
	\nbput[npos=.5,labelsep=.5]{\bset{\pairc{\nomips}{\arangeZ}}}
\ncline{->}{c11}{n11}
	\naput[npos=.35,labelsep=.5]{$e_2$}
\nbput[npos=.37,labelsep=1]{true}
	\nbput[npos=.9,labelsep=.5]{\bset{\pairc{\nomips}{\arangeZ}}}
\nccurve[doubleline=true,angleA=300,angleB=60]{->}{c11}{n31}
	\naput[npos=.3,labelsep=.5]{$e_3$}
		\naput[npos=.15,labelsep=1]{false}
	\naput[npos=.5,labelsep=.5]{\bset{\pairc{\set{\symmips}}{\arangeZ}}}
\ncline{->}{n11}{n31}
	\naput[npos=.6,labelsep=.5]{$e_4$}
	\nbput[npos=.3,labelsep=.5]{\bset{\pairc{\nomips}{\arangeF}}}

\ncline[doubleline=true]{->}{n31}{c21}
	\nbput[labelsep=1]{$e_5$}
	\naput[npos=.5,labelsep=1]{$\begin{array}{@{}l@{}}
	\big\{\pairc{\nomips}{\arangeF},\\
	\ \pairc{\set{\symmips}}{\arangeZ}\big\}\end{array}$}
\ncline[doubleline=true]{->}{c21}{n51}
	\naput[npos=.4,labelsep=.2]{$e_6$}
	\nbput[npos=.47,labelsep=1]{true}
	\nbput[npos=.99,labelsep=.2]{\bset{\pairc{\nomips}{\arangeF}}}
\nccurve[angleA=300,angleB=60]{->}{c21}{exit1}
	\naput[npos=.4,labelsep=1]{$e_7$}
	\naput[npos=.17,labelsep=1]{false}
	\naput[npos=.55,labelsep=.5]{\bset{\pairc{\nomips}{\arangeZF}}}
\ncline{->}{n51}{exit1}
	\naput[npos=.6,labelsep=.5]{$e_8$}
	\nbput[npos=.25,labelsep=.5]{\bset{\pairc{\nomips}{\arangeF}}}
	
\end{pspicture}
\end{tabular}
\end{tabular}
\caption{Computing \fpmfp solution for the example of Figure~\ref{motivating_example}.
The lifted data flow analysis computes a set of pairs $\bIn_n/\bOut_n$ at each node $n$ such that 
      a \pairname \pairc{\setm}{d}, indicates $d$  flows through each \mips $m\in \setm$. 
      if $\setm=\nomips$ then $d$ does not flow through any \mips at node $n$. Figure~\ref{motivating_example} contains 
			single \mips \symmips, hence, two pairs are computed (corresponding to $\set{\symmips}$ and $\set{}$) at each program 
			point, for brevity, we have only shown pairs where $d\not=\top$.
}
\label{fig:motivating_example2}

\end{figure*}

\begin{example}
 In Figure \ref{fig:motivating_example2}, the data flow value \arangeZ flows through \mips \symmips at edges $e_3,\ e_5,$ and $e_6$. 
On the other hand, the data flow value \arangeF does not flow through any \mips. (Recall that flow through a \mips \symmips means 
flow along \symmips\textemdash from start edge of \symmips till end edge of \symmips).
\end{example}

Section~\ref{sec:Computation} defines the data flow equations to compute $\bIn_n/\bOut_n$.
The \fpmfp solution is a set of values in \lat, represented by the data flow variables $\bbIn_n/\bbOut_n$ (Equation \ref{eq:dfe01} 
and \ref{eq:dfe02}) for every node $n$ in the \cfg of the program. 
They are computed from $\bIn_n/\bOut_n$ using the following operations.
\begin{align}
 \bbIn_n&=\foldmeet(\bIn_n) \label{eq:dfe01}\\
 \bbOut_n&=\foldmeet(\bOut_n)\label{eq:dfe02}\\\label{foldeq}
 where,\ \foldmeet(\setS) &= \bigsqcap_{\pairc{\setm}{d}\in \setS} d 
\end{align}
The $\bbIn_n/\bbOut_n$ values represent the \fpmfp solution  which represents the data flow information reaching node $n$ along 
all \cfps excluding the ones that are known to be infeasible \cfps and are given as input.

\section{Computing the \fpmfp Solution}\label{sec:Computation}

We now illustrate how the data flow equations for computing the \fpmfp solution are derived from that of the \mfp specifications of 
a data flow analysis. 
Section~\ref{sec:psmfp.dfe} explains this lifting by defining the data flow equations, node flow functions, and the meet operator.
A key enabler for this lifting is a set of  specially crafted edge flow functions\footnote{The edge flow function for an edge 
$m\rightarrow n$ uses $\Out_{m}$ to compute $\In_{n}$ in a forward analysis, and vice-versa in a backward analysis.} 
(Section~\ref{sec:psmfp.eff}) that  discard data flow values that reach a program point from infeasible \cfps. 

\subsection{\fpmfp Solution Specification}
\label{sec:psmfp.dfe}
Let $\In_n,\Out_n \in \lat$ be the data flow values computed by a data flow analysis at node $n$, where 
\lat is a meet semi-lattice satisfying the descending chain condition. Additionally, it contains a $\top$ element\textemdash 	we add 
an artificial $\top$ value, if there is no natural $\top$.

Equations \ref{eq:dfe11} and \ref{eq:dfe12} (given below) represent the data flow equations for computing \mfp solution over the \cfg of 
a procedure, say \emph{p}. For simplicity we
assume a forward analysis. \Boundary represents the boundary information reaching procedure \emph{p} from its callers. The meet operator 
\meet computes the \emph{glb} (greatest lower bound) of elements in \lat. Function \emph{pred}{$(n)$} returns the predecessor nodes of 
node \emph{n} in the \cfg of \emph{p}. The node flow function $\fun_n$ and edge flow function $\fune_{\mn}$ compute the effect of node 
$n$ and edge \mn in the \cfg respectively. Usually the edge flow functions are identity.
\begin{align}
	\In_n & = \begin{cases}
	 \Boundary& n=Start_p
	\\
 	 \bigsqcap\limits_{m\in pred(n)} \!\!\!\!\fune_{\mn}(\Out_m)& otherwise
	\end{cases}	\label{eq:dfe11}\\
	\Out_n & = \fun_n(\In_n)\label{eq:dfe12}
\end{align}
We lift Equations~\ref{eq:dfe11} and~\ref{eq:dfe12} to define
a data flow analysis that computes data flow variables $\bIn_n,\bOut_n$  each of which is a set of \pairnames 
of the form \pairc{\setm}{d}. As explained in Section~\ref{sec:psmfp.definition}, $\bIn_n,\bOut_n$ are
values in \hlat, where 
\text{$ \hlat: \mpowerset \rightarrow \lat$} such that \setmips is the \emph{set of all \mips} in the program, and
\lat is the lattice of values computed by 
Equations~\ref{eq:dfe11} and~\ref{eq:dfe12}. 
\begin{align}
	\bIn_n & = \begin{cases}
	 \bBoundary& n=Start_p\\
 	 \bigsqcap\limits_{m\in pred(n)}^{\text{\textemdash}}\bFe_{\mn}(\bOut_m)& otherwise
	\end{cases}	\label{eq:dfe21}
	\\\label{eq:dfe22}
	\bOut_n & = \bF_n(\bIn_n)\\ \label{eq:bBoundary}
	\bBoundary&=\bset{\pairc{\nomips}{\Boundary}}\  \cup\  \bset{\pairc{\setm}{\top}\mid \setm\not=\nomips, \setm\subseteq \setmips}
\end{align}
The top value that is used for initialization is  $\bset{\pairc{\setm}{\top}\mid \setm\subseteq \setmips}$.
The node flow function ($\bF_n$) is pointwise application of $\fun_n$ to the \pairnames in the input set:
\begin{align}\label{eq:nff}
\bF_n(\setS)&=\{\pairc{\setm}{\fun_n(d)} \mid \pairc{\setm}{d}\in \setS\}
\end{align}
The meet operator (\bMeet) is pointwise application of $\sqcap$ to the \pairnames that contain the same set of \mips  i.e.,
\begin{align}\label{eq:meet}
\setS \bMeet \setS' = \{ \pairc{\setm}{d\meet d'} \mid  \pairc{\setm}{d}\in \setS, \pairc{\setm}{d'}\in \setS'\}
\end{align}
Note that \bMeet is defined for all values in \bIn/\bOut because  \bIn/\bOut each contain one pair corresponding to each subset of \setmips. 

\newcommand{\bSqsubseteq}{\text{$\overline{\sqsubseteq}$}\xspace}
If the  partial order between elements in \lat is $\sqsubseteq$ then the partial order between elements in \hlat 
is \bSqsubseteq and is given as follows.
\begin{align}\label{eq:partial.order}
\setS \bSqsubseteq \setS' \iff    (\pairc{\setm}{d}\in \setS\land \pairc{\setm}{d'}\in \setS' \implies d\sqsubseteq d')
\end{align}

\newcommand{\enof}[2]{\text{\emph{endof}$(#1,#2)$}\xspace}
\subsection{Defining the Edge Flow Function}\label{sec:psmfp.eff}
We now formally define the edge flow function $\bFe(\setS)$. The edge flow function performs the following two operations over an input 
\pairname \pairc{\setm}{d} in \setS.
\begin{enumerate}
 \item If $e$ is end edge of some \mips in \setm, then the data flow value $d$ is blocked,
 \item Otherwise $d$ is associated with a set of \mips \setmD where $\setmD=\cpob{e}{\setm}$ 
(described in Section 3).
\end{enumerate}

The data flow value $d$ from multiple \pairnames in \setS can get associated with the same set of \mips \setmD. Hence in the 
output \pairname $\pairc{\setm'}{d'}$, $d'$ is effectively computed by taking meet of values of $d$ from individual 
\pairnames $\pairc{\setm}{d}\in \setS$ where $\setmD=\cpob{e}{\setm}$, as shown in Equation~\ref{eq:eff.eff}. Let the function \enof{\setm}{e} 
be a boolean function that denotes if $e$ is end edge of some \mips in \setm or not i.e.,
\begin{align}
\enof{\setm}{e}=&(\exists \symmips\in \setm.\, e\in \mipsend(\symmips))\label{eq:enof}
\end{align}
The first term in the right hand side of the following edge flow function corresponds to the Item (1) in the list above, while the 
second term corresponds to the Item (2) of the list.
\begin{align}
\label{eq:eff.eff}
\bFe_{e}(\setS)=& \Bigg\{\pairc{\setm'}{\top}\mid \setm'\subseteq \setmips,\, \enof{\setm'}{e}\Bigg\}  \\ \nonumber
   &\bigcup\\ \nonumber
&\Bigg\{\pairc{\setm'}{d'}\mid \setm'\subseteq \setmips,\, \neg\enof{\setm'}{e},\,
  d'=\bigsqcap_{\pairc{\setm}{d}\in \setS,\ \cpob{e}{\setm}=\setm'} \!\!\!\!\!\!\!\!\!\!\!\!\!\!\!d\ \ \ \ \ \ \ \ \ \ \Bigg\}
\end{align}
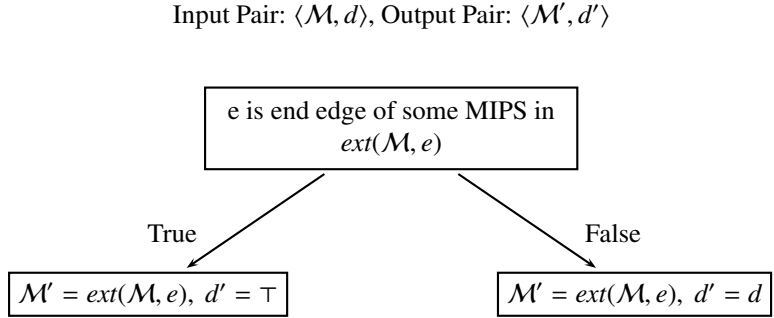
\begin{figure}[t!]
\begin{pspicture}(0,0)(15,45)
\putnode{n0}{origin}{63}{45}{Input Pair: $\pairc{\setm}{d}$, Output Pair: $\pairc{\setm'}{d'}$}

\putnode{n1}{n0}{0}{-15}{\psframebox{$\begin{array}{c} $e$\text{ is end edge of some \mips in} \\  \cpob{e}{\setm} \end{array}$}}
\putnode{n1l}{n1}{-32}{-22}{\psframebox{$\setm'=\cpob{e}{\setm},\ d'=\top$}}
\putnode{n1r}{n1}{32}{-22}{\psframebox{$\setm'=\cpob{e}{\setm},\ d'=d$}}
\ncline[nodesep=0.5]{->}{n1}{n1l}
	\nbput[npos=0.9]{True}
\ncline[nodesep=0.5]{->}{n1}{n1r}
	\naput[npos=0.9]{False}
\end{pspicture}
\caption{Computing output pair $\pairc{\setm'}{d'}$ from input pair $\pairc{\setm}{d}$ in edge flow function 
$\bFe_e$}\label{fig:illustrate.eff}
\end{figure}

\newcommand{\setSo}{\text{$\mathcal{S}_1$}\xspace}
\newcommand{\setSt}{\text{$\mathcal{S}_2$}\xspace}
\newcommand{\setD}{\text{$\mathcal{D}$}\xspace}
\newcommand{\setU}{\text{$\mathbb{U}$}\xspace}

\section{Extending the \fpmfp Computation to the Inter-procedural Level}\label{sec:interprocedural}
In this section, we explain the \fpmfp computation for \emph{inter-procedural data flow analysis}~\cite{sharir1978two} by describing the call 
node handling. Specifically, we formalize the \fpmfp computation for the \emph{functional approach}~\cite{sharir1978two} of 
inter-procedural data flow analysis. First, we briefly explain the functional approach below. Next, we describe the class of 
\emph{inter-procedural \mips} (\mips that span across multiple procedures) that we admit for the \fpmfp computation 
(Section~\ref{sec:interprocedural.mips}). Finally, we explain how \fpmfp handles call nodes 
(Section~\ref{sec:interprocedural.call.handling}).

The functional approach consists of two phases. The first phase computes summaries of procedures
in a program, by traversing the callgraph in a bottom-up order (i.e., summaries for callee procedures are computed before computing the 
summaries of the caller procedures). In this phase, a non-recursive procedure is analyzed once, while a recursive procedure is analyzed 
multiple times until a fix-point of procedure summaries is reached.
After the summary computation, a top-down traversal of the call-graph is performed to 
propagate data flow values from callers to callees. In this phase, a procedure is analyzed in
an intra-procedural manner because the effect of call nodes is represented by the summaries of
the callee procedures. Therefore, explaining call node handling and procedure summary computation is sufficient to extend 
the FPMFP computation at the inter-procedural level using the functional approach.




We first state the types of inter-procedural \mips that we admit (Section~\ref{sec:interprocedural.mips}), 
followed by the description of call node handling (Section~\ref{sec:interprocedural.call.handling}), and 
procedure summary computation (Section~\ref{sec:interprocedural.summary}).

\subsection{Inter-procedural \mips}\label{sec:interprocedural.mips} 
An inter-procedural \mips can span across multiple procedures, unlike an intra-procedural \mips that remains inside a single procedure.
We allow all intra-procedural \mips as input for a \fpmfp computation. Additionally, we allow a specific class of inter-procedural \mips 
in the \fpmfp computation. We describe this class of inter-procedural \mips below.

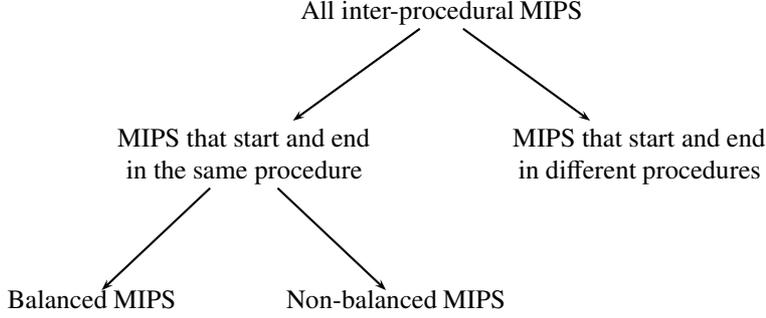
\begin{figure}[t!]
\begin{pspicture}(0,20)(25,62)
\putnode{n00}{origin}{60}{59}{All inter-procedural \mips}
\putnode{n0}{n00}{26}{-19}{$\begin{array}{c}\text{\mips that start and end}\\ \text{in different procedures}\end{array}$}
\putnode{n1}{n00}{-26}{-19}{$\begin{array}{c}\text{\mips that start and end}\\ \text{in the same procedure}\end{array}$}
\putnode{n1l}{n1}{-20}{-19}{Balanced \mips}
\putnode{n1r}{n1}{20}{-19}{Non-balanced \mips}
\ncline[nodesep=0.5]{->}{n00}{n0}
\ncline[nodesep=0.5]{->}{n00}{n1}
\ncline{->}{n1}{n1l}
\ncline{->}{n1}{n1r}
\end{pspicture}
\vspace{5pt}
\caption{Types of Inter-procedural \mips}\label{fig:interprocedural.mips.types}
\end{figure}

We categorize an inter-procedural \mips as shown in Figure~\ref{fig:interprocedural.mips.types}. Here, we distinguish between the 
two types of inter-procedural \mips depending on if they start and end in the same procedure or not. We admit a subclass 
of \mips that belong to the former type; this subclass is defined below. 
We refer these \mips as \emph{balanced inter-procedural \mips}, 
because each \emph{call edge} is matched by a corresponding \emph{call-return edge}.


\begin{definition}(Balanced Inter-procedural \mips)\label{def:balanced.mips}
Let a \mips \symmips be an inter-procedural \mips and \setV be the set of variables present in the condition on the end edge of \symmips.
We say \symmips is a balanced inter-procedural \mips if it starts and ends in 
the same procedure in a non-recursive manner\footnote{For each call edge there is a corresponding call return edge in the \mips.}, and 
the variables in \setV are not modified inside the procedures called through 
the intermediate nodes of \symmips \footnote{In general, this may not hold for all inter-procedural \mips that start and end in the same 
procedure. Specifically, 
there could exist an inter-procedural \mips \symmipsO such that the variables in the condition on the end edge of \symmipsO are 
modified along some paths inside the procedures called from intermediate nodes of the \mips.}.
\end{definition}

\begin{figure*}[t!]
\psset{unit=1mm}
\centering
\small
\begin{tabular}{cc|c}
\setlength{\tabcolsep}{1.2pt}
\renewcommand{\arraystretch}{1.3}
\begin{tabular}{@{}c@{}}
\begin{pspicture}(-1,0)(50,84)
\putnode{n0}{origin}{18}{84}{\framebox{$l\!=\!2$}}
	\putnode{w}{n0}{-7}{0}{$S_p$}

\putnode{n1}{n0}{-8}{-13}{\framebox{$a\!=\!0$}}
	\putnode{w}{n1}{-7}{0}{$n_1$}

\putnode{n2}{n1}{8}{-13}{\framebox{\em use a}}
	\putnode{w}{n2}{-7}{0}{$n_2$}

\putnode{n3}{n2}{0}{-13}{\framebox{{\em q}()}}
	\putnode{w}{n3}{-7}{0}{$n_3$}
	
\putnode{n4}{n3}{0}{-13}{\framebox{$a\!>\!5$}}
	\putnode{w}{n4}{-7}{0}{$n_4$}

\putnode{n5}{n4}{-8}{-15}{\framebox{\em use a}}
	\putnode{w}{n5}{-7}{0}{$n_5$}

\putnode{n6}{n4}{0}{-28}{\framebox{\white\em use a}}
	\putnode{w}{n6}{-7}{0}{$E_p$}

\putnode{q0}{n2}{25}{0}{\framebox{\white\em use a}}
	\putnode{w}{q0}{7}{0}{$S_q$}

\putnode{q1}{q0}{0}{-13}{\framebox{$l\!=\!0$}}
	\putnode{w}{q1}{7}{0}{$q_1$}
	
\putnode{q2}{q1}{0}{-13}{\framebox{\white\em use a}}
	\putnode{w}{q2}{7}{0}{$E_q$}
	
\ncline{->}{n0}{n1}
\nbput[npos=0.5,labelsep=.5]{$e_0$}
\ncline[doubleline=true,doublesep=.2]{->}{n1}{n2}
\nbput[npos=0.35,labelsep=0]{$e_2$}
\nccurve[angleA=300,angleB=60]{->}{n0}{n2}
\naput[npos=0.5,labelsep=.5]{$e_1$}
\ncline[doubleline=true,doublesep=.2]{->}{n2}{n3}
\nbput[npos=0.5,labelsep=.5]{$e_3$}
\ncline[doubleline=true,doublesep=.2]{->}{n3}{n4}
\nbput[npos=0.5,labelsep=.5]{$e_4$}
\ncline[doubleline=true,doublesep=.2]{->}{n4}{n5}
\naput[npos=0.5,labelsep=0.2]{$e_5$}
\nbput[npos=0.7,labelsep=0.2]{\true}
\nccurve[angleA=300,angleB=60]{->}{n4}{n6}
\naput[npos=0.53,labelsep=.2]{$e_6$}
\naput[npos=0.8,labelsep=.2]{\false}
\ncline{->}{n5}{n6}
\nbput[npos=0.53,labelsep=.2]{$e_7$}

\nccurve[linestyle=dashed,angleA=15,angleB=90,armB=0.9]{->}{n3}{q0}
\naput[npos=0.43,labelsep=.2]{$C_{pq}$}

\ncline{->}{q0}{q1}
\nbput[npos=0.53,labelsep=.2]{$e_8$}

\ncline{->}{q1}{q2}
\nbput[npos=0.53,labelsep=.2]{$e_9$}
\nccurve[linestyle=dashed,angleA=270,angleB=345,armA=0.5]{->}{q2}{n3}
\naput[npos=0.43,labelsep=.2]{$R_{qp}$}

\end{pspicture}
\end{tabular}
&
&
\begin{tabular}{l}
\begin{tabular}{@{}l@{}}  
$\begin{array}{l}
 \bullet\ \mips\ \symmips_1:e_2\rightarrow e_3\rightarrow e_4\rightarrow e_5
\end{array}$
\end{tabular}
\\
\\
\\
\begin{tabular}{@{}l@{}}  
$\begin{array}{l}
\text{The \fpmfp solution}
\end{array}$
\end{tabular}
\\
\begin{tabular}{|c|c|c|c|}

\cline{1-4}
	\multicolumn{1}{|c|}{\multirow{2}{*}{\rotatebox{90}{Edges}} }
	& \multicolumn{2}{|c|}{Sets of \mipss}
	& \fpmfp 
	\\ \cline{2-3}
	\multicolumn{1}{|c|}{ }
	& \set{}
	& \set{$\symmips_1$}
	& $\sqcap$
	\\ \cline{1-4}
	$e_3$ 
	& \rangeT
	& \rangeT
	& \rangeT 
	\\ \cline{1-4}
	$e_4$
	& \rangeZ
	& \rangeZ 
	& \rangeZ
	\\ \cline{1-4}
	$e_8$
	& \rangeT
	& 
	& \rangeT
	\\ \cline{1-4}
	$e_9$
	& \rangeZ
	& 
	& \rangeZ
	\\ \cline{1-4}
\end{tabular}
\end{tabular}
\\
(a)&&(b)
\end{tabular}
\caption{Computing \fpmfp solution using the functional approach of inter-procedural analysis: 
$S, E$ represent the start and end nodes of a procedure respectively. $C_{pq}$ represents the transfer of 
control from procedure $p$ to procedure $q$ at a call node and $R_{qp}$ represents the return of control
 from $q$ to $p$.}
\label{fig:interprocedural.summarybased}
\end{figure*}
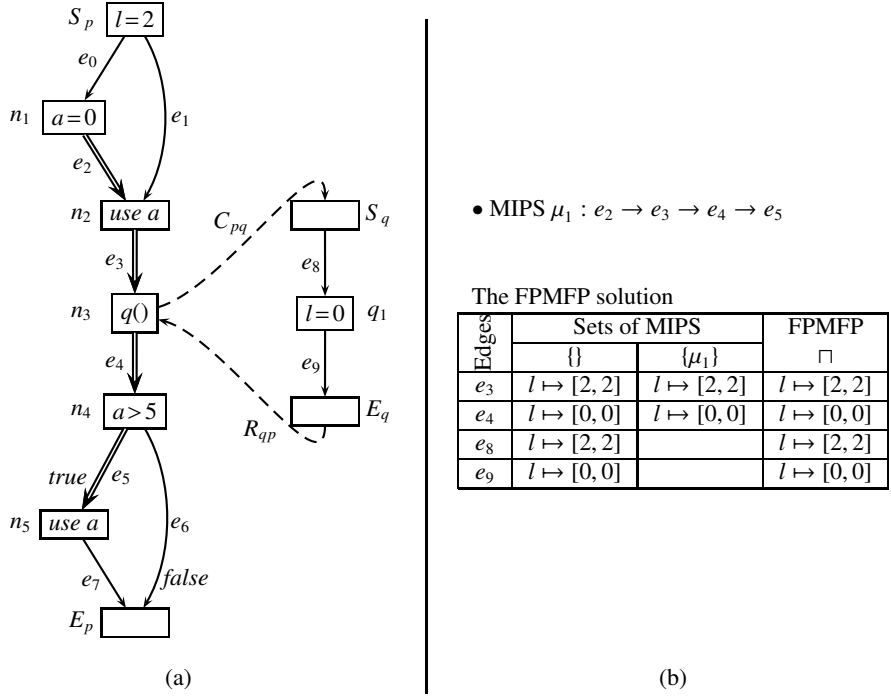

For example in Figure~\ref{fig:interprocedural.summarybased}, \mips 
$\symmips_1: n_1\xrightarrow{e_2}n_2\xrightarrow{e_3}n_3\xrightarrow{e_4}n_4\xrightarrow{e_5}n_5$  is a balanced inter-procedural \mips. 
A balanced \mips has the following interesting property.
\begin{quote}
For a balanced \mips \symmips that starts and ends in a procedure $p$, the control flow inside the procedures called from 
intermediate nodes of $\symmips$ can be abstracted out during the \fpmfp computation of $p$. 
\end{quote}

\begin{example}
 In Figure 
\ref{fig:interprocedural.summarybased}, \mips  
$\symmips_1: n_1\xrightarrow{e_2}n_2\xrightarrow{e_3}n_3\xrightarrow{e_4}n_4\xrightarrow{e_5}n_5$ 
 goes through the call node $n_3$ that calls procedure $q$. However, the control flow inside $q$ is abstracted out during a \fpmfp
computation of the caller procedure $p$. In particular, the edges in procedure $q$ are not marked as inner edges of   
$\symmips_1$ (even though the call $q()$ is part of \symmipsO).
\end{example}

Henceforth, we assume all inter-procedural \mips that are input to our analysis are balanced. We now explain the handling of a call node
followed by the procedure summary computation for bit-vector frameworks.

\subsection{Handling the Call Nodes}\label{sec:interprocedural.call.handling}

We now describe how the effect of a call node is incorporated in a \fpmfp computation. For this, we assume summaries of all 
procedures are already computed. We explain the procedure summary computation for bit-vector frameworks in 
Section~\ref{sec:interprocedural.summary}.

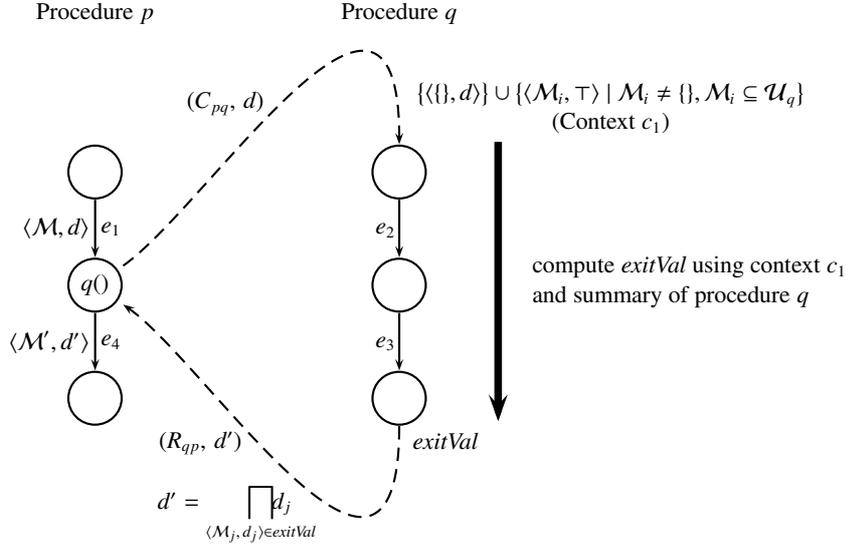
\begin{figure*}[t!]
\psset{unit=1mm}
\centering
\small
\begin{pspicture}(-1,0)(55,64)
\small
\putnode{n1}{origin}{-20}{48}{\pscirclebox{\white{\em q}()}}
	\putnode{w1}{n1}{0}{21}{Procedure $p$}

\putnode{n2}{n1}{0}{-15}{\pscirclebox{{\em q}()}}
	
\putnode{n6}{n2}{0}{-15}{\pscirclebox{\white{\em q}()}}

\putnode{n3}{n1}{40}{0}{\pscirclebox{\white{\em q}()}}
  \putnode{w}{n3}{0}{21}{Procedure $q$}
	\putnode{n3arrow}{n3}{13}{4}{}
	
\putnode{n4}{n3}{0}{-15}{\pscirclebox{\white{\em q}()}}
	
\putnode{n5}{n4}{0}{-15}{\pscirclebox{\white{\em q}()}}
	\putnode{n5arrow}{n5}{13}{-3}{}

\ncline{->}{n3arrow}{n5arrow}
	\ncline[linewidth=1]{->}{n3arrow}{n5arrow}
	\naput{$\begin{array}{l}\text{compute $\text{\emph{exitVal}}$ using context $c_1$}\\ 
				\text{and summary of procedure $q$}\end{array}$}
\ncline{->}{n1}{n2}
	\naput[npos=0.5,labelsep=.5]{$e_1$}
	\nbput[npos=0.5,labelsep=.5]{$\pairc{\setm}{d}$}

\ncline{->}{n2}{n6}
	\naput[npos=0.5,labelsep=.5]{$e_4$}
  \nbput[npos=0.5,labelsep=.5]{$\pairc{\setm'}{d'}$}
	
\nccurve[linestyle=dashed,angleA=35,angleB=90,ncurvB=1.6]{->}{n2}{n3}
		\naput[npos=0.43,labelsep=.2]{$(C_{pq},\,d)$}
    \naput[npos=0.95,labelsep=.2]{$\begin{array}{c}\bset{\pairc{\set{}}{d}}\cup\bset{\pairc{\setm_i}{\top}\mid \setm_i\not=\set{}, \setm_i\subseteq 
		\setmips_q}\\ (\text{Context}\ c_1)\end{array}$}

\ncline{->}{n3}{n4}
\nbput[npos=0.53,labelsep=.2]{$e_2$}

\ncline{->}{n4}{n5}
	\nbput[npos=0.53,labelsep=.2]{$e_3$}

\nccurve[linestyle=dashed,angleA=270,angleB=325,armA=0.5,ncurvA=1.6]{->}{n5}{n2}
	\naput[npos=0.63,labelsep=.2]{$(R_{qp},\,d')$}
	\naput[npos=0.02,labelsep=1.5]{$\text{\emph{exitVal}}$}
	\naput[npos=0.5,labelsep=.4]{$d'=\displaystyle\bigsqcap_{\pairc{\setm_j}{d_j}\in \emph{exitVal}}\!\!\!\!\!\!\!\!\!\!\!\!d_j$}
\end{pspicture}
\caption{Generic view of call node handling in a \fpmfp computation. $C_{pq}$ represents the transfer of control from procedure $p$ 
to procedure $q$ at a call node and $R_{qp}$ represents the return of control from $q$ to $p$. $\setmips_q$ is the set containing
 all i) intra-procedural \mips of q, and ii) balanced inter-procedural \mips that start and end in $q$.}
\label{fig:interprocedural.summarybased.generic}
\end{figure*}

Figure~\ref{fig:interprocedural.summarybased.generic} shows an abstract view of the handling of a call node in a \fpmfp computation. At a 
a call node $q()$, the output \pairname $\pairc{\setm'}{d'}$ corresponding to an input \pairname $\pairc{\setm}{d}$ is computed as follows:
\begin{itemize}
\item $\setm'=\cpob{e_4}{\setm}$, as explained in the edge flow function in Section~\ref{sec:psmfp.eff}.
\item $d'$ is obtained by using $d$ and the procedure summary of $q$. 
\end{itemize}

We now explain the complete effect of a procedure call on the caller and callee procedures using the example in 
Figure~\ref{fig:interprocedural.summarybased} as running example.

\subsubsection{Effect of a Caller Procedure on its Callee Procedure}

The boundary information ($\bBoundary_q$) for a callee procedure $q$ is computed by taking the meet of data flow values present at all 
call sites of $q$. Next, $\bBoundary_q$ is used for solving the callee procedure $q$. If $C_q$ is the set of all call sites of $q$, and 
$\setmips_q$ is the set containing 1) intra-procedural \mips of $q$, and 2) balanced inter-procedural \mips that start and end in 
$q$ then $\bBoundary_q$ is computed as follows: 
\begin{align*}
\bBoundary_q&=\bset{\pairc{\set{}}{x}} \cup \bset{\pairc{\setm}{\top}\mid \setm\subseteq \setmips_q,
 \setm\not=\set{}}\ \text{,where } \\
x&=\bigsqcap_{n\in C_q}\foldmeet(\bIn_n) 
\end{align*}
In Figure~\ref{fig:interprocedural.summarybased}, the data flow information in $\bIn_{n3}$ is used to compute the boundary 
information ($\bBoundary_q$) for procedure $q$ i.e., $\bBoundary_q=\bset{\pairc{\set{}}{\rangeT}}$. 

Moreover, a single boundary is maintained for each procedure. If there is a change in the data flow information at any call site of 
a procedure $q$ then the new boundary $\bBoundary_q'$ is merged with the existing boundary information of $q$ ($\bBoundary_q$). 
If this leads to a change in the resulting boundary information of $q$, the procedure $q$ is added to the worklist for solving. Later, 
the procedure is solved in an intra-procedural manner using the new boundary information by the worklist algorithm~\cite{sharir1978two}.

\subsubsection{Effect of a Callee Procedure on its Caller Procedure}
The \mfp computation (specified by equations \ref{eq:dfe11} and  \ref{eq:dfe12} from Section~\ref{sec:psmfp.definition}) defines the node 
flow functions ($f_n$) for all nodes \textemdash including nodes that call other procedures. 
In the functional approach of inter-procedural analysis~\cite{sharir1978two}, the call node flow functions are defined using summaries of 
the corresponding callee procedures. We use these node flow functions to compute 
the effect of a procedure call on the caller procedure using equation~\ref{eq:nff}. For example, 
in Figure \ref{fig:interprocedural.summarybased} the \mfp analysis defines the node flow 
function $f_{n3}$ which incorporates the effect of call $q()$ i.e., $q$ replaces the original value of $l$ by 0. The \fpmfp computation 
uses $f_{n3}$ to compute $\bF_{n3}$ using equation \ref{eq:nff} as follows: 
\begin{align*}
&\bF_{n3}\bbrace{\bset{\pairc{\set{}}{\rangeT}, \pairc{\set{\symmips_1}}{\rangeT}}}\\
&= \bset{\pairc{\set{}}{f_{n3}(\rangeT)}, \pairc{\set{\symmips_1}}{f_{n3}(\rangeT)}}\\
&= \bset{\pairc{\set{}}{\rangeZ}, \pairc{\set{\symmips_1}}{\rangeZ}}
\end{align*}
We elaborate the procedure summary computation for bit-vector frameworks in Section~\ref{sec:interprocedural.summary}

\newcommand{\genin}[1]{\text{$\text{GIn}_{#1}$}\xspace}
\newcommand{\genout}[1]{\text{$\text{GOut}_{#1}$}\xspace}
\newcommand{\genf}[1]{\text{\emph{genfunc}$_{#1}$}\xspace}
\newcommand{\genBoundary}{\text{$\set{}$}\xspace}
\newcommand{\gensum}[1]{\text{$\text{GENSum}_{#1}$}\xspace}

\newcommand{\setX}{\text{$X$}\xspace}
\newcommand{\killn}{\text{$\text{KILL}_n$}\xspace}
\newcommand{\killx}[1]{\text{$\text{KILL}_{#1}$}\xspace}
\newcommand{\genn}{\text{$\text{GEN}_n$}\xspace}
\newcommand{\genx}[1]{\text{$\text{GEN}_{#1}$}\xspace}
\newcommand{\genmeet}{\text{$\cup$}\xspace}

\newcommand{\bgenin}[1]{\text{$\overline{\text{GIn}}_{#1}$}\xspace}
\newcommand{\bgenout}[1]{\text{$\overline{\text{GOut}}_{#1}$}\xspace}
\newcommand{\bgenf}[1]{\text{$\overline{\text{\emph{genfunc}}}_{#1}$}\xspace}
\newcommand{\bgenBoundary}{\text{$\overline{\genBoundary}$}\xspace}
\newcommand{\bgensum}[1]{\text{$\overline{\text{GENSum}_{#1}}$}\xspace}
\newcommand{\bgenmeet}{\text{$\displaystyle\sqcap^{\overline{gen}}$}\xspace}

\newcommand{\killin}[1]{\text{$\text{KIn}_{#1}$}\xspace}
\newcommand{\killout}[1]{\text{$\text{KOut}_{#1}$}\xspace}
\newcommand{\killf}[1]{\text{\emph{killfunc}$_{#1}$}\xspace}
\newcommand{\killBoundary}{\text{$\top$}\xspace}
\newcommand{\killmeet}{\text{$\displaystyle\cap$}\xspace}
\newcommand{\msymbols}{\!\!\!}

\newcommand{\bkillin}[1]{\text{$\overline{\text{KIn}}_{#1}$}\xspace}
\newcommand{\bkillout}[1]{\text{$\overline{\text{KOut}}_{#1}$}\xspace}
\newcommand{\bkillf}[1]{\text{$\overline{\text{\emph{killfunc}}}_{#1}$}\xspace}
\newcommand{\bkillBoundary}{\text{$\overline{\text{KILLBI}}$}\xspace}
\newcommand{\bkillmeet}{\text{$\displaystyle\sqcap^{\overline{kill}}$}\xspace}

\subsection{Computing the feasible path GEN and KILL summaries for bit-vector frameworks}
\label{sec:interprocedural.summary}
In this section, we state the \mfp specifications used for the computation of procedure summaries for bit-vector 
frameworks~\cite{khedker2009data}. Next, we lift these specifications to the corresponding \fpmfp specifications. 

We use the standard notions of a procedure summary computation for bit-vector frameworks from~\cite{khedker2009data}. In these frameworks, 
the node flow functions are of the following form $\displaystyle f_n(\setX)=(\setX-\killn)\cup \genn$, where \genn and \killn are constant 
for each node. 
Hence the procedure summaries can be constructed by composing individual node flow functions, independent of the calling context information. 

The meet operator in bit-vector frameworks is either the set-union or set-intersection. For simplicity of exposition, we explain the 
procedure summary computation assuming the meet operator is union, henceforth we refer these data flow problems as \emph{union problems}.
We explain the computation for forward data flow analysis, a dual modeling exists for backward analysis (like live variables
analysis~\cite{khedker2009data}).

Let \genn, \killn be the constant GEN and KILL information at a non-call node $n$ respectively. Then GEN and KILL summaries 
for procedures are computed by solving corresponding GEN and KILL data flow problems as described below. 

\newcommand{\ote}[1]{\text{$\overline{\text{\emph{#1}}}$}\xspace}
\newcommand{\nodekill}[1]{\text{$\text{\emph{nodekill}($#1$)}$}\xspace}
\newcommand{\callees}[1]{\text{$\text{\emph{callees}($#1$)}$}\xspace}
\newcommand{\atexit}[1]{\text{\emph{Exit}$_{#1}$}\xspace}

\newcommand{\bnodekill}[1]{\text{$\ote{nodekill}(#1)$}\xspace}
\newcommand{\bkillSumm}[1]{\text{KSUM$_{#1}$}\xspace}
\newcommand{\batexit}[1]{\text{$\ote{Exit}_{#1}$}\xspace}

\newcommand{\nodegen}[1]{\text{$\text{\emph{nodegen}}(#1)$}\xspace}
\newcommand{\bnodegen}[1]{\text{$\ote{nodegen}(#1)$}\xspace}
\newcommand{\bgenSumm}[1]{\text{GSUM$_{#1}$}\xspace}

\textbf{Solving the KILL data flow problem.} 
The kill summary for union problems is computed using \killmeet as the meet operator, indicating that a value is added to the kill 
summary of a procedure iff the value is killed along all \cfps that reach the exit of the procedure. Initially,
kill summary of all procedures is set to $\top$ which is the top value of the kill data flow lattice. 
\set{} is used as the boundary value. The \mfp specifications of a KILL data flow problem for a procedure $p$ are as follows. 
\begin{align}
	\killin{n} & = \begin{cases}
									\set{} & n=Start_p\\
									\bigcap\limits_{m\in pred(n)}  \killout{m}& otherwise
								 \end{cases}	\label{eq:kill.dfe11}\\
	\killout{n} & = \killin{n}\cup\nodekill{n}\\
	\nodekill{n}&=\begin{cases}
									\killn & \twotb\text{\emph{n is not a call node}}\\
									\killout{\atexit{q}} & \twotb\text{\emph{n calls a procequre q}}
								\end{cases}\label{eq:kill.dfe12}
\end{align} 
\nodekill{n} represents the kill information for the node $n$; the information is constant if $n$ is not a call node. Otherwise, \nodekill{n} 
is the information at the exit the procedure called from $n$. For simplicity of exposition,
we assume at most one procedure can be called from a node (in practice, more than one procedures can be called from a node 
through function pointers, in which case the kill summary of the node is computed by doing intersection of kill information at 
exit of all procedures that can be called from $n$).

Equations \ref{eq:kill.dfe11} to \ref{eq:kill.dfe12} are solved for each procedure chosen in a bottom up traversal of the  
call graph of a program. In this, each non-recursive procedure is solved once, while a recursive procedure is solved multiple times 
until the data flow values at all points saturate. 

The final value in \killout{} at the exit of a procedure $p$ is the 
KILL summary for procedure $p$, as given by the \mfp solution. This summary may include the KILL information reaching from infeasible 
\cfps. Hence, to get a more precise summary, 
we lift the \mfp specifications to the \fpmfp specifications for each procedure $p$ in the program as follows. Let $\setmips_p$ be 
the set containing 1) intra-procedural \mips of $p$, and 2) balanced inter-procedural \mips that start and end in procedure $p$.  

The \fpmfp specifications (obtained by lifting \mfp specifications from equations \ref{eq:kill.dfe11} to \ref{eq:kill.dfe12}):
\begin{align}
	\bkillin{n} & = \begin{cases}
										\bkillBoundary & n=Start_p\\
											&\\
										\mathop{\overline{\bigsqcap}}\limits_{m\in pred(n)} \bFe_{m\rightarrow n}(\bkillout{m})& otherwise
									\end{cases}	\label{eq:kill.dfe11.fpmfp}\\
	\bkillBoundary&=\bset{\pairc{\set{}}{\set{}}}\cup\bset{\pairc{\setm}{\top}\mid \setm\subseteq\setmips_p,\ \setm\not=\set{}}\\
	\bkillout{n} & = \bset{\pairc{\setm}{\setD\cup\bnodekill{n}}\mid \pairc{\setm}{\setD}\in \bkillin{n}}\label{eq:kill.dfe.fpmfpO}\\
	\bnodekill{n}&=\begin{cases}
												\killn & \twotb\text{\emph{n is not a call node}}\\
												&\\
												\bigcap\limits_{\pairc{\setm}{\setD}\in \bkillout{\atexit{q}}}\setD & 
												\twotb\text{\emph{n calls a procequre q}}
									\end{cases}\label{eq:kill.dfe12.fpmfp}
\end{align}
	The $\overline{\sqcap}$ does a point-wise intersection of the kill data 
	flow values in the input set of \pairnames as follows. 
\begin{align}
	\displaystyle\setSo \overline{\sqcap} \setSt &= \bset{\pairc{\setm}{\setD_1\cap \setD_2}\mid \pairc{\setm}{\setD_1}\in \setSo,\ 
	\pairc{\setm}{\setD_2}\in \setSt}
\end{align}
	The final kill summary of a procedure $p$ is computed by doing intersection of the data flow values at the
	exit of $p$ i.e.,
	\begin{align}
	\bkillSumm{p}&=\bigcap\limits_{\pairc{\setm}{\setD}\in \bkillout{\atexit{p}}}\setD
	\end{align}
We now explain the GEN summary computation of a procedure below. 

\textbf{Solving the GEN data flow problem.}
The GEN summary for union problems is computed using \genmeet as the meet operator, indicating that a data flow value is added to GEN 
summary if it is generated along atleast one \cfp that reaches the exit of the procedure. Next, \genBoundary is used as the boundary value. 
The corresponding \mfp specifications are as follows.

The \mfp specifications:
\begin{align}
	\genin{n} & = \begin{cases}
										\set{} & n=Start_p\\ 
										& \\
										\bigcup\limits_{m\in pred(n)} \genout{m}& otherwise
								\end{cases}	\label{eq:gen.dfe11.mfp}\\
	\genout{n} & =(\genin{n}-\nodekill{n}) \cup \nodegen{n}\label{eq:gen.dfe.mfp}\\
	\nodegen{n}&=\begin{cases}
									\genn & \twotb\text{\emph{n is not a call node}}\\
									\genout{\atexit{q}} & \twotb\text{\emph{n calls a procedure q}}
							\end{cases}\label{eq:gen.dfe12}
\end{align}
Based on these \mfp specifications, we derive the \fpmfp specifications for each procedure $p$ in the program as follows. Let $\setmips_p$ 
be the set containing 1) intra-procedural \mips of $p$, and 2) balanced inter-procedural \mips that start and end in $p$.

The \fpmfp specifications:
\begin{align}
	\bgenin{n} & = \begin{cases}
									\bset{\pairc{\setm}{\set{}}\mid \setm\subseteq\setmips_p} & n=Start_p\\
										&\\
									\displaystyle\mathop{\overline{\bigsqcap}'}\limits_{m\in pred(n)} \bFe_{m\rightarrow n}(\bgenout{m})& otherwise
								 \end{cases}	\label{eq:gen.dfe11.fpmfp}\\
	\bgenout{n} & = \bset{\pairc{\setm}{(\setD-\bnodekill{n})\cup \bnodegen{n}}\mid \pairc{\setm}{\setD}\in \bgenin{n}}
	\label{eq:gen.dfe.fpmfp}\\
\bnodegen{n}&=\begin{cases}
								\genn & \twotb\text{\emph{n is not a call node}}\\
								&\\
								\displaystyle\bigcup_{\pairc{\setm}{\setD}\in \bgenout{Exit_q}} \setD & \twotb\text{\emph{n calls a procedure q}}
							\end{cases}\label{eq:gen.dfe12.fpmfp}
\end{align}	
	$\overline{\sqcap}'$ does a point-wise union of data flow values in the set of input \pairnames as follows. 
\begin{align}
	\displaystyle\setSo \overline{\sqcap}' \setSt &= \bset{\pairc{\setm}{\setD_1\cup \setD_2}\mid \pairc{\setm}{\setD_1}\in \setSo,\ 
	\pairc{\setm}{\setD_2}\in \setSt}
\end{align}
	The final gen summary of a procedure $p$ is computed by doing the union of data flow values at the exit of 
	the procedure $p$ i.e.,
	\begin{align}
		\bgenSumm{p}&= \bigcup_{\pairc{\setm}{\setD}\in \bgenout{Exit_p}} \setD
	\end{align}

\textbf{Substituting GEN and KILL summaries to compute the effect of procedure calls.}

The GEN and KILL summaries are computed for all procedures using the corresponding constant boundary values. We call this summary 
computation phase. Later, these GEN and KILL summaries are used to compute the effect of procedure $p$ at all callsites of $p$ during 
the actual \fpmfp computation (using actual boundary values) as follows:
Let $n: p()$ be a callsite of $p$ then 
\begin{align}
f_n(\setX)= (\setX-\bkillSumm{p})\cup\bgenSumm{p}\\
\bF_n(\setS)=\set{\pairc{\setm}{f_n(\setD)}\mid \pairc{\setm}{\setD}\in\setS}
\end{align}


\newcommand{\rangeZPositive}{\text{$l\mapsto[0,\infty]$}\xspace}
\newcommand{\rangePositive}{\text{$l\mapsto[1,\infty]$}\xspace}
\begin{figure}[t!]
\begin{pspicture}(0,20)(25,70)
\small
\putnode{n00}{origin}{62}{68}{All Optimizations}
\putnode{n0}{n00}{35}{-20}{$\begin{array}{c}\text{Optimizations that involve}\\ \text{ignoring a \pairname}\end{array}$}
\putnode{n01}{n0}{0}{-23}{$\begin{array}{c}\text{Optimization 3} \\ \text{(Section 6.3)}\end{array}$}
\putnode{n1}{n00}{-27}{-20}{$\begin{array}{c}\text{Optimizations that involve}\\ \text{merging \pairnames }\end{array}$}
\putnode{n1l}{n1}{-22}{-23}{$\begin{array}{c}\text{Optimization 1}\\ \text{(Section 6.1)}\end{array}$}
\putnode{n1r}{n1}{22}{-23}{$\begin{array}{c}\text{Optimization 2} \\ \text{(Section 6.2)}\end{array}$}

\ncline[nodesep=0.5]{->}{n00}{n0}
\ncline[nodesep=0.5]{->}{n00}{n1}
\ncline{->}{n1}{n1l}
\ncline{->}{n1}{n1r}
\ncline{->}{n0}{n01}
\end{pspicture}
\caption{Optimizations for scalability of the \fpmfp computation: The leaf nodes describe a particular optimization, and rest of the nodes 
describe the type of the optimization.}\label{fig:optimizations.overview}
\end{figure}
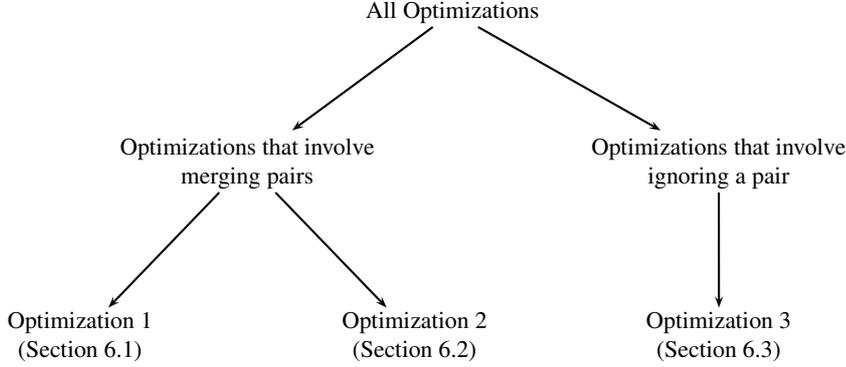

\section{Optimizations for Computing \fpmfp Solutions}\label{sec:optimizations}
We now describe few optimizations that improve the scalability of a \fpmfp computation. 
First, we explicate a relation between efficiency of the \fpmfp computation and the number of \pairnames in the \fpmfp computation. 
Next, we propose ideas to reduce the number of \pairnames. Our empirical data shows the reduction obtained is significant.

As evident from equations~\ref{eq:dfe21} to \ref{eq:eff.eff}, the efficiency of a \fpmfp computation is inversely proportional to the 
number of \pairnames. Theoretically, the number of \pairnames needed is exponential in the number of \mips in \setmips (set of all 
\mips) because a 
different \pairname is created corresponding to each subset of \setmips. However, at any program point the number of \pairnames that 
are evaluated is bounded by $|\setmips|+1$ (explained in Section~\ref{sec:complexity.analysis}). 
Below, we describe the ideas to reduce the number of \pairnames by either merging them or ignoring the ones that have no effect on
the final \fpmfp solution. These optimizations are shown in Figure~\ref{fig:optimizations.overview}. We describe each of these 
optimizations below. 

\subsection{Merging Pairs}\label{subsection:optimization.one} 
We now describe an optimization that allows us to merge some of the \pairnames present at a program point. We observed that a large
class of \mips found in our benchmarks were satisfying the following property.
\begin{quote}
\symp:  The condition on the end edge of a \mips \symmips evaluates to \emph{false} at the 
start edge of \symmips, and the variables present in the condition are not modified at the intermediate nodes of \symmips. 
\end{quote}


We observe the following peculiarity of \mips that satisfy \symp.

\begin{observation}\label{observation:merge.pairs}
We do not need to distinguish between \mips that satisfy \symp and have the same end edge, because the condition on the end edge 
does not hold at the starts of these \mips and data flow values associated with them are blocked at the same end edge. Hence, the data 
flow values associated with them can be merged.
\end{observation}

We support Observation~\ref{observation:merge.pairs} with a proof in Section~\ref{sec:proof.optimizations}. 
Observation~\ref{observation:merge.pairs} allows us to merge \pairnames at a program point as described below.
\begin{table}[tp]%
\small
\center
\begin{tabular}{|c|c|c|c|c|c|c|}
\cline{3-7}
	\multicolumn{1}{c}{}
  &\multicolumn{1}{c}{}
	& \multicolumn{4}{|c|}{Sets of \mipss}
	& \fpmfp 
	\\ \cline{3-6}
	\multicolumn{1}{c}{}
	& \multicolumn{1}{c|}{}
	& \set{}
	& $\set{\symmips_1}$
	& $\set{\symmips_2}$
	& $\set{\symmips_1,\symmips_2}$
	& $\sqcap$
	\\ \hline
	\multirow{4}{*}{\rotatebox{90}{Edges}}
	& $e_3$ 
	& \zrangeO
	& $\top$
	& $\top$
	& $\top$ 
	& \zrangeO
	\\ \cline{2-7}
	&$e_5$
	& $\top$
	& \zrangeZ
	& $\top$
	& $\top$
	& \zrangeZ
	\\ \cline{2-7}
	&$e_6$
	& $\top$
	& $\top$
	& \zrangeT
	& $\top$
	& \zrangeT
	\\ \cline{2-7}
	&$e_7$
	& \zrangeO
	& \zrangeZ
	& \zrangeT
	& $\top$
	& \zrangeZT
	\\ \cline{2-7}
	&$e_9$
	& \zrangeO
	& $\top$
	& $\top$
	& $\top$
	& \zrangeO
	\\ \hline
\end{tabular}
\caption{The \fpmfp solution for example in Figure \ref{fig:equivalent.mips}a. For brevity, only values of z are shown.}
\label{tab:equivalent.mips}
\end{table}

\newcommand{\tsymmipsO}{\text{$\tilde{\symmips}_1$}\xspace}

\emph{Optimization 1.} We merge two \pairnames $\pairc{\setm}{d}$ and $\pairc{\setm'}{d'}$ into a single \pairname $\pairc{\setm\cup 
\setm'}{d\sqcap d'}$ \emph{iff}
the sets $\setm$ and $\setm'$ are \emph{end edge equivalent} (denoted by $\eequal\ $) as defined below:
\begin{align}
	\setm\eequal \setm'
	\Leftrightarrow
\bigcup_{\symmips\in \setm}\mipsend(\symmips)=\bigcup_{\symmips'\in \setm'}\mipsend(\symmips')
\label{eq:e.equiv}
\end{align}

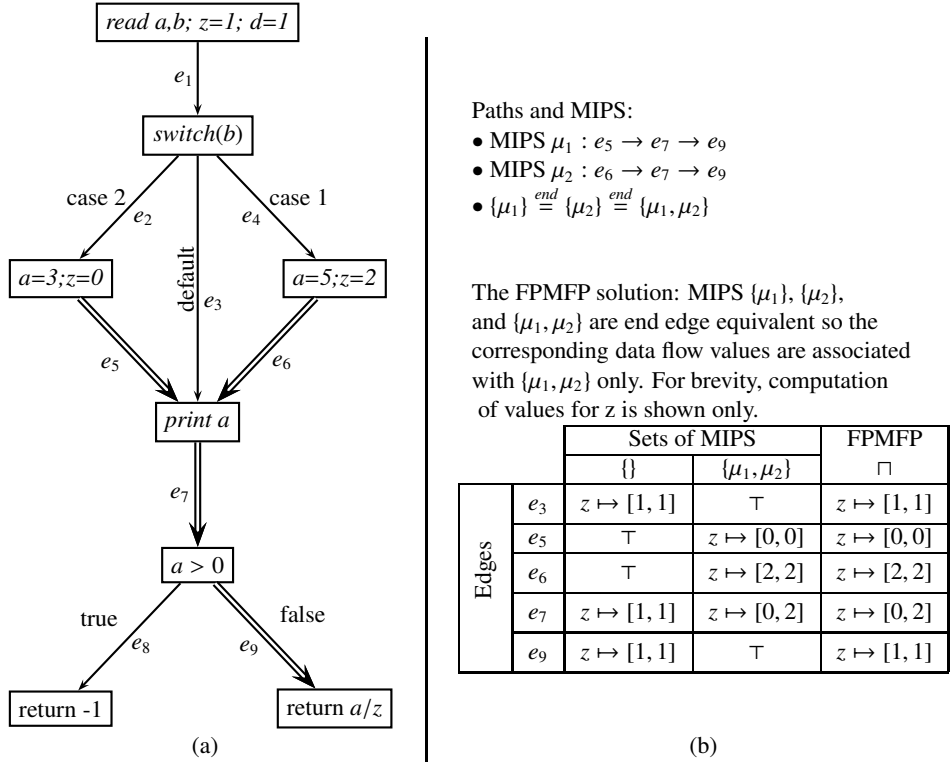
\begin{figure*}[t]
\psset{unit=1mm}
\centering
\small
\begin{tabular}{c|c}
\begin{tabular}{@{}c@{}}
\begin{pspicture}(-126,-83)(-72,8)
\putnode{n01}{origin}{-100}{10}{\psframebox{\emph{read a,b; z=1; d=1}}}
\putnode{n02}{n01}{0}{-15}{\psframebox{\emph{\text{$switch(b)$}}}}

\putnode{n03l}{n02}{-18}{-19}{\psframebox{\emph{a=3;z=0}}}
\putnode{n03r}{n02}{18}{-19}{\psframebox{\emph{a=5;z=2}}}
	
\putnode{n04}{n03l}{18}{-19}{\psframebox{\emph{print a}}}
\putnode{n05}{n04}{0}{-19}{\psframebox{\emph{$a>0$}}}

\putnode{n06l}{n05}{-18}{-19}{\psframebox{return -1}}
\putnode{n06r}{n05}{18}{-19}{\psframebox{return $a/z$}}
		

\ncline{->}{n01}{n02}
	\nbput[labelsep=.2]{$e_1$}
	
\ncline{->}{n02}{n03l}
	\nbput[labelsep=.2]{case 2}
	\naput[labelsep=.5]{$e_2$}
		
\ncline{->}{n02}{n04}
	\nbput[labelsep=.2]{\rotatebox{90}{default}}
	\naput[npos=.6,labelsep=.5]{$e_3$}

\ncline{->}{n02}{n03r}
	\naput[labelsep=.2]{case 1}
	\nbput[labelsep=.5]{$e_4$}
	
\ncline[doubleline=true]{->}{n03l}{n04}
	\nbput[labelsep=.5]{$e_5$}
	
\ncline[doubleline=true]{->}{n03r}{n04}
	\naput[labelsep=.1]{$e_6$}
	                            
\ncline[doubleline=true]{->}{n04}{n05}
	\nbput[labelsep=.1]{$e_7$}
		                            
\ncline{->}{n05}{n06l}
	\naput[labelsep=.1]{$e_8$}
	\nbput{true}

\ncline[doubleline=true]{->}{n05}{n06r}
	\nbput[labelsep=.1]{$e_9$}
	\naput{false}
	
\end{pspicture}
\end{tabular}
&
\begin{tabular}{l}
\begin{tabular}{@{}l@{}}  
$\begin{array}{l}
  \text{Paths and \mipss:}\\
 \bullet\ \mips\ \symmips_1:e_5\rightarrow e_7\rightarrow e_9\\
 \bullet\ \mips\ \symmips_2:e_6\rightarrow e_7\rightarrow e_9\\
 \bullet\ \setmuO\eequal\setmuT\eequal\setmuOT\\
 \end{array}$
\end{tabular}
\\
\\
\\
\begin{tabular}{@{}l@{}}  
$\begin{array}{l}
\text{The \fpmfp solution: \mips $\setmuO$, $\setmuT$, }\\
\text{and \setmuOT are end edge equivalent so the }\\
\text{corresponding  data flow values are associated }\\
\text{with $\setmuOT$ only. For brevity, computation}\\
\text{ of values for z is shown only. }\\
\end{array}$
\end{tabular}
\\
\begin{tabular}{|c|c|c|c|c|}
\cline{3-5}
	\multicolumn{1}{c}{ }
  &\multicolumn{1}{c|}{ }
	& \multicolumn{2}{|c|}{Sets of \mipss}
	& \fpmfp 
	\\ \cline{3-4}
	\multicolumn{1}{c}{ }
	&\multicolumn{1}{c|}{ }
	& \set{}
	& \setmuOT
	& $\sqcap$
	\\ \cline{1-5}
	\multirow{5}{*}{\rotatebox{90}{\hspace*{-.3cm}Edges}}
	&$e_3$ 
	& \zrangeO
	& $\top$
	& \zrangeO 
	\\ \cline{2-5}
	&$e_5$
	& $\top$
	& \zrangeZ 
	& \zrangeZ
	\\ \cline{2-5}
	&$e_6$
	& $\top$
	& \zrangeT
	& \zrangeT
	\\ \cline{2-5}
	&$e_7$
	& \zrangeO
	& \zrangeZT
	& \zrangeZT
	\\ \cline{2-5}
	&$e_9$
	& \zrangeO
	& $\top$
	& \zrangeO
	\\ \cline{1-5}
\end{tabular}
\end{tabular}
\\
(a)&(b)
\end{tabular}
\caption{Equivalent \mipss}
\label{fig:equivalent.mips}
\end{figure*}

The \fpmfp computation for the example in Figure~\ref{fig:equivalent.mips} is shown in Table~\ref{tab:equivalent.mips}. Four \pairnames 
are computed at each edge corresponding to each subset of \setmuOT. However, \mipss $\symmipsO$, \symmipsT have the same end edge 
$e_9$ and hence the following sets are end edge equivalent: \setmuO, \setmuT, \setmuOT. We obtain the \fpmfp solution by merging these 
\pairnames as shown in Figure~\ref{fig:equivalent.mips}b where only two \pairnames are maintained at each edge. 


\newcommand{\normalize}[1]{\text{$\tilde{#1}$}}
\newcommand{\bmipsend}[1]{\text{$\overline{\mipsend}(#1)$}}
\newcommand{\csof}{contains-suffix-of\xspace}
\newcommand{\cson}{\text{$\text{CSO}$}\xspace}
\newcommand{\pone}{\text{$P_1$}\xspace}
 \newcommand{\ptwo}{\text{$P_2$}\xspace}
 \newcommand{\setmT}{\text{$\setm_2$}\xspace}
 \newcommand{\setmTh}{\text{$\setm_3$}\xspace}
\subsection{Removing Duplicate Data Flow Values Across Pairs} \label{sec:optimization.cso}
In this section, we explain how the same data flow values can appear in more than one \pairnames at a program point. Further, 
we propose an idea to remove this duplication of values while retaining the precision and soundness of the \fpmfp solution. This 
optimization has helped us to improve the scalability of our approach to 150KLOC, whereas without this optimization scalability 
was 73KLOC.

In a \fpmfp computation, at an edge $e$, two \pairnames can contain the same data flow value $d$ iff $d$ reaches $e$ along two 
or more \cfps (because in this case $d$ may flow through two different \mips which are not related by \cpon relation). 
In such case, if one of the \cfp is a feasible \cfp then the two \pairnames can be merged into one as illustrated in 
Example~\ref{ex:computation.outside.edge} below. 

\begin{figure}[t!]
\begin{pspicture}(0,0)(15,100)

\small
\putnode{n00}{origin}{65}{100}{\framebox{d=1}}

\putnode{n01}{n00}{0}{-15}{\framebox{$c\!>\!5$}}

\putnode{n0}{n01}{7}{-17}{\framebox{$a\!=\!0$}}

\putnode{n1}{n01}{0}{-33}{\framebox{$b\!>\!5$}}

\putnode{n2}{n1}{7}{-15}{\framebox{\em read a}}
	
\putnode{n6}{n2}{-7}{-15}{\framebox{$a\!>\!5$}}

\putnode{n7}{n6}{-8}{-17}{\framebox{\em use a}}

\putnode{n8}{n7}{15}{0}{\framebox{\white\em use a}}

\ncline{->}{n00}{n01}
\naput[npos=0.4,labelsep=0]{$e_0$}

\ncline{->}{n01}{n0}
\naput[npos=0.85,labelsep=0.2]{\false}
\naput[npos=0.4,labelsep=0]{$e_1$}

\nccurve[angleA=240,angleB=120]{->}{n01}{n1}
\nbput[npos=0.45,labelsep=0.2]{\true}
\nbput[npos=0.2,labelsep=0]{$e_2$}
\nbput[npos=0.8,labelsep=2]{$\bset{\pairc{\set{}}{\drangeO}}$}

\ncline[doubleline=true]{->}{n0}{n1}
\nbput[doubleline=true, npos=0.4,labelsep=.5]{$e_3$}
\naput[labelsep=1,npos=0.2]{$\bset{\pairc{\setmu}{\drangeO}}$}

\ncline[doublesep=.2]{->}{n1}{n2}
\nbput[npos=0.4,labelsep=0]{$e_4$}
\naput[labelsep=1,npos=0.9]{$\bset{\pairc{\set{}}{\drangeO}}$}

\ncline{->}{n2}{n6}
\nbput[npos=0.45,labelsep=0]{$e_6$}
\naput[labelsep=1,npos=0.4]{$\bset{\pairc{\set{}}{\drangeO}}$}

\nccurve[doubleline=true, angleA=240,angleB=120]{->}{n1}{n6}

\nbput[labelsep=1,npos=0.6]{$\bset{\pairc{\set{}}{\drangeO},\ \pairc{\setmu}{\drangeO}}$}
\nbput[npos=0.3,labelsep=.5]{$e_5$}

\ncline[doubleline=true,doublesep=.2]{->}{n6}{n7}
\nbput[npos=0.3,labelsep=.5]{$e_7$}
\nbput[npos=0.85,labelsep=0.2]{\true}
\nbput[labelsep=2,npos=0.6]{$\bset{\pairc{\set{}}{\drangeO},\ \pairc{\setmu}{\text{\sout{\drangeO}}}}$}

\ncline[doublesep=.2]{->}{n6}{n8}
\naput[npos=0.4,labelsep=0.2]{$e_8$}
\naput[npos=0.85,labelsep=0.2]{\false}
\naput[labelsep=5,npos=0.6]{$\bset{\pairc{\set{}}{\drangeO}}$}

\end{pspicture}
\caption{An example illustrating the \fpmfp computation. The path segment $\symmips: e_3\rightarrow e_5\rightarrow e_7$ is a \mips. 
For simplicity, we have only shown value range of variable d, and  \pairnames $\pairc{\setm}{x}$ where $x\not=\top$.}
\label{fig:computation.outside.edge}
\end{figure}
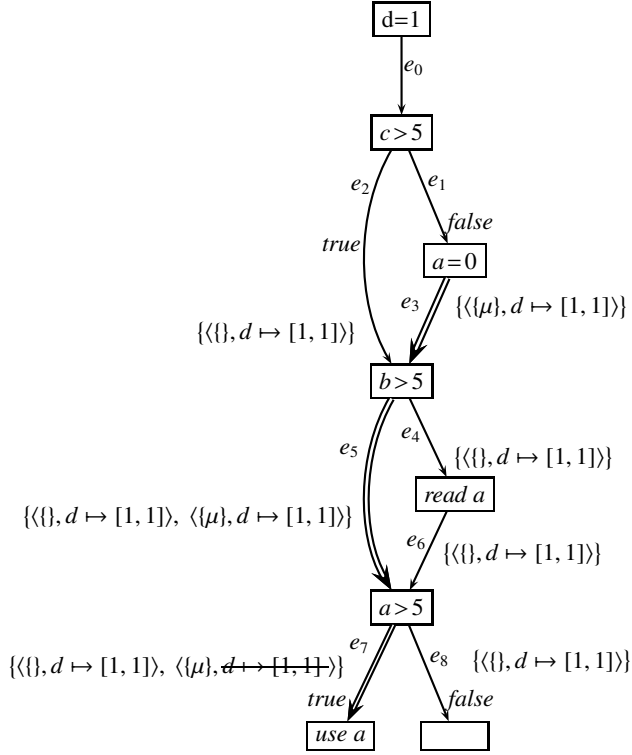

\begin{example}\label{ex:computation.outside.edge}
For example in Figure~\ref{fig:computation.outside.edge}, the data flow value \drangeO flows through \mips \symmips, hence it is 
associated with \setmu at edges in \symmips i.e., $e_3, e_5, e_7$. Next, the value associated with \setmu is blocked at the end 
edge $e_7$ of \symmips, because this value is reaching along the infeasible \cfp that contains \symmips 
i.e., $e_0\rightarrow e_1\rightarrow e_3\rightarrow e_5\rightarrow e_7$.
  
On the other hand, \drangeO also reaches $e_7$ along the following \cfp that is not infeasible $\sigma: e_0\rightarrow e_2\rightarrow 
e_4\rightarrow e_6\rightarrow e_7$, hence \drangeO is also associated with \set{}. Effectively, \drangeO is included at $e_7$ in the 
\fpmfp solution.
		
 In such case, we can get the same \fpmfp solution without associating \drangeO with  \setmu at $e_5, e_7$ as shown in Table~
\ref{tab:computation.outside.edge}. Observe that the corresponding \fpmfp solution is same to that in 
Figure~\ref{fig:computation.outside.edge}.
 \end{example}

We use the following general criteria to treat the \pairnames that contain the same data flow value at a program point. 

\begin{quote}
If a data flow value $d$ reaches the end edge of a \mips \symmips along a feasible \cfp then $d$ need not be associated with \symmips, 
because even if $d$ is blocked within \symmips, it reaches along the feasible \cfp and hence its association with \symmips is 
redundant.
\end{quote}

\begin{table}[tp]
\center
\begin{tabular}{|c|c|c|c|c|}
\cline{3-5}
	\multicolumn{1}{c}{ }
  &\multicolumn{1}{c|}{ }
	& \multicolumn{2}{|c|}{Sets of \mipss}
	& \fpmfp 
	\\ \cline{3-4}
	\multicolumn{1}{c}{ }
	&\multicolumn{1}{c|}{ }
	& \set{}
	& \set{$\symmips_1$}
	& $\sqcap$
	\\ \cline{1-5}
	\multirow{5}{*}{\rotatebox{90}{\hspace*{-.3cm}Edges}}
	&$e_3$
	& $\top$
	& \drangeO
	& \drangeO
	\\ \cline{2-5}
	&$e_5$ 
	& \drangeO
	& $\top$
	& \drangeO 
	\\ \cline{2-5}
	&$e_6$
	& \drangeO
	& $\top$ 
	& \drangeO
	\\ \cline{2-5}
	&$e_7$
	& \drangeO
	& $\top$
	& \drangeO
	\\ \cline{1-5}
\end{tabular}
\caption{Removing Duplicates}
\label{tab:computation.outside.edge}
\end{table}

To find out if a data flow value reaches the end edge of a \mips along a feasible \cfp, we define the \emph{\csof} relation below. 

\begin{definition}(Contains-suffix-of (\cson)) 
For a \mips \symmips and an edge $e$ in \symmips, let \suffix{\symmips}{e} be the sub-segment of \symmips from $e$ to $\mipsend(\symmips)$.  
Then, we say that  ``\mips \symmipsO \csof  \mips \symmipsT at $e$'', iff \symmipsO contains \suffix{\symmipsT}{e}.
\begin{align}\label{eq:cso.def}
  \cso{e}{\symmipsO}&=\set{\symmipsT\mid \symmipsT\in \setmips,\ \symmipsO\ \text{contains}\ \suffix{\symmipsT}{e}}
  \end{align}
\end{definition}	
Observe that the \cson is a reflexive relation, and a dual of the \cpon relation defined in section \ref{sec:fpmfp.overview} in the following 
sense: at an edge $e$, \symmipsO \cpon \symmipsT means \symmipsO and \symmipsT follow the same \cfp till edge $e$ whereas, 
\symmipsO \cson \symmipsT means \symmipsO and \symmipsT follow the same \cfp after edge $e$ however they may or may not follow the same 
\cfp before edge $e$. 
 
The \cson relation has the following properties. 
\begin{itemize}
\item \emph{Property 1}: if a \mips \symmipsO contains suffix of a \mips \symmipsT at an edge $e$, then the data flow value  
associated with \symmipsT is blocked at or before reaching the end of \symmipsO (because after $e$, \symmipsO and \symmipsT follow the
same \cfp on which \symmipsT ends at or before the end of \symmipsO). 

\begin{example}
For example in Figure~\ref{fig:cso.one}a, \symmipsO \cson \symmipsT at $e_5$. Hence, \rangeZ is blocked at $e_6$ in \pairname
\pairc{\set{\symmipsT}}{\rangeZ} as well as at $e_7$ in \pairc{\set{\symmipsO}}{\rangeZ}. Thus \rangeZ does not reach $e_7$.
\end{example}


\item \emph{Property 2}: if two \pairnames \pairc{\setmO}{d} and \pairc{\setmT}{d} contain the 
same data flow value $d$ at an edge $e$, and $d$ is not killed along a \mips \symmipsT in \setmT then the following holds.
\begin{quote}
$d$ reaches the end of \symmipsT along atleast one feasible \cfp that passes through $e$, provided that \symmipsT does not contain 
suffix of any \mips in \setmO at e. Specifically in this case, the data flow value $d$ from \pairname \pairc{\setmO}{d} reaches the end
 of \symmipsT (proof in Section~\ref{sec:proof.optimizations}). Hence the association of $d$ with \symmipsT can be discarded. 
\end{quote}
\begin{example}
For example in Figure~\ref{fig:cso.one}a, \symmipsT does not contain suffix of \symmipsO. Further, \rangeZ flows through \symmipsO and 
\symmipsT and is blocked at corresponding end edges as shown in Figure~\ref{fig:cso.one}b. However, \rangeZ is included in the \fpmfp 
solution at the end edge $e_6$ of \symmipsT because it reaches along the following feasible \cfp: $e_0\rightarrow e_3\rightarrow e_5
\rightarrow e_6\rightarrow e_9$. 
\end{example}

In such case, we can get the same \fpmfp solution without maintaining an association with \symmipsT. 
\end{itemize}

Based on the above properties, we now formally describe the following criteria to merge two \pairnames that have the same 
data flow value.  

\emph{Optimization 2.} At an edge e, we shift the common data flow value $d$ 
from two \pairnames $\pairc{\setm_1}{d}$ and $\pairc{\setm_2}{d}$ to a single \pairname $\pairc{\setm_3}{d}$ where $\setm_3$ 
contains only those \mips in $\setmO\cup\setmT$ that contain suffix of at least one \mips in \setmO and at least one \mips in \setmT
\footnote{Note that \cson is a reflexive relation, hence if $\symmipsO \in \setmO$ then ``\symmipsO contains suffix of at least one 
\mips in \setmO'' is trivially true.} i.e., 
\begin{align*}
\setm_3=&\set{\symmipsO\mid \symmipsO\in \setm_1,\ \exists \symmipsT\in \setm_2.\, \symmipsO\ \cson\ \symmipsT\text{ at }e}\\
	     &\cup\\
			 &\set{\symmipsT\mid \symmipsT\in \setm_2,\ \exists \symmipsO\in \setm_1.\, \symmipsT\ \cson\ \symmipsO\text{ at }e}
\end{align*}

\begin{example}
For example in Figure~\ref{fig:cso.one}a, at edges $e_5$ and $e_6$ \symmipsO \cson \symmipsT, hence we associate \rangeZ with 
\symmipsO only as shown in Figure~\ref{fig:cso.one}c. Observe the corresponding \fpmfp solution is same as in Figure~\ref{fig:cso.one}b.
\end{example}

\begin{example}
For example in Figure~\ref{fig:cso.two}a, the data flow value \rangeZ flows though two \mips \symmipsO and \symmipsT. Further neither 
\symmipsO \cson \symmipsT nor \symmipsT \cson \symmipsO, hence \rangeZ reaches $\mipsend(\symmipsO)$ and $\mipsend(\symmipsT)$ along 
the following feasible \cfps $\sigma_1$ and $\sigma_2$ respectively, 
$\sigma_1: e_0\rightarrow e_1\rightarrow e_3\rightarrow e_4\rightarrow e_7$, 
$\sigma_2: e_0\rightarrow e_2\rightarrow e_4\rightarrow e_5$. Hence, \rangeZ is included in the \fpmfp solution  at edges $e_7$ 
and $e_5$ as shown in Figure~\ref{fig:cso.two}b.

We obtain the same \fpmfp solution without associating \rangeZ with \symmipsO or \symmipsT as shown in Figure~\ref{fig:cso.two}c.
\end{example}

\begin{figure*}[t!] 
\psset{unit=1mm}
\centering
\small
\begin{tabular}{l|l}
\begin{tabular}{@{}c@{}}
\begin{pspicture}(-126,-83)(-87,8)
\putnode{n001}{origin}{-115}{13}{\psframebox{\emph{if $(l==0)$}}}

\putnode{n01}{n001}{0}{-14}{\psframebox{\emph{if $(d==2)$}}}
				
\putnode{n02}{n01}{19}{-14}{\psframebox{\emph{c=2, read d}}}
	
\putnode{n03}{n01}{0}{-25}{\psframebox{\emph{print l}}}

\putnode{n04}{n03}{0}{-16}{\psframebox{\emph{if $(c==0)$}}}
	
\putnode{n05}{n04}{18}{-15}{\psframebox{\emph{if $(d==2)$}}}

\putnode{n06}{n05}{0}{-20}{\psframebox{\emph{print l}}}
				
\putnode{exit01}{n04}{0}{-35}{\psframebox{\emph{exit}}}
	

\ncline{->}{n001}{n01}
	\naput[npos=.4,labelsep=.2]{$e_0$}
	\naput[npos=.7,labelsep=.5]{true}

\nccurve[angleA=210,angleB=150]{->}{n001}{n03}
	\naput[npos=.5,labelsep=.2]{$e_1$}
	\naput[npos=.6,labelsep=.5]{false}

\ncline{->}{n01}{n02}
	\naput[npos=.4,labelsep=.2]{$e_2$}
	\naput[npos=.7,labelsep=.5]{true}

\ncline[doubleline=true, linecolor=blue]{->}{n01}{n03}
	\naput[npos=.5,labelsep=.2]{false}
	\naput[npos=.35,labelsep=.5]{$e_3$}
	
\ncline[doubleline=true,linecolor=red]{->}{n02}{n03}
	\naput[npos=.6,labelsep=.5]{$e_4$}
	
\ncline[doubleline=true,linecolor=blue]{->}{n03}{n04}	
\ncline[doubleline=true,linecolor=red,offset=5pt]{->}{n03}{n04}
\naput[npos=.6,labelsep=.5]{$e_5$}
	
\ncline[doubleline=true,linecolor=blue]{->}{n04}{n05}	
\ncline[doubleline=true,linecolor=red,offset=5pt,nodesepA=3pt,nodesepB=-3pt]{->}{n04}{n05}
	\naput[npos=.4,labelsep=.1]{$e_6$}
	\naput[npos=.7,labelsep=.1]{true}
	                            
\ncline[doubleline=true,linecolor=blue]{->}{n05}{n06}
	\naput[npos=.3,labelsep=.1]{$e_7$}
	\naput[npos=.45,labelsep=.01,nrot=0]{true}

\ncline{->}{n06}{exit01}
	\naput[npos=.5,labelsep=.2]{$e_{10}$}
	                            
\ncline{->}{n04}{exit01}
	\nbput[npos=.35,labelsep=.5]{$e_8$}
	\nbput[npos=.5,labelsep=.5]{false}
	                              
\ncline{->}{n05}{exit01}
	\nbput[npos=.45,labelsep=.5]{$e_9$}
	\naput[npos=.5,labelsep=-1.5]{\rotatebox{45}{false}}
	
\end{pspicture}
\end{tabular}
&
\hspace*{-10pt}
\begin{tabular}{l}
\begin{tabular}{@{}l@{}}  
$\begin{array}{l}
  \text{Paths and \mipss:}\\
 \bullet\ \mips\ \symmips_1:e_3\rightarrow e_5\rightarrow e_6\rightarrow e_7 (\text{marked in blue})\\
 \bullet\ \mipsstart(\symmips_1)=\set{e_3},\ \mipsinner(\symmips_1)=\set{e_5,e_6},\\
 \bullet\ \mipsend(\symmips_1)=\set{e_7}\\ 
 \bullet\ \mips\ \symmips_2:e_4\rightarrow e_5\rightarrow e_6 (\text{marked in red})\\
\bullet\ \mipsstart(\symmips_2)=\set{e_4},\ \mipsinner(\symmips_2)=\set{e_5},\ \mipsend(\symmips_2)=\set{e_6}\\
\bullet\ \text{At }e_5\text{ and } e_6,\ \symmipsO\ \cson\ \symmipsT
\end{array}$
\end{tabular}
\\
\\
\setlength{\tabcolsep}{1.2pt}
\begin{tabular}{|c|c|c|c|c|c|}
\hline
	\multicolumn{1}{|c|}{\multirow{2}{*}{\!\rotatebox{90}{Edges}\!}}
	& \multicolumn{4}{|c|}{Sets of \mipss}
	& \fpmfp 
	\\ \cline{2-5}
	 \multicolumn{1}{|c|}{}
	& \set{}
	& $\set{\symmips_1}$
	& $\set{\symmips_2}$
	& $\set{\symmips_1,\symmips_2}$
	& $\sqcap$
	\\ \hline
	 $e_1$ 
	& \rangePositive
	& $\top$
	& $\top$
	& $\top$ 
	& \rangePositive
	\\ \hline
	$e_3$
	& $\top$
	& \rangeZ
	& $\top$
	& $\top$
	& \rangeZ
	\\ \hline
	$e_4$
	& $\top$
	& $\top$
	& \rangeZ
	& $\top$
	& \rangeZ
	\\ \hline
	$e_5$
	& \rangePositive
	& \rangeZ
	& \rangeZ
	& $\top$
	& \rangeZPositive
	\\ \hline
	$e_6$
	& \rangePositive
	& \rangeZ
	& \text{\sout{\rangeZ}}
	& $\top$
	& \rangeZPositive
	\\ \hline
	$e_7$
	& \rangePositive
	& \text{\sout{\rangeZ}}
	& $\top$
	& $\top$
	& \rangePositive
	\\ \hline
\end{tabular}
\\\multicolumn{1}{c}{(b)}
\\
\\
\setlength{\tabcolsep}{1.2pt}
\begin{tabular}{|c|c|c|c|c|c|}
\hline
	\multicolumn{1}{|c|}{\multirow{2}{*}{\!\rotatebox{90}{Edges}\!}}
	& \multicolumn{4}{|c|}{Sets of \mipss}
	& \fpmfp 
	\\ \cline{2-5}
	\multicolumn{1}{|c|}{}
	& \set{}
	& $\set{\symmips_1}$
	& $\set{\symmips_2}$
	& $\set{\symmips_1,\symmips_2}$
	& $\sqcap$
	\\ \hline
	 $e_1$ 
	& \rangePositive
	& $\top$
	& $\top$
	& $\top$ 
	& \rangePositive
	\\ \hline
	$e_3$
	& $\top$
	& \rangeZ
	& $\top$
	& $\top$
	& \rangeZ
	\\ \hline
	$e_4$
	& $\top$
	& $\top$
	& \rangeZ
	& $\top$
	& \rangeZ
	\\ \hline
	$e_5$
	& \rangePositive
	& \rangeZ
	& $\top$
	& $\top$
	& \rangeZPositive
	\\ \hline
	$e_6$
	& \rangePositive
	& \rangeZ
	& $\top$
	& $\top$
	& \rangeZPositive
	\\ \hline
	$e_7$
	& \rangePositive
	& \text{\sout{\rangeZ}}
	& $\top$
	& $\top$
	& \rangePositive
	\\ \hline
\end{tabular}
\end{tabular}
\\
\multicolumn{1}{c}{(a)}&\multicolumn{1}{c}{(c)}
\end{tabular}
\vspace{-10pt}
\caption{Illustrating \pairnames that have the same data flow value and have \cson relation between constituent \mips.
Here $l$ is an unsigned int variable.}
\vspace{-10pt}
\label{fig:cso.one}
\end{figure*}

\begin{figure*}[t!] 
\psset{unit=1mm}
\centering
\small
\begin{tabular}{l|l}
\begin{tabular}{@{}c@{}}
\begin{pspicture}(-126,-83)(-87,8)
\putnode{n01}{origin}{-118}{-7}{\psframebox{\emph{if} $(c\,!\!=0)$}}
\putnode{n001}{n01}{0}{12}{\psframebox{\emph{l=0}}}
				
\putnode{n02}{n01}{20}{-14}{\psframebox{\emph{c=2}}}
	
\putnode{n03}{n01}{0}{-25}{\psframebox{\emph{print l}}}

\putnode{n04}{n03}{0}{-16}{\psframebox{\emph{if $(c==0)$}}}
	
\putnode{n05}{n04}{18}{-15}{\psframebox{\emph{assert$(l!\!=0)$}}}
				
\putnode{exit01}{n04}{0}{-25}{\psframebox{\emph{exit}}}
	

\ncline{->}{n001}{n01}
	\nbput[npos=.4,labelsep=.2]{$e_0$}
	
\ncline{->}{n01}{n02}
	\naput[npos=.4,labelsep=.2]{$e_1$}
	\naput[npos=.7,labelsep=.5]{true}
	
\ncline[doubleline=true, linecolor=blue]{->}{n01}{n03}
	\nbput[npos=.5,labelsep=.2]{false}
	\nbput[npos=.35,labelsep=.5]{$e_2$}
	
\ncline[doubleline=true,linecolor=red]{->}{n02}{n03}
	\naput[npos=.6,labelsep=.5]{$e_3$}
	
\ncline[doubleline=true,linecolor=blue]{->}{n03}{n04}	
\nbput[npos=.6,labelsep=.5]{$e_4$}
\ncline[doubleline=true,linecolor=red,offset=5pt]{->}{n03}{n04}
	
\ncline[doubleline=true,linecolor=red]{->}{n04}{n05}
	\naput[npos=.4,labelsep=.1]{$e_5$}
	\naput[npos=.7,labelsep=.1]{true}
	                            	                            
\ncline[doubleline=true,linecolor=blue]{->}{n04}{exit01}
	\nbput[npos=.35,labelsep=.5]{$e_7$}
	\nbput[npos=.5,labelsep=.5]{false}
	                              
\ncline{->}{n05}{exit01}
	\naput[npos=.6,labelsep=.5]{$e_6$}
	
\end{pspicture}
\end{tabular}
&
\hspace*{-10pt}
\begin{tabular}{l}
\begin{tabular}{@{}l@{}}  
$\begin{array}{l}
  \text{Paths and \mipss:}\\
 \bullet\ \mips\ \symmips_1:e_2\rightarrow e_4\rightarrow e_7\ (\text{marked in blue}) \\
 \bullet\ \mips\ \symmips_2:e_3\rightarrow e_4\rightarrow e_5\ (\text{marked in red})\\
 \bullet\ \mipsstart(\symmips_1)=e_2\\
 \bullet\ \mipsstart(\symmips_2)=e_3\\
 \bullet\ \mipsinner(\symmips_1)=\mipsinner(\symmips_2)=e_4\\
 \bullet\ \mipsend(\symmips_1)=e_7, \mipsend(\symmips_2)=e_5\\
\end{array}$
\end{tabular}
\\
\\
\setlength{\tabcolsep}{1.2pt}
\begin{tabular}{|c|c|c|c|c|c|}
\hline
   \multicolumn{1}{|c|}{\multirow{2}{*}{\rotatebox{90}{Edges}}}
	& \multicolumn{4}{|c|}{Sets of \mipss}
	& \fpmfp 
	\\ \cline{2-5}
	 \multicolumn{1}{|c|}{}
	& \set{}
	& $\set{\symmips_1}$
	& $\set{\symmips_2}$
	& $\set{\symmips_1,\symmips_2}$
	& $\sqcap$
	\\ \hline
	  $e_2$ 
	& $\top$
	& \rangeZ
	& $\top$
	& $\top$ 
	& \rangeZ
	\\ \hline
	$e_3$
	& $\top$
	& $\top$
	& \rangeZ
	& $\top$
	& \rangeZ
	\\ \hline
	$e_4$
	& $\top$
	& \rangeZ
	& \rangeZ
	& $\top$
	& \rangeZ
	\\ \hline
	$e_5$
	& \rangeZ
	& $\top$
	& \text{\sout{\rangeZ}}
	& $\top$
	& \rangeZ
	\\ \hline
	$e_7$
	& \rangeZ
	& \text{\sout{\rangeZ}}
	& $\top$
	& $\top$
	& \rangeZ
	\\ \hline
\end{tabular}
\\\multicolumn{1}{c}{(b)}
\\
\\
\setlength{\tabcolsep}{1.6pt}
\begin{tabular}{|c|c|c|c|c|c|}
\hline
	\multicolumn{1}{|c|}{\multirow{2}{*}{\rotatebox{90}{Edges}}}
	& \multicolumn{4}{|c|}{Sets of \mipss}
	& \fpmfp 
	\\ \cline{2-5}
	 \multicolumn{1}{|c|}{}
	& \set{}
	& $\set{\symmips_1}$
	& $\set{\symmips_2}$
	& $\set{\symmips_1,\symmips_2}$
	& $\sqcap$
	\\ \hline
	 $e_2$ 
	& \rangeZ
	& $\top$
	& $\top$
	& $\top$ 
	& \rangeZ
	\\ \hline
	$e_3$
	& \rangeZ
	& $\top$
	& $\top$
	& $\top$
	& \rangeZ
	\\ \hline
	$e_4$
	& \rangeZ
	& $\top$
	& $\top$
	& $\top$
	& \rangeZ
	\\ \hline
	$e_5$
	& \rangeZ
	& $\top$
	& $\top$
	& $\top$
	& \rangeZ
	\\ \hline
	$e_7$
	& \rangeZ
	& $\top$
	& $\top$
	& $\top$
	& \rangeZ
	\\ \hline
\end{tabular}
\end{tabular}
\\
\multicolumn{1}{c}{(a)}&\multicolumn{1}{c}{(c)}
\end{tabular}
\vspace{-10pt}
\caption{Illustrating \pairnames that have the same data flow value but do not have \cson relation between constituent \mips.
\rangeZ reaches $\mipsend(\symmipsO)$ and $\mipsend(\symmipsT)$ along the following feasible \cfps $\sigma_1$ and $\sigma_2$ 
respectively, $\sigma_1: e_0\rightarrow e_1\rightarrow e_3\rightarrow e_4\rightarrow e_7$, 
$\sigma_2: e_0\rightarrow e_2\rightarrow e_4\rightarrow e_5$.}
\vspace{-10pt}
\label{fig:cso.two}
\end{figure*}

\subsection{Removing the \pairnames that contain $\top$ data flow value} \label{sec:remove.top}
In a \pairname $\pairc{\setm}{d}$ the data flow value $d$ remains $\top$ at edges that are outside a \mips in \setm. We use the 
following observation to eliminate such \pairnames. 

\begin{observation}\label{observation:empty.pairs}
Ignoring the \pairnames that contain $\top$ as data flow value does not affect the precision or the soundness of the
\fpmfp solution because the \fpmfp solution is computed by taking meet of the data flow values in each \pairname at a program point. 
\end{observation}

\emph{Optimization 3}. We eliminate the \pairnames that have $\top$ as the associated data 
flow value, if the total number of remaining \pairnames at the program point is more than one.  

For example in Figure~\ref{fig:equivalent.mips}b, the \pairnames corresponding to \set{} at edges $e_5$ and $e_6$ 
contain $\top$ value hence these \pairnames can be eliminated. Similarly, \pairnames 
corresponding to \setmuOT at edges $e_3$ and $e_9$ can be eliminated.

With this modification the node and edge flow functions (Equations~\ref{eq:nff}, \ref{eq:eff.eff}) remain same except that  the meet 
operator (Equation~\ref{eq:meet}) now assumes absence of a \pairname corresponding to some $\setm\subseteq \setmips$ in the input sets 
as presence of a \pairname $\pairc{\setm}{\top}$.

\subsection{Complexity Analysis}\label{sec:complexity.analysis}
The cost of the \fpmfp computation is equivalent to computing $k+1$ parallel \mfp solutions, where $k$ is bounded by the maximum number of 
\pairnames at a program point that do not contain $\top$ as the associated data flow value. 

At an edge $e$, the maximum number of \pairnames that do not contain $\top$ value is same as the 
maximum number of distinct sets given by \cpo{e}{\symmips} over all \mips \symmips, plus 1 for the \pairname corresponding to  
\set{}. In worst case, \cpo{e}{\symmips} set for each 
\symmips is distinct. Hence the total number of such \pairnames (i.e., $k$ value from previous paragraph) is bounded by the number of \mips 
in the \cfg. In practice we found the $k$ value to be much smaller where the \fpmfp computation time was $2.9\times$ of the 
\mfp computation time on average. This is because not all \mips in a program overlap with each other.

\newcommand{\meanRdef}{2.5\%}
\newcommand{\medianRdef}{2.5\%}
\newcommand{\avgRdef}{4.1\%}
\newcommand{\uptoRdef}{19\%}

\section{Experiments and Results}
\label{sec:Experiments}
In this section, we empirically compare \fpmfp solution with \mfp solution with respect to the precision of the solutions and the 
efficiency of their computation. Indeed, we did not find any other existing technique that is as generic as the \mfp computation, and 
eliminates the effect of infeasible paths from exhaustive data flow analysis
(existing techniques are explained in Section~\ref{sec:Relatedwork}).

The Section is organized as follows. First, we explain the experimental setup and benchmark characteristics 
(Section~\ref{sec:experiments.setup}). Next, we describe two client analyses on which we compared precision of \fpmfp and \mfp solutions
(Section~\ref{sec:experiments.precision}), followed by comparing the performance of \fpmfp and \mfp solutions 
(Section~\ref{sec:experiments.performance}). 
Lastly, we discuss infeasible path patterns observed in our benchmarks
(Section~\ref{sec:experiments.typesofmips})
 
\subsection{Experimental Setup}\label{sec:experiments.setup}
We performed our experiments on a 64 bit machine with 16GB RAM running Windows 7 with Intel core i7-5600U processor having a clock speed of 
2.6 GHz.
We used 30 C programs of up to 150KLoC size which contained 7 SPEC CPU-2006 benchmarks, 5 industry codesets, and 18 open source
 codesets~\cite{lee2012sound,muske2018repositioning,zhang2013diagnosis}. These programs were randomly chosen 
to cover a range of application areas with different program characteristics. The number of distinct 
\mipss range from 17 for the \emph{acpid} benchmark to 2.1k for the \emph{h264ref} benchmark. In these benchmarks, up to 61\% 
(geomertic mean 29\%) functions had at least one \mips in their \cfg. On average 77\% \mipss were overlapping with at least one 
other \mips.



\begin{table}[t!]
\centering
\small
\caption{Benchmark properties: three types of benchmarks that we used include 
  Open source~\cite{lee2012sound,muske2018repositioning,zhang2013diagnosis}, Industry, and SPEC CPU 2006 benchmarks.
}
\label{tab:precision1}
\setlength{\tabcolsep}{1.2pt}
\renewcommand{\arraystretch}{1.2}
\begin{tabular}{|l|l|r|r|r|r|r|}
\hline
\multicolumn{7}{|c|}{\multirow{1}{*}{Benchmark Properties}}  \\ 
\hline 
\multicolumn{1}{|c|}{\multirow{4}{*}{Type}} &
\multicolumn{1}{|c|}{\multirow{4}{*}{Name}} & \multirow{3}{*}{KLOC} & \multirow{3}{*}{
\#\mipss} &\multirow{3}{*}{$\begin{array}{c}\#\text{Overlapping}\\ \mipss \end{array}$}& 
\multirow{3}{*}{$\begin{array}{c}\#\text{Balanced} \\ \mipss\end{array}$} &
\multicolumn{1}{|c|}{\multirow{3}{*}{$\begin{array}{c}\%\text{Funcs}\\ \text{Impacted}\end{array}$}} 
\\  
\multicolumn{1}{|c|}{}&\multicolumn{1}{|c|}{}&
\multicolumn{1}{|c|}{}&\multicolumn{1}{c|}{}&\multicolumn{1}{c|}{}&\multicolumn{1}{c|}{}&\multicolumn{1}{c|}{} 
\\
& &&& & & \multicolumn{1}{c|}{}\\ 
\hline
\multicolumn{1}{|c|}{\multirow{18}{*}{\rotatebox{90}{Open Source}}}
&1.acpid & 0.3 & 6 & 6 & 6&7  \\ \cline{2-7}
&2.polymorph & 0.4 & 24 & 24 & 17&11  \\ \cline{2-7}
&3.nlkain & 0.5 & 33 & 8 & 3&58  \\ \cline{2-7}
&4.spell & 0.6 & 14 & 11 & 12&23  \\ \cline{2-7}
&5.ncompress & 1 & 30 & 27 & 16&30  \\ \cline{2-7}
&6.gzip & 1.3 & 49 & 46 & 40&25  \\ \cline{2-7}
&7.stripcc & 2.1 & 107 & 97 & 74&30  \\ \cline{2-7}
&8.barcode-nc & 2.3 & 60 & 49 & 22&25  \\ \cline{2-7}
&9.barcode & 2.8 & 76 & 66 & 28&24  \\ \cline{2-7}
&10.archimedes & 5 & 118 & 86 & 53&38  \\ \cline{2-7}
&11.combine & 5.7 & 234 & 210 & 124&29  \\ \cline{2-7}
&12.httpd & 5.7 & 226 & 211 & 88&16  \\ \cline{2-7}
&13.sphinxbase & 6.6 & 213 & 199 & 109&16  \\ \cline{2-7}
&14.chess & 7.5 & 262 & 186 & 112&39  \\ \cline{2-7}
&15.antiword & 19.3 & 669 & 570 & 306&32  \\ \cline{2-7}
&16.sendmail & 20.3 & 726 & 681 & 349&32  \\ \cline{2-7}
&17.sudo & 45.8 & 250 & 223 & 148&15  \\ \cline{2-7}
&18.ffmpeg & 80 & 1363 & 1201 & 891&9  \\ \hline
\multicolumn{1}{|c|}{\multirow{7}{*}{SPEC}}
&19.mcf & 5.4 & 19 & 17 & 4&20  \\ \cline{2-7}
&20.bzip2 & 16 & 501 & 485 & 184&40  \\ \cline{2-7}
&21.hmmer & 16.5 & 595 & 522 & 270&20  \\ \cline{2-7}
&22.sjeng & 27 & 438 & 379 & 198&33  \\ \cline{2-7}
&23.milc & 30 & 497 & 375 & 141&38  \\ \cline{2-7}
&24.h264ref & 35.9 & 2189 & 1696 & 526&31  \\ \cline{2-7}
&445.gobmk & 150 & 1834 & 1435 & 756&13  \\ \hline 
\multicolumn{1}{|c|}{\multirow{3}{*}{Industry}}
&26.NZM & 2.6 & 52 & 43 & 34&43  \\ \cline{2-7}
&27.IC & 3.7 & 50 & 29 & 13&15  \\ \cline{2-7}
&28.AIS & 6 & 83 & 56 & 21&14  \\ \cline{2-7}
&29.FRSC & 6 & 42 & 21 & 1&15  \\ \cline{2-7}
&30.FX & 7.5 & 140 & 78 & 39&34  \\ \hline
\end{tabular}
\end{table}

We used the iterative fix-point based data flow analysis  
algorithm~\cite{aho2003compilers,khedker2009data} implemented in Java in a commercial 
static analysis tool \emph{TCS Embedded Code Analyzer}~\cite{TECA} for our data flow analysis. 
 Initially, the tool transforms the source code to an intermediate representation, and constructs \cfgs for procedures in the program. 
Later, it runs some basic inbuilt analyses like pointer analysis, alias analysis in flow sensitive and context insensitive manner, 
followed by constructing the program call graph. Since \fpmfp just lifts the abstract operations given in  the input \mfp specification, 
it remains oblivious to and respects the analysis specific choices  like use of aliasing, pointers, and field sensitivity. In our 
evaluation, we implemented the functional approach of inter-procedural analysis (described in Section~\ref{sec:interprocedural})
 for computing \mfp and \fpmfp solutions for 
potentially uninitialized variable analysis and reaching 
definitions analysis. The details of the same are discussed in 
Section~\ref{sec:experiments.precision}.


\begin{table}[ht]
\centering
\setlength{\tabcolsep}{8pt}
\renewcommand{\arraystretch}{1}
\caption{Reduction in the number of potentially uninitialized variable Alarms}
\label{tab:clientanalysis}
\begin{tabular}{|l|l|r|r|r|}
\hline
\multicolumn{2}{|c|}{\multirow{3}{*}{Benchmarks}}	   &\multicolumn{3}{c|}{Uninitialized} \\ 
	\multicolumn{1}{|c}{}			&\multicolumn{1}{c|}{}       &\multicolumn{3}{c|}{Variable Alarms}  \\ \cline{3-5}
	\multicolumn{1}{|c}{}			&\multicolumn{1}{c|}{}&\multicolumn{1}{c|}{\mfp} & \multicolumn{1}{c|}{\fpmfp} &\multicolumn{1}{c|}{reduction(\%)}
					 \\ \hline
\multicolumn{1}{|c|}{\multirow{16}{*}{\rotatebox{90}{Open Source}}}
 & 1.acpid & 1 & 1 & 0(0)\\ \cline{2-5}
 & 2.polymorph & 4 & 4 & 0(0)\\ \cline{2-5}
 & 3.nlkain & 4 & 0 & 4(100)\\ \cline{2-5}
 & 4.spell & 3 & 3 & 0(0)\\ \cline{2-5}
 & 5.ncompress & 7 & 7 & 0(0)\\ \cline{2-5}
 & 6.gzip & 0 & 0 & - \\ \cline{2-5}
 & 7.stripcc & 27 & 1 & 26(96.30)\\ \cline{2-5}
 & 8.barcode-nc & 0 & 0 & - \\ \cline{2-5}
 & 9.barcode & 2 & 0 & 2(100)\\ \cline{2-5}
 & 10.archimedes & 61 & 56 & 5(8.20)\\ \cline{2-5}
 & 11.combine & 63 & 63 & 0(0)\\ \cline{2-5}
 & 12.httpd & 117 & 117 & 0(0)\\ \cline{2-5}
 & 13.sphinxbase & 46 & 43 & 3(6.52)\\ \cline{2-5}
 & 14.chess & 16 & 16 & 0(0)\\ \cline{2-5}
 & 15.antiword & 18 & 18 & 0(0)\\ \cline{2-5}
 & 16.sendmail & 103 & 102 & 1(0.97)\\ \cline{2-5}
 & 17.sudo & 62 & 58 & 4(6.45)\\ \cline{2-5}
 & 18.ffmpeg & 124 & 112 & 12(9.68)\\ \hline 
\multicolumn{1}{|c|}{\multirow{7}{*}{\rotatebox{90}{SPEC}}} 
 & 19.mcf & 5 & 5 & 0(0)\\ \cline{2-5}
 & 20.bzip2 & 72 & 35 & 37(51.39)\\ \cline{2-5}
 & 21.hmmer & 636 & 629 & 7(1.10)\\ \cline{2-5}
 & 22.sjeng & 19 & 19 & 0(0)\\ \cline{2-5}
 & 23.milc & 188 & 168 & 20(10.64)\\ \cline{2-5}
 & 24.h264ref & 52 & 48 & 4(7.69)\\ \cline{2-5} 
 & 25.gobmk & 1216 & 1199 & 17(1.39)\\ \hline 
\multicolumn{1}{|c|}{\multirow{5}{*}{\rotatebox{90}{Industry}}}
 & 26.NZ & 3 & 0 & 3(100)\\ \cline{2-5}
 & 27.IC & 5 & 5 & 0(0)\\ \cline{2-5}
 & 28.AIS & 0 & 0 & - \\ \cline{2-5}
 & 29.FRSC & 3 & 3 & 0(0)\\ \cline{2-5}
 & 30.FX & 84 & 84 & 0(0)\\ \hline
\end{tabular}
\end{table}

\begin{table}[htp]
\centering
\small
\caption{Reduction in the number of def-use pairs:  1) \mfp (resp. \fpmfp) column shows the sum of number of def-use pairs 
 at all program nodes $n$ in the \cfg of a program for global and local variables. 
}
\label{tab:precision1}
\setlength{\tabcolsep}{5.8pt}
\renewcommand{\arraystretch}{1.3}
\begin{tabular}{|l|l|r|r|r|}
\hline
\multicolumn{2}{|c|}{\multirow{1}{*}{Benchmarks}} & \multicolumn{3}{c|}{Reaching Definition Analysis}  \\ 
\hline 
\multicolumn{1}{|c|}{\multirow{2}{*}{Type}} &
\multicolumn{1}{|c|}{\multirow{2}{*}{Name}} & \multicolumn{2}{c|}{\#def-use pairs} & 
\multicolumn{1}{c|}{\multirow{2}{*}{reduction(\%)}} 
\\  \cline{3-4}
\multicolumn{1}{|c|}{}&\multicolumn{1}{|c|}{}&
\multicolumn{1}{c|}{}&\multicolumn{1}{c|}{}& \multicolumn{1}{c|}{}\\
 &&  \mfp & \multicolumn{1}{c|}{\fpmfp} & \multicolumn{1}{c|}{} \\ 
\hline
\multicolumn{1}{|c|}{\multirow{18}{*}{\rotatebox{90}{Open Source}}}
  &1.acpid & 156 & 156 & 0(0.00) \\ \cline{2-5}
  &2.polymorph & 228 & 228 & 0(0.00) \\ \cline{2-5}
  &3.nlkain & 1042 & 965 & 77(7.39) \\ \cline{2-5}
  &4.spell & 516 & 515 & 1(0.19) \\ \cline{2-5}
  &5.ncompress & 1201 & 1175 & 26(2.16) \\ \cline{2-5}
  &6.gzip & 3423 & 3401 & 22(0.64) \\ \cline{2-5}
  &7.stripcc & 2703 & 2645 & 58(2.15) \\ \cline{2-5}
  &8.barcode-nc & 3051 & 3007 & 44(1.44) \\ \cline{2-5}
  &9.barcode & 3709 & 3653 & 56(1.51) \\ \cline{2-5}
  &10.archimedes & 44337 & 44216 & 121(0.27) \\ \cline{2-5}
  &11.combine & 16618 & 15859 & 759(4.57) \\ \cline{2-5}
  &12.httpd & 10475 & 10072 & 403(3.85) \\ \cline{2-5}
  &13.sphinxbase & 9641 & 8482 & 1159(12.02) \\ \cline{2-5}
  &14.chess & 31386 & 31303 & 83(0.26) \\ \cline{2-5}
  &15.antiword & 68889 & 60144 & 8745(12.69) \\ \cline{2-5}
  &16.sendmail & 102812 & 101470 & 1342(1.31) \\ \cline{2-5}
  &17.sudo & 14391 & 14211 & 180(1.25) \\ \cline{2-5}
  &18.ffmpeg & 89148 & 86607 & 2541(2.85) \\ \hline
\multicolumn{1}{|c|}{\multirow{7}{*}{SPEC}}
  &19.mcf & 3570 & 3565 & 5(0.14) \\ \cline{2-5}
  &20.bzip2 & 82548 & 77967 & 4581(5.55) \\ \cline{2-5}
  &21.hmmer & 70597 & 70021 & 576(0.82) \\ \cline{2-5}
  &22.sjeng & 77602 & 76959 & 643(0.83) \\ \cline{2-5}
  &23.milc & 18133 & 17828 & 305(1.68) \\ \cline{2-5}
  &24.h264ref & 426346 & 409325 & 17021(3.99) \\ \cline{2-5}
  &25.gobmk &   213243 & 184142 & 29101(13.65)\\ \hline
\multicolumn{1}{|c|}{\multirow{3}{*}{Industry}}
  &26.NZ & 2652 & 2651 & 1(0.04) \\ \cline{2-5}
  &27.IC & 3111 & 3063 & 48(1.54) \\ \cline{2-5}
  &28.AIS & 2475 & 2414 & 61(2.46) \\ \cline{2-5}
  &29.FRSC & 1297 & 1295 & 2(0.15) \\ \cline{2-5}
  &30.FX & 4971 & 4925 & 46(0.93) \\ \hline
\end{tabular}
\end{table}


\begin{table}[htp]
\centering
\setlength{\tabcolsep}{1.2pt}
\renewcommand{\arraystretch}{1.3}
\caption{Comparing performance of \mfp and \fpmfp computation. \emph{Prep.} represents the \mips detection phase. Note the analysis times are
not comparable with gcc or clang, because our implementation is in Java.}
\label{tab:perform}
\begin{tabular}{|l|l|r|r|r|r|r|r|r|r|}
\hline
\multicolumn{2}{|c|}{\multirow{2}{*}{Benchmarks}}
				&\multicolumn{4}{c|}{Analysis Time (Sec)}     &\multicolumn{4}{c|}{Memory Consumption (MB)}\\ \cline{3-10}
\multicolumn{1}{|c}{}&\multicolumn{1}{c|}{}&\multicolumn{1}{c|}{Prep} &\multicolumn{1}{c|}{\mfp} & 
\multicolumn{1}{c|}{\fpmfp}&
\multicolumn{1}{c|}{$\begin{array}{c}\text{Incre-}\\ \text{ase(x)}\end{array}$}
				&\multicolumn{1}{c|}{Prep} &\multicolumn{1}{c|}{\mfp} & \multicolumn{1}{c|}{\fpmfp}&
				\multicolumn{1}{c|}{$\begin{array}{c}\text{Incre-}\\ \text{ase(x)}\end{array}$}
					\\ \hline
\multicolumn{1}{|c|}{\multirow{16}{*}{\rotatebox{90}{Open Source}}}
&1.acpid &0 & 0 & 0 & 0		         &0 & 0 & 0 & 0\\ \cline{2-10}
&2.polymorph &0 & 0 & 0 & 0   &1 & 0 & 0 & 0\\ \cline{2-10}
&3.nlkain &0 & 0 & 0 & 0           &1 & 1 & 1 & 0\\ \cline{2-10}
&4.spell &0 & 0 & 0 & 0            &1 & 1 & 0 & 0\\ \cline{2-10}
&5.ncompress &1 & 0 & 1 & 1(1)        &2 & 1 & 4 & 2(2)\\ \cline{2-10}
&6.gzip &1 & 1 & 5 & 4(4)             &5 & 10 & 24 & 14(1.4)\\ \cline{2-10}
&7.stripcc &3 & 1 & 6 & 5(5)          &8 & 5 & 21 & 15(3)\\ \cline{2-10}
&8.barcode-nc  &2 & 1 & 2 & 1(1)      &6 & 8 & 6 & -1(-0.1)\\ \cline{2-10}
&9.barcode &2 & 2 & 2 & 0          &7 & 10 & 8 & -1(-0.1)\\ \cline{2-10}
&10.archimedes &6 & 3 & 32 & 29(10)    &24 & 44 & 156 & 112(2.5)\\ \cline{2-10}
&11.combine &5 & 3 & 5 & 2(0.6)         &16 & 23 & 19 & -3(-0.1)\\ \cline{2-10}
&12.httpd &40 & 14 & 19 & 5(0.3)        &28 & 39 & 21 & -18(-0.46)\\ \cline{2-10}
&13.sphinxbase &7 & 3 & 3 & 0      &13 & 16 & 7 & -9(-0.56)\\ \cline{2-10}
&14.chess  &13 & 7 & 30 & 23(3)       &31 & 93 & 191 & 98(1)\\ \cline{2-10}
&15.antiword &13 & 16 & 82 & 66(4)    &47 & 101 & 218 & 117(1)\\ \cline{2-10}
&16.sendmail &110 & 142 & 2060 & 1918(13)  &81 & 340 & 1548 & 1208(3.5)\\ \cline{2-10}
&17.sudo &15 & 9 & 17 & 8(1)          &35 & 71 & 19 & -52(-0.7)\\ \cline{2-10}
&18.ffmpeg &234 & 51 & 80 & 29 (0.5)    &158 & 266 & 90 & -175(-0.6)\\ \hline
\multicolumn{1}{|c|}{\multirow{8}{*}{\rotatebox{90}{SPEC 2006}}}
&19.mcf     &1 & 1 & 1 & 0        &0 & 3 & 3 & 0\\ \cline{2-10}
&20.bzip2 &8 & 7 & 69 & 62 (9)        &28 & 44 & 90 & 46(1)\\ \cline{2-10}
&21.hmmer &14 & 9 & 23 & 14(1.5)       &50 & 67 & 58 & -8(-0.1)\\ \cline{2-10}
&22.sjeng  &10 & 16 & 62 & 46(2.8)     &27 & 122 & 215 & 92(0.75)\\ \cline{2-10}
&23.milc   &6 & 4 & 7 & 3(0.7)         &25 & 25 & 14 & -11(-0.4)\\ \cline{2-10}
&24.h264ref &60 & 115 & 451 & 336(2.9)   &131 & 1003 & 1644 & 640(0.6)\\ \cline{2-10}
&25.gobmk &4218&1696&9818&8122(4.7)      &390&9365&17392&8032(0.85)\\ \hline
\multicolumn{1}{|c|}{\multirow{5}{*}{\rotatebox{90}{Industry}}}
&26.NZM   &1 & 1 & 2 & 1(1)    &3 & 8 & 7 & -1(-0.1)    \\ \cline{2-10}
&27.IC   &5 & 4 & 3 & -1(-0.25)     &11 & 20 & 9 & -11(-0.5)\\ \cline{2-10}
&28.FRSC   &3 & 2 & 1 & -1(-0.5)   &-15 & 10 & 4 & -6(-0.6)  \\ \cline{2-10}
&29.AIS  &4 & 3 & 4 & 1(0.3)     &9  & 20 & 16 & -4(-0.2)   \\ \cline{2-10}
&30.FX &7 & 3 & 7 & 4(1.3)       &20 & 35 & 47 & 11(0.3)\\ \hline
\end{tabular}
\vspace{-10pt}
\end{table}

\newcommand{\uptoOptO}{99\%}
\newcommand{\avgOptO}{85\%}
\newcommand{\meanOptO}{78\%}

\begin{table}[htp]
\centering
\setlength{\tabcolsep}{1.2pt}
\small
\renewcommand{\arraystretch}{1.3}
\caption{Impact of Optimization 1 on the \fpmfp computation (in Optimization 1, we consider two \mips with the same end edge
as equivalent, and subsequently merge the pairs that contain equivalent \mips).}
\label{tab:performance.optO}
\begin{tabular}{|l|l|r|r|r|}
\hline
\multicolumn{2}{|c|}{\multirow{4}{*}{Benchmarks}}
				&\multicolumn{3}{c|}{Number of Distinct \mips}    \\ 
\multicolumn{2}{|c|}{\multirow{2}{*}{}}  
				&\multicolumn{3}{c|}{Before and After Optimization 1}     \\ \cline{3-5}
\multicolumn{1}{|c}{}&\multicolumn{1}{c|}{} &\multicolumn{1}{c|}{Before} & \multicolumn{1}{c|}{After} &
\multicolumn{1}{c|}{Reduction(\%)}
					\\ \hline
\multicolumn{1}{|c|}{\multirow{16}{*}{\rotatebox{90}{Open Source}}}
&1.acpid & 83 & 6 & 77(92.77)     \\ \cline{2-5}
&2.polymorph 0.96 & 65 & 24 & 41(63.08)         \\ \cline{2-5}
&3.nlkain & 35 & 33 & 2(5.71)                 \\ \cline{2-5}
&4.spell & 99 & 14 & 85(85.86)                 \\ \cline{2-5}
&5.ncompress & 14163 & 30 & 14133(99.79)       \\ \cline{2-5}
&6.gzip & 23524624 & 49 & 23524575(99.99)     \\ \cline{2-5}
&7.stripcc & 774751 & 107 & 774644(99.99)      \\ \cline{2-5}
&8.barcode-nc & 208 & 60 & 148(71.15)          \\ \cline{2-5}
&9.barcode & 3851 & 76 & 3775(98.03)          \\ \cline{2-5}
&10.archimedes & 1054292 & 118 & 1054174(99.99)  \\ \cline{2-5}
&11.combine & 27061 & 234 & 26827(99.14)         \\ \cline{2-5}
&12.httpd & 76591463106 & 226 & 76591462880(99.99)   \\ \cline{2-5}
&13.sphinxbase & 6463 & 213 & 6250(96.70)           \\ \cline{2-5}
&14.chess  & 19701157845 & 262 & 19701157583(99.99)   \\ \cline{2-5}
&15.antiword & 1446759232593972 & 669 & 1446759232593303(99.99)    \\ \cline{2-5}
&16.sendmail & 94082730655046 & 726 & 94082730654320(99.99)        \\ \cline{2-5}
&17.sudo & 549565 & 250 & 549315(99.95)        \\ \cline{2-5}
&18.ffmpeg & 5019442696182 & 1363 & 5019442694819(99.99)      \\ \hline
\multicolumn{1}{|c|}{\multirow{8}{*}{\rotatebox{90}{SPEC 2006}}}
&19.mcf     & 44 & 19 & 25(56.82)            \\ \cline{2-5}
&20.bzip2 & 115451494 & 501 & 115450993(99.99)   \\ \cline{2-5}
&21.hmmer & 8606721 & 595 & 8606126(99.99)        \\ \cline{2-5}
&22.sjeng  & 1508084297532741 & 438 & 1508084297532303(99.99)    \\ \cline{2-5}
&23.milc   & 2322 & 497 & 1825(78.60)         \\ \cline{2-5}
&24.h264ref & 51269021916 & 2189 & 51269019727(99.99)   \\ \cline{2-5}
&25.gobmk &$>2^{64}$&1834&$>2^{64}$(99.99)    \\ \hline
\multicolumn{1}{|c|}{\multirow{5}{*}{\rotatebox{90}{Industry}}}
&26.NZM       & 1064 & 52 & 1012(95.11)     \\ \cline{2-5}
&27.IC        & 158 & 50 & 108(68.35)           \\ \cline{2-5}
&28.AIS       & 176 & 83 & 93(52.84)           \\ \cline{2-5}
&29.FRSC      & 89 & 42 & 47(52.81)          \\ \cline{2-5}
&30.FX        & 704 & 140 & 564(80.11)            \\ \hline
\end{tabular}
\vspace{-10pt}
\end{table}

\newcommand{\uptoOptT}{99\%}
\newcommand{\avgOptT}{48.9\%}
\newcommand{\meanOptT}{19.8\%}

\begin{table}[htp]
\centering
\setlength{\tabcolsep}{3pt}
\small
\renewcommand{\arraystretch}{1.3}
\caption{Impact of Optimization 2 on the \fpmfp computation. Timeout indicates the analysis did not finish within
10K seconds.}
\label{tab:performance.optT}
\begin{tabular}{|l|l|r|r|r|}
\hline
\multicolumn{2}{|c|}{\multirow{4}{*}{Benchmarks}}&
				\multicolumn{3}{c|}{Analysis Time (in Seconds) }\\ 
\multicolumn{2}{|c|}{\multirow{2}{*}{}}  
				   &\multicolumn{3}{c|}{Before and After Optimization 2}\\ \cline{3-5}
\multicolumn{1}{|c}{}&\multicolumn{1}{c|}{} 
				 &\multicolumn{1}{c|}{Before} & \multicolumn{1}{c|}{After} 
				  &\multicolumn{1}{c|}{Reduction(\%)}
					\\ \hline
\multicolumn{1}{|c|}{\multirow{16}{*}{\rotatebox{90}{Open Source}}}
&1.acpid                  &0&0&0 \\ \cline{2-5}
&2.polymorph 0.96          &0&0&0 \\ \cline{2-5}
&3.nlkain                  &0&0&0 \\ \cline{2-5}
&4.spell                  &0&0&0 \\ \cline{2-5}
&5.ncompress           &3&1&2(66) \\ \cline{2-5}
&6.gzip                &22&5&17(77) \\ \cline{2-5}
&7.stripcc             &55&6&49(89) \\ \cline{2-5}
&8.barcode-nc        &2&2&0 \\ \cline{2-5}
&9.barcode              &2&2&0 \\ \cline{2-5}
&10.archimedes          &53&32&21(39) \\ \cline{2-5}
&11.combine            &19&5&14(73) \\ \cline{2-5}
&12.httpd              &290&19&271 (93) \\ \cline{2-5}
&13.sphinxbase        &4&3&1(25) \\ \cline{2-5}
&14.chess              &76&30&46(60) \\ \cline{2-5}
&15.antiword          &705&82&623(88) \\ \cline{2-5}
&16.sendmail          &Timeout&2060&- \\ \cline{2-5}
&17.sudo         &64&17&47(73) \\ \cline{2-5}
&18.ffmpeg            &Timeout&80&- \\ \hline
\multicolumn{1}{|c|}{\multirow{8}{*}{\rotatebox{90}{SPEC 2006}}}
&19.mcf              &1&1&0 \\ \cline{2-5}
&20.bzip2        &931&69&862 (92)\\ \cline{2-5}
&21.hmmer          &143&23&120(83) \\ \cline{2-5}
&22.sjeng            &501&62&439(87) \\ \cline{2-5}
&23.milc             &11&7&4(36) \\ \cline{2-5}
&24.h264ref         &2977&451&2526(84) \\ \cline{2-5}
&25.gobmk          &Timeout&9818&- \\ \hline
\multicolumn{1}{|c|}{\multirow{5}{*}{\rotatebox{90}{Industry}}}
&26.NZM            &4&2&2(50) \\ \cline{2-5}
&27.IC             &6&3&3(50)) \\ \cline{2-5}
&28.AIS            &7&4&3(42) \\ \cline{2-5}
&29.FRSC           &3&1&2(66) \\ \cline{2-5}
&30.FX             &9&7&2(22) \\ \hline
\end{tabular}
\vspace{-10pt}
\end{table}


\subsection{Comparing the Precision of \fpmfp and \mfp Solutions} \label{sec:experiments.precision}

\subsection{Reduction in the Number of Potentially Uninitialized Variable Alarms}
A potentially uninitialized variable analysis reports the variables that
are used at a program point but are not initialized along at least 
one path that reaches the program point. We computed the potentially uninitialized variables as follows: first,
we implemented the \emph{must defined variables} analysis that computes the set of non-array variables that are 
defined along all \cfps reaching a program point. Next, we take 
the complement of the \emph{must defined} data flow results to report the variables that are used at a program point 
$p$ but are not initialized along at least one \cfp that reaches $p$. We call such \cfps as \emph{witness \cfps}.

As shown in Table~\ref{tab:clientanalysis}, we computed two sets of potentially uninitialized variable alarms corresponding
to the \mfp and \fpmfp solutions of must defined variable analysis. In particular, the potentially uninitialized variable analysis 
that used the \fpmfp solution of \emph{must defined} analysis reported up to \uptoUninit\ (average \avgUninit, geo. mean \meanUninit) 
fewer alarms compared to when it used the \mfp solution. In the reduced alarm cases, all the witness \cfps were found infeasible. 

More specifically, in 14 out of 27 benchmarks the \fpmfp solution helped achieve positive reduction in the uninitialized 
variable alarms, while in 13 benchmarks there was no reduction in the alarms. In the latter case the following two scenarios 
were observed: in the first scenario, the addresses of uninitialized variables are passed as parameter 
to library functions whose source code is not available and hence its behavior is conservatively approximated
(i.e, the analysis assumed that the variables are not initialized inside the functions), 
and in the second scenario a 
variable is conditionally initialized inside the callee procedure where the evaluation of the condition depends on 
the calling context thereby necessitating a context sensitive analysis to eliminate the false positive alarm 
(our implementation is context insensitive). 

\subsection{Reduction in the Number of Def-Use Pairs}
We implemented the classic reaching definitions analysis~\cite{aho2003compilers,khedker2009data} 
(non SSA based) in which for each program point, we compute the set of definitions that reach the point.  We measured the effect of the 
reduction in the number of reaching definitions on Def-Use pair computation, which is a client analysis of the reaching definitions analysis. 
In this, we pair the definition of a variable with its reachable use point. 

In the \fpmfp computation, 
the def and use points that are connected with each other by only infeasible paths are eliminated. However, 
we found that most def-use points are connected by at least one feasible \cfp, and hence a 100\% reduction is  
not achievable by any technique.

As shown in Table~\ref{tab:precision1}, we found that up to \uptoDefUse\ (average \avgDefUse, geometric mean \meanDefUse) fewer def-use pairs were reported when using 
\oursol in place of the \mfp solution. In particular,
 in 28 out of 30 benchmarks the \fpmfp solution lead to fewer def-use pairs than the \mfp solution. 
In general, we observed that the precision increased with increase in the number of \mipss. For example, the \emph{gobmk} benchmark 
contains a large number of \mips (1.9k), consequently we observed significant (13\%) reduction in the number of def-use pairs. On the 
other hand, the \emph{acpid} benchmark had only 17 \mips, consequently we did not see any reduction in the number of def-use pairs. 
As an outlier case, on the \emph{sphinxbase} benchmark that had only 216  \mips we observed 12\% reduction in def-use pairs. Here, we 
observed that the end edges of many \mips are reachable only along infeasible \cfps, hence in \fpmfp computation the data flow along
these edges is completely blocked, thus all the corresponding def-use pairs (at the program points pre-dominated by the end edges) are eliminated.

The def-use pairs 
have multifold applications; in particular, they are useful for program slicing, program dependency graph 
creation, program debugging, and test case generation. For example in program testing, some coverage criteria 
mandate that each def-use pair should be covered by at least one test case. Here, up to \uptoDefUse\ fewer def-use pairs means fewer manual 
test cases need to be written to cover these pairs, which is a significant reduction in the manual time and efforts.

\subsection{Comparing the Performance of \fpmfp and \mfp Computations}\label{sec:experiments.performance}
Computation of the \oursol was less efficient than the corresponding \mfp solutions. The details of the relative memory and time 
consumption of both the analyses for reaching definitions computation are presented in Table \ref{tab:perform}. The \emph{Prep.} 
column represents the time and memory consumption for infeasible path detection phase. This is performed once for each program.

The \fpmfp analysis took \avgTime times more time than \mfp analysis on average (geo. mean \meanTime). In general, the analysis
 time increased with increase in the number of \mips in the program. Apart from this, the \fpmfp analysis consumed
\avgMemory times more memory than the \mfp analysis on average. However, on 50\%  of the bechmarks the \fpmfp analysis consumed less 
memory than the \mfp analysis because the  definitions reaching along infeasible \cfps are ignored during the analysis. 
 
\textbf{The effect of optimizations for scalability on analysis time}

Table~\ref{tab:performance.optO} shows the reduction in the number of distinct \mips after applying Optimization 1 
(Section~\ref{subsection:optimization.one}). Here, two 
\mips are considered equivalent if they have the same end edge, in this case we can merge two pairs if they contain 
equivalent \mips. We found that if we count \mips with same end edge single time, then the total number of \mips 
reduces by up to \uptoOptO (average \avgOptO, geometric mean \meanOptO).  This leads to reduction in number of pairs 
at various program points.  


Table~\ref{tab:performance.optT} shows the effectiveness of optimization 2 (Section~\ref{sec:optimization.cso}) on top of optimization 1. 
In particular, it shows the analysis time before and after 
applying the optimization 2. We found that 3 out of 30 applications timed out (i.e., did not finish within 10K seconds) 
when we disabled optimization 2. Similarly, for remaining applications the analysis time increased by up to \uptoOptT 
 (average \avgOptT, geometric mean \meanOptT) after disabling optimization 2.

\begin{figure*}[htp]
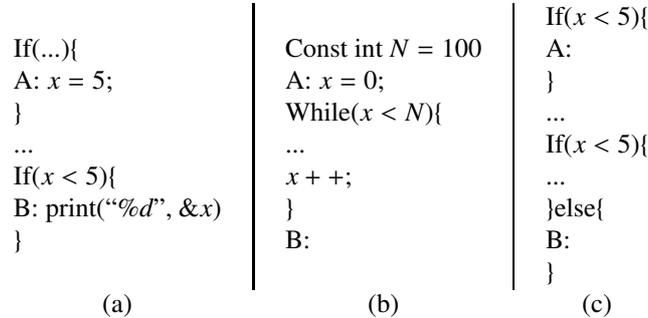
 
\psset{unit=1mm}
\centering
\begin{tabular}{l|l|l}
\begin{tabular}{l}
If(...)\{\\
A:    $x=5$;\\
\}\\
...\\
If($x<5$)\{\\
B:    print(``\%$d$'', \&$x$)\\
\}\\
\end{tabular}
&
\begin{tabular}{l}
Const int $N=100$\\
A:    $x=0$;\\
While($x<N$)\{\\
      ... \\
       $x++$;\\
\}\\
B: \\
\end{tabular}
&
\begin{tabular}{l}
If($x<5$)\{\\
A:    \\
\}\\
...\\
If($x< 5$)\{    \\
...\\
\}else\{\\
B:\\
\}\\
\end{tabular}
\\
\multicolumn{1}{c}{(a)}&\multicolumn{1}{c}{(b)}&\multicolumn{1}{c}{(c)}
\end{tabular}
\vspace{-10pt}
\caption{Illustrating the types of \mips observed. In each of the example above there is a \mips 
going from point A to point B. The variable $x$ is not modified in 
the code portion represented by three dots ($\ldots$) .}
\vspace{-10pt}
\label{fig:mips.observed}
\end{figure*}

\subsection{Types of \mips Observed}\label{sec:experiments.typesofmips}
In general case, knowing whether a path is infeasible or not is undecidable. Nevertheless, a large number 
of infeasible paths in a program can be identified using statically detectable correlations between two 
branch conditions or between a branch condition and an assignment statement. Indeed, in our work, we 
detected a large number of \mips using Bodik's approach (described in Appendix A). 

We observe that a lot of \mips in our benchmarks belong to one of the following two types resulting
from correlation between an assignment statement and a branch statement or two branch statements.
\begin{enumerate}
\item The first type involves assignment of constant value to a variable along some node in a path 
followed by the variable being used in some conditional expression that performs either \emph{equality} or 
\emph{less than} or \emph{greater than} check on the variable later in the path.
Two typical examples of this type are shown Figure~\ref{fig:mips.observed}a and \ref{fig:mips.observed}b.
In the first example a variable is assigned a constant value inside
the if branch (point A) and subsequently used in other branch condition that appears later in the path (point B).
The second example contains a loop that executes at least once hence the path that goes from point A to point B without 
executing the loop is infeasible. 

\item The second type involves  two syntactically same conditions used in two branches at different places
in a path and the variables in the conditions are not modified along the path that connects the 
branches. For example in Figure~\ref{fig:mips.observed}c the path that goes from point A to point B is infeasible.
\end{enumerate}

\subsection{Summary of Empirical Observations}
Previous empirical evidence~\cite{bodik1997refining} on linux kernel code shows 9-40\% of conditional statements contribute to at least 
one infeasible \cfp. In our benchmarks up to 61\% (geometric mean 29\%) functions had at least one \mips in their \cfg. Precise elimination of 
data flow values reaching from infeasible \cfps has allowed us to reduce the number of def-use pairs by up to \uptoDefUse\   
with an average of \avgDefUse\ and a geometric mean of \meanDefUse\ over \mfp solution. Similarly, the reduction in the  
potentially uninitialized variable alarms was up to \uptoUninit\ (average \avgUninit, geo. mean \meanUninit).


\section{Related Work}\label{sec:Relatedwork}
In this section, we compare \fpmfp with existing approaches of improving data flow precision.
In particular, we
begin by classifying the existing approaches that improve the  precision of data flow analysis into two categories.
The approaches in the first category avoid the effect of infeasible paths from data flow analysis 
(Section~\ref{sec:relatedwork.avoidip}), while the
approaches in second category avoid the join points (of two or more \cfps) that lead to 
imprecision in the analysis results 
(Section~\ref{sec:relatedwork.avoidij}). Lastly, we explain how \fpmfp differs from approaches in both the categories.

\begin{figure}[t]
\begin{pspicture}(0,20)(25,130)

\putnode{n00}{origin}{73}{123}{$\begin{array}{c}\text{Approaches to Improve}\\ \text{Data Flow Precision}
														   \end{array}$}
\putnode{n00l}{n00}{-38}{-30}{$\begin{array}{c}\text{Avoid}\\ 
																	 \text{Infeasible Paths}\end{array}$}
		\putnode{n0}{n00l}{-22}{-30}{$\begin{array}{c}\text{Analysis}\\ \text{Dependent}\\ \text{Approach}\end{array}$}
		\putnode{n01}{n0}{0}{-30}{$\begin{array}{c}\text{Selective} \\ \text{Path Sensitivity}\end{array}$}

		\putnode{n1}{n00l}{21}{-30}{$\begin{array}{c}\text{Analysis}\\ \text{Independent}\\ \text{Approach}\end{array}$}
		\putnode{n1l}{n1}{-12}{-30}{$\begin{array}{c}\text{\cfg}\\ \text{Restructuring} \end{array}$}
		\putnode{n1r}{n1}{12}{-30}{$\begin{array}{c}\text{Our}\\ \text{Approach}\end{array}$}

\putnode{n00r}{n00}{27}{-30}{$\begin{array}{c}\text{Avoid}\\ 
																	 \text{Imprecise Joins}\end{array}$}
		\putnode{n00r1}{n00r}{-13}{-30}{$\begin{array}{c}\text{\cfg} \\ \text{Restructuring}\end{array}$}
		\putnode{n00r2}{n00r}{13}{-30}{$\begin{array}{c}\text{Trace}\\ \text{Partitioning} \end{array}$}

\ncline[nodesep=0.5]{->}{n00}{n00l}
\ncline[nodesep=0.5]{->}{n00}{n00r}
	\ncline[nodesep=0.5]{->}{n00l}{n0}
	\ncline[nodesep=0.5]{->}{n00l}{n1}
	\ncline{->}{n1}{n1l}
	\ncline{->}{n1}{n1r}
	\ncline{->}{n0}{n01}
	
\ncline[nodesep=0.5]{->}{n00r}{n00r1}
\ncline[nodesep=0.5]{->}{n00r}{n00r2}
\end{pspicture}
\caption{Categorization of approaches that improve data flow precision by 
either avoiding infeasible paths or imprecise joins. 
}
\label{fig:relatedwork.overview}
\end{figure}
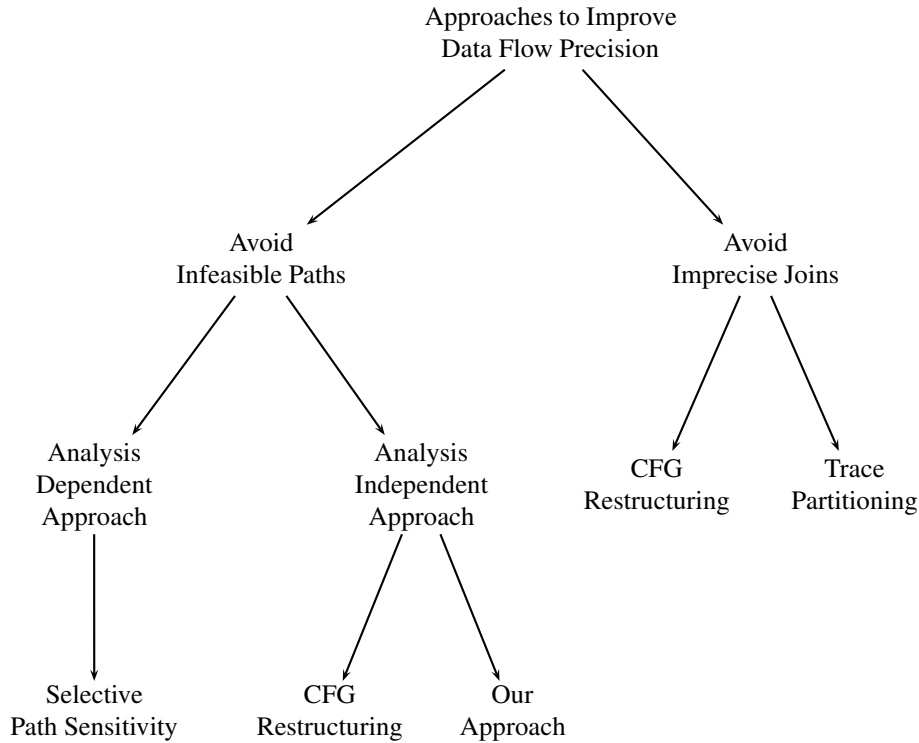

\subsection{Approaches that Avoid Infeasible Paths}\label{sec:relatedwork.avoidip}


In general, presence of infeasible paths in programs is well known~\cite{zhu2019detecting,hedley1985causes,
malevris1990predictive,lee2019infeasible,jiang2019approach,gong2019detecting,sewell2016complete,zeng2017type}. 
Hedley et al.~\cite{hedley1985causes} presented a detailed analysis of the causes and effects of infeasible paths in 
programs. Malevris et al.~\cite{malevris1990predictive} observed that the greater the number of conditional 
statements contained in a path, the greater the probability of the path being 
infeasible. Bodik et al.~\cite{bodik1997refining} found that around 9-40\% of the conditional statements 
in programs show statically detectable correlation with infeasible control flow paths.

Many approaches have been proposed to eliminate the effect of infeasible paths from data flow analysis. 
We classify these approaches in two categories as shown in Figure~\ref{fig:relatedwork.overview}. The first 
category includes approaches 
that are analysis dependent (Section~\ref{sec:relatedwork.analysisspecific}), while the second category 
includes the approaches that are analysis independent (Section~\ref{sec:relatedwork.analysisindependent}).  
We describe and relate approaches in each of these categories with \fpmfp below.

\subsection{Analysis Dependent Approaches}\label{sec:relatedwork.analysisspecific}
In this section, we describe approaches that eliminate effect of infeasible paths from data flow analysis
in analysis dependent way. In particular, these approaches either depend on analysis specific heuristics, 
 or they detect infeasible paths on the fly during a data flow analysis (and hence they detect infeasible 
paths separately for each different data flow analysis over the same program thereby increasing the 
analysis cost). Specifically, these approaches use the idea of path sensitive analysis to 
distinguish between the data flow values reaching along different classes of \cfps
~\cite{das2002esp,dhurjati2006path,hampapuram2005symbolic,xie2003archer,dillig2008sound,dor2004software}.



In a completely path sensitive analysis each path information 
is separately tracked along with the corresponding path constraints. 
If the path constraint is found unsatisfiable, the path information 
is not propagated further. This approach does not scale to most practical 
programs due to presence of loops and recursion leading to an unbounded 
number of program paths combined with data flow lattices containing possibly infinite values.

The above limitations of a completely path sensitive analysis are avoided by adapting a 
selective approach to path sensitive analysis~\cite{dor2004software,dillig2008sound,xie2003archer,
hampapuram2005symbolic,dhurjati2006path,das2002esp,thakur2008comprehensive,chen2007exploiting}. 
Here, they 
use selective path constraints governed by various 
heuristics for deciding which path information should be kept separately and which path information can be approximated at join 
points. 
These heuristics are often aligned with the end goal of the analysis. 
Our work differs from 
these approaches in that our work does not depend on analysis specific heuristics for eliminating 
the effect of infeasible paths, and therefore can be generically applied to all \mfp based data flow analyses.
Below, we describe the approaches that use selective path sensitive analyses.


Dor et al~\cite{dor2004software} proposed a precise path sensitive value flow analysis approach using value bit 
vectors. Here, they symbolically evaluate a program by
generating symbolic states that include both the execution
state and the value alias set. At merge points in the control
flow graph, if two symbolic states have the same value
alias set, they produce a single symbolic state by merging
their execution states. Otherwise, they process the symbolic
states separately. Like our work, they avoid the cost of full path sensitive
analysis, yet capture relevant correlations. Their work differs from our work in 
that their work is applicable only to separable data flow problems , while our work is applicable 
to any data flow problem that can be solved using \mfp computation.

Chen et al.~\cite{rui2006infeasible} proposed an algorithm for detecting infeasible paths by incorporating 
the branch correlations analysis into the data flow analysis technique.
Next, the detected infeasible paths are then used to improve the precision of structural testing by ignoring infeasible paths. 

Hampapuram et al.~\cite{hampapuram2005symbolic} performed a symbolic path simulation approach in which they executed 
each path symbolically and propagated the information along these paths. If a path is found infeasible in simulation its 
information is not propagated. A lightweight decision procedure is used to check the feasibility of a path during path simulation.

Dillig et al.~\cite{dillig2008sound} proposed a selective path sensitive analysis for modular programs. In this, they 
ignored the unobservable  variables that are the variables representing  the results of unknown functions (functions
unavailable for analysis); hidden system states (e.g., the garbage collector’s free list, operating system’s
process queue, etc.),
or imprecision in the analysis (e.g., selecting an unknown element of
an array). In particular, they observed that unobservable variables are useful within their natural
scope, for example, tests on unobservable variables can possibly be proven
mutually exclusive (e.g., testing whether the result of a malloc call
is null, and then testing whether it is non-null). However, outside of their natural scope 
unobservable variables provide no additional information  
and can be eliminated. 
Ignoring unobservable variables allowed them to reduce the number of distinctions that needs
to be maintained in the path sensitive analysis, for example, they did not distinguish between
paths that differ only in the values of unobservable variables. 
They have applied their work for proving a fixed set of program properties. In that, they
eliminate the effect of infeasible paths while proving the properties because the analysis
is path sensitive. However, they detect infeasible paths separately for
each different data flow analysis on the same underlying program, which does 
not happen in our approach.

Xie et al.~\cite{xie2003archer} focused on memory access errors (like
\emph{array index out of bounds} and \emph{null pointer dereferences}) because they may lead to non-deterministic and elusive 
crashes. They proposed a demand-driven path sensitive inter-procedural analysis to bound the values of both variables and 
memory sizes. For achieving efficiency, they restrict the computation to only those variables and memory sizes that directly 
or indirectly affect the memory accesses of interest. They use the following 
observation about \setc programs: most of the times memory sizes are passed as parameters to functions, along with the pointers to 
corresponding memory addresses; further, these sizes are used in conditions that guard the memory accesses. 
Hence, by tracking the actual values of the memory sizes the corresponding memory accesses can be evaluated as safe or unsafe. 
Xie et al. were able to detect many memory access errors in real code sets using this method. Moreover, they could eliminate the 
effect of some of the infeasible paths that reach the memory accesses of interest because they used a path sensitive analysis. 
Their work differs from our work on two fronts: first, they do a demand driven analysis while our work is exhaustive; second
in their work the infeasible paths are detected on the fly during a data flow analysis, so the same path may be detected 
multiple times for the same underlying program, while in our work infeasible paths are detected only once for a program.



Das et al.~\cite{das2002esp} proposed a property specific path sensitive analysis. In this at first, a 
set of program properties of interest are identified. Next, a partial 
path sensitive analysis is performed where two paths are separated iff they contain 
different values for the variables related to properties of interest. 

On somewhat similar lines, Dhurjati et al.~\cite{das2002esp} proposed an iterative refinement based 
approach to path sensitive data flow analysis. Here, at first, a path insensitive analysis 
is performed. Next, if the analysis is not able to prove the properties of interest then 
the analysis results are used to compute a set of predicates that are related to the 
property of interest . In the next iteration of the analysis, these predicates are 
used to differentiate paths along which predicate evaluation is different. If this partially 
path sensitive analysis proves the properties of interest then the analysis is terminated, 
else the analysis is repeated with a different set of predicates discovered in the 
recent iteration of the analysis. Thus, the analysis continues in multiple iterations.

Infeasible control flow paths is a property of programs instead of any particular data flow analysis. 
We use this observation to first detect infeasible paths in programs and propose an approach to improve 
the precision of any data flow analysis over the programs, unlike above approaches that remove the 
impact of infeasible paths from a particular data flow analysis.


\subsection{Analysis Independent Approaches}\label{sec:relatedwork.analysisindependent}
We now describe approaches that eliminate effect of infeasible paths from data flow analysis
in analysis independent way. In particular, these approaches do not depend on analysis specific heuristics, 
and they identify infeasible paths only once for a program. These approaches are classified into two 
categories depending on whether they use \cfg restructuring or not, as explained below.

\subsubsection{Approaches that Use \cfg Restructuring}\label{sec:relatedwork.cfgrestructuring}

Bodik et al.~\cite{bodik1997interprocedural} proposed an inter-procedural version of infeasible path 
detection and its application to improve the data flow precision. However, they use control flow graph 
restructuring which is not done in our approach. \cfg restructuring 
may blow up the size of the \cfg. Moreover, it does not take the 
advantage of analysis specific information, unlike \fpmfp that uses 
analysis specific information to dynamically discard the distinctions 
between \mips, where they do not lead to precision improvement as 
described in Optimization 2 (Section~\ref{sec:optimization.cso}).

Balakrishna et al.~\cite{balakrishnan2008slr} 
presented a technique for detecting infeasible
paths in programs using abstract interpretation. The technique uses a
sequence of path-insensitive forward and backward runs of an abstract
interpreter to infer paths in the control flow graph that cannot be exercised
in concrete executions of the program.  Next, they refined program \cfg by 
removing detected infeasible paths in successive iterations of a property verification.

Marashdih et al.~\cite{marashdih2017web} proposed a \cfg restructuring approach to eliminate 
infeasible paths for PHP programs. 
In particular, they proposed a methodology for the detection of XSS (Cross Site Scripting) 
from the PHP web application using static analysis. The methodology  eliminates the infeasible 
paths from the \cfg thereby reducing the false positive rate in the 
outcomes. On similar lines, we have found reduction in false positives in possibly uninitialized 
variable alarms but without using \cfg restructuring.

\subsubsection{Our Approach}
We used the work done by Bodik et al.~\cite{bodik1997refining} for detecting infeasible paths from 
the \cfg of a program. Next, we automatically lift any data flow analysis to an analysis that separates 
the data flow values reaching along known infeasible \cfps from the values that do not reach along
the infeasible \cfps. This allows us to block the data flow values that reach along infeasible \cfps
at a program point, thereby eliminating the effect of detected infeasible paths from data flow analyses 
without using \cfg restructuring and in a generic manner.

\subsection{Approaches that Avoid Imprecise Joins}\label{sec:relatedwork.avoidij}
In this section, we discuss approaches that improve data flow precision by avoiding imprecise joins i.e., 
the merging of information at the nodes that are shared by two or more paths, and that may lead to loss of 
precision.

\subsection{Approaches that Use Trace Partitioning}

In abstract interpretation, the trace partitioning approaches~\cite{handjieva1998refining,holley1981qualified,mauborgne2005trace} 
partition program traces where values of some variable or conditional expression differ in two selected traces. 
The variables or expressions are selected based on the alarms to be analyzed or some heuristic. 

Mauborgne et al.~\cite{mauborgne2005trace} used trace partitioning to delay the merging of information at join points
in \cfg where the join points lead to decrease in precision. They separate sets of traces before the join points and 
merge traces after the join points. The separation and merge points for traces are specified as input to the program in 
the form of pre-processing directives. Handjieva et al.~\cite{handjieva1998refining} proposed a restricted version of
trace partitioning where merging of parts in partitions was not allowed (Unlike Mauborgne et al.'s work).

The trace partitioning is similar to our approach because it keeps data flow values separately for 
each part (representing a set of traces) in partition. However, our approach does not rely on alarms, does not need the designer of an 
analysis to decide a suitable heuristic, and is not restricted to a particular analysis.
We lift partitioning from within an analysis to the infeasibility of control flow paths which is a fundamental
property of control flow paths independent of an analysis. This allows us to
devise an automatic approach to implement a practical trace partitioning.
An interesting aspect of our approach is that although it is oblivious to any analysis, it can be seen as dynamic
partitioning that uses 
infeasibility of control flow paths as criteria for partitioning.

\subsection{Approaches that Use \cfg Restructuring}
\cfg restructuring has also been used to avoid imprecise joins.
In particular, Aditya et al.~\cite{thakur2008comprehensive} proposed an approach in 
which at first they identify join points that lead to loss of precision in data flow analysis. 
Next, they improve the data flow precision by
eliminating these join points from control flow graph through control flow graph restructuring. 

Ammon et al.~\cite{ammons1998improving,ammons2004improving} proposed a \cfg restructuring for 
avoiding information merging along hot paths i.e., the paths that are more frequently executed in program 
compared to other paths. They showed that this improves precision of various data flow analysis and increases 
the opportunities of program optimizations. Their work differs from our work in that they did not address the 
imprecision added by infeasible control flow paths.

\section{Conclusions and Future Work}\label{sec:Conclusion}

An exhaustive data flow analysis answers all queries of a particular type over all possible executions of a program. However, in 
practice it is hard to see an exhaustive analysis that is efficient and computes a precise result, because of the following 
issues that often have conflicting solutions.
\begin{itemize}
\item Issue 1: to achieve efficiency, the analysis may need to merge the information reaching at a program point along different paths, 
because the number of paths as well as the corresponding data flow information may be very large or even unbounded.
\item Issue 2: to achieve precision, the analysis may need to distinguish between the information reaching from different paths, in order to 
1) discard the information reaching along infeasible paths, and 2) avoid potential loss of information because of merging that 
happens at join points of two or more paths.
\end{itemize}

These issues have been handled in past as follows: the first issue is handled 
by merging the information reaching along different paths at their shared path segments in a \cfg 
(instead of keeping the information separate). The approach achieves efficiency because it reduces the 
 amount of information computed. 
However, the computed information is usually weaker than the actual possible information  
because of the information merging, and inclusion of information reaching along infeasible paths.

To handle the second issue, two types of approaches have been proposed. 
\begin{enumerate} 
\item The first type of approaches eliminate the effect of infeasible paths from analysis 
in analysis dependent or analysis independent way. In particular, analysis dependent approaches 
use selective path sensitive analysis  that uses heuristics to selectively distinguish information 
reaching along some paths, and subsequently discard the information along infeasible paths thereby 
eliminating effect of infeasible paths from the specific analysis. On the other hand, analysis 
independent approaches use \cfg restructuring to eliminate infeasible paths from \cfg itself 
thereby eliminating the effect of infeasible paths from all analyses performed on restructured \cfg.


\item The second type of approaches eliminate the effect of imprecise joins by using 
either \cfg restructuring or trace partitioning. More specifically, in \cfg restructuring 
they  eliminate known imprecise joins points from \cfg. On the other hand, in trace 
partitioning they create equivalence classes of program traces using user specified 
criteria. Next, at imprecise join points, only the information corresponding to traces in 
same class are merged, and information from traces belonging to different classes is kept 
separate. Thus, they reduce the effect of imprecise joins to some extent. Both the approaches 
of removing imprecise joins rely on analysis specific heuristics because imprecise joins 
are analysis specific. 
\end{enumerate}

In this work, we proposed a generic approach to remove the effect of infeasible paths from 
exhaustive data flow analyses. 
In particular, we introduced the notion of 
feasible path \mfp solutions that separate and discard the data flow values reaching along infeasible paths. 
A key insight that enabled \fpmfp is the following.
\begin{quote}
Infeasible paths is a property of programs and not of 
any particular data flow analysis over the programs. Hence, we can separate the identification 
of infeasible paths in the \cfg of a program from discarding the corresponding data flow values 
during a data flow analysis. 
\end{quote}

The above insight allows us to devise a generic technique that intuitively lifts a data flow analysis 
to multiple parallel and interacting data flow analyses each of which eliminates the effect 
of a class of known infeasible paths. 
In particular, this is realized in \fpmfp through a two phase approach: 
in the first phase, we 
detect minimal infeasible path 
segments from the input program. In the second phase, we lift the input data flow analysis so as to 
separate 
the values that flow through a \mips from the values that do not flow 
through the \mips. Further, the analysis blocks the values that flow through the \mips at the 
end of the \mips thereby eliminating the effect of infeasible paths from data flow analysis.

In our experiments on 30 Benchmarks (selected from Open Source, Industry, and SPEC CPU 2006),
 we compared precision and performance of  \fpmfp solutions with that of \mfp solutions. Here,
we observed  up to \uptoDefUse\ (average \avgDefUse, geo. mean \meanDefUse) reduction in the number 
of def-use pairs in the reaching definitions analysis, and up to \uptoUninit\ (average \avgUninit, geo. mean \meanUninit) 
reduction in the number of alarms in the possibly-uninitialized variables analysis, 
when using \fpmfp solution in place of \mfp solution. 
Further, the \fpmfp computation took $2.9\times$ more time compared to the \mfp computation.
In our experience, this cost-precision trade-off is acceptable considering the corresponding reduction in the 
manual efforts. 
Specifically, in program testing testcases need to be written to ensure each def-use pair is 
covered by at least one testcase. Similarly, for validating the possibly- uninitialized variable 
alarms requires manual efforts and is error prone.


In future, we see that the \fpmfp solutions can be further improved both in terms of scalability 
and precision. First, the scalability can be improved by identifying 
more optimizations that can discard the pairs that do not lead to precision improvement. 
Second, the precision can be improved by adding handling for wider class of 
inter-procedural \mips. However, this should be complemented by corresponding 
optimizations to keep the distinctions to bare minimum to retain the efficiency of the 
approach.

Moreover, we see a possibility that  the  precision gain and scalability of \fpmfp on a 
program can be anticipated using a lightweight pre-analysis of the program. This 
can help decide whether to apply \fpmfp to the program or not.
Similar approaches have been proved helpful in earlier attempts 
of precise and scalable context sensitive program 
analyses~\cite{oh2014selective,oh2015selective,wei2015adaptive,hassanshahi2017efficient}. 
We believe this will help for \fpmfp also.

\bibliography{scp}
\allowdisplaybreaks
\newcommand{\newin}[1]{\text{$\overline{\overline{\In_{#1}}}$}}
\newcommand{\newout}[1]{\text{$\overline{\overline{\Out_{#1}}}$}}
\newcommand{\dset}{\{\pairc{\setm_1}{d_1},\pairc{\setm_2}{d_2},...\}}
\newcommand{\foldset}{\text{$\foldmeet(\dset)$}}

\newcommand{\fournspace}{\!\!\!\!}
\newcommand{\mipsactive}{\emph active }

\newcommand{\flow}{\text{\sf\em adjustFlow}\xspace}
\newcommand{\setmi}{\text{$\setm_i$}\xspace}

\newcommand{\edgestart}{\emph{startMipsAt}}
\newcommand{\edgeinner}{\emph{innerMipsAt}}
\newcommand{\edgeend}{\emph{endMipsAt}}
\newcommand{\destlabel}{h}
\newcommand{\conde}{\text{\emph{cond$(e)$}} }
\newcommand{\conden}{\text{\emph{cond$(e_n)$}} }
\newcommand{\condat}[1]{\text{$\llbracket cond(e)\rrbracket_{#1}$}}
\newcommand{\condenat}[1]{\text{$\llbracket cond(e_n)\rrbracket_{#1}$}}
\newcommand{\written}{\text{\emph{modified}} }
\newcommand{\symmipsF}{\text{$\mu_4$}\xspace}
	
\newcommand{\ntwotb}{\!\!}
\newcommand{\nfourtb}{\ntwotb\ntwotb}
\newcommand{\nthreetb}{\!\!\!}
\newcommand{\nfivetb}{\!\!\!\!\!}
\newcommand{\ntentb}{\nfivetb\nfivetb}
\newcommand{\ntwentytb}{\ntentb\ntentb}
\newcommand{\edgesp}[1]{\text{\emph{edges}$(#1)$}\xspace}
\newcommand{\equ}{=}

\newcommand{\lemath}{\text{\emph{Lemma 3}}\xspace}
\newcommand{\strt}{\text{\emph{Start}}\xspace}
\newcommand{\inEdges}[1]{\text{\emph{inEdges}$(#1)$}\xspace}
\newcommand{\inEdgesNoEmph}[1]{\text{inEdges$(#1)$}\xspace}

\newcommand{\apaths}[1]{\text{\emph{paths}$(#1)$}\xspace}
\newcommand{\fpaths}[1]{\text{\emph{fpaths}$(#1)$}\xspace}
\newcommand{\fpathsNoEmph}[1]{\text{fpaths$(#1)$}\xspace}
\newcommand{\pwithms}[1]{\text{\emph{mpaths}$(#1)$}\xspace}
\newcommand{\pwms}[1]{\text{\emph{mpaths}$^c(#1)$}\xspace}
\newcommand{\pwmsNoEmph}[1]{\text{mpaths$^c(#1)$}\xspace}
\newcommand{\pwmsl}[2]{\text{$\text{\emph{mpaths}}^c_{#1}(#2)$}\xspace}
\newcommand{\pwmslNoEmph}[2]{\text{$\text{mpaths}^c_{#1}(#2)$}\xspace}

\newcommand{\fsigma}[2]{\text{$f_{#1}(#2)$}\xspace}
\newcommand{\sigt}{\text{$\sigma$}\xspace}
\newcommand{\lemmaS}{\text{Lemma 2}\xspace}
\newcommand{\setN}{\text{$\mathcal{N}$}\xspace}

\section{Proof of Soundness:}\label{sec:pos}
In this section, we formally define the \emph{meet over feasible paths} (\mofp) solution, and prove that \fpmfp computes an 
sound-approximation of \mofp solution. At first, we state some auxiliary definitions which are useful for the  proof of soundness.

\subsection{Definitions}

\begin{definition} For an edge $e:m\rightarrow n$, we refer the data flow value $\bFe_e(\bOut_m)$ obtained by applying 
the edge flow function $\bFe_e$ to the data flow value $\bOut_m$ as ``the data flow value at e''.
\end{definition}

\begin{definition}\label{definition:cfps.that.reach.n}
Recall that all \cfps referred in our analysis are intra-procedural. We refer \cfps that begin at start node of a \cfg and 
end at a particular node $n$ as ``\cfps that reach $n$''.
\end{definition}

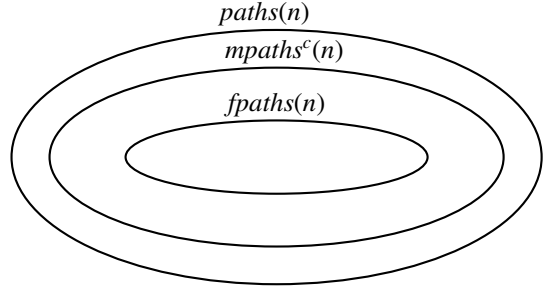
\begin{figure}[t!]
\begin{pspicture}(0,20)(25,58)
\putnode{n00}{origin}{55}{59}{\ \apaths{n}}
\putnode{n10}{n00}{3}{-5}{\pwms{n}}
\putnode{n01}{n00}{2}{-12}{\fpaths{n}}
\psellipse(57,40)(35,17)
\psellipse(57,40)(30,12)
\psellipse(57,40)(20,5)
\end{pspicture}
\caption{Paths that reach node $n$: $\apaths{n}\supseteq\pwms{n}\supseteq\fpaths{n}$}\label{fig:paths.reaching.n}
\end{figure}

Let \setmips be the set of all \mips that are input to a \fpmfp computation, then we create the following classes of \cfps 
that reach a node $n$; these \cfps are illustrated in Figure~\ref{fig:paths.reaching.n}. 
\begin{itemize}
\item \apaths{n}: All \cfps that reach node $n$. This includes feasible as well as infeasible \cfps that reach $n$.
\item \pwithms{n}: All \cfps that reach node $n$ and also contain atleast one \mips of \setmips (\pwithms{n} is a subset of \apaths{n}).
All \cfps in \pwithms{n} are infeasible \cfps because they contain a \mips.
\item \pwms{n}: All \cfps that reach node $n$ and do not contain any \mips of \setmips. This represents a set complement of \pwithms{n} 
i.e., $\pwms{n} = \apaths{n}-\pwithms{n}$. 
\item \fpaths{n}: All \cfps that reach node $n$ and are feasible. \fpaths{n} is a subset of \pwms{n}.
\end{itemize}

\begin{definition}\label{definition:cfps.that.reach.e}
We refer \cfps that begin at start node of a \cfg and end at a 
particular edge $e$ as ``\cfps that reach $e$''. These are the same \cfps that reach the destination node of $e$ and have $e$
as the last edge.
\end{definition}

We create the following classes of \cfps that reach an edge $e$ 
namely \apaths{e}, \pwithms{e}, \pwms{e}, \fpaths{e}. These classes are defined 
similar to the corresponding classes of \cfps that reach a node $n$. Also, if $n$ is the destination node of an edge $e$ then the following
holds.
\begin{align*}
\apaths{e}\subseteq\apaths{n}\\
\pwithms{e}\subseteq\pwithms{n}\\
\pwms{e}\subseteq\pwms{n}\\
\fpaths{e}\subseteq\fpaths{n}
\end{align*}
We defined the notion of contains-prefix-of for a \mips in definition~\ref{definition.cpo}, we now extend this to \cfps as follows.

\begin{definition} For a \cfp $\sigma$ that reaches an edge $e$, \cpo{e}{\sigma} contains all \mips \symmips such that \prefix{\symmips}{e} is 
contained in \sigt i.e.,
\begin{align}
\cpo{e}{\sigma}=\set{\symmips\mid\symmips\in\setmips,\, \sigma\text{ contains }\prefix{\symmips}{e}} \label{eq:cpo.path}
\end{align}
\end{definition}
\begin{observation}\label{observation.cpo.path.extend} Let $\sigma.e'$ be a path obtained by extending a path $\sigma$ that reaches an edge 
$e$ to its successor edge $e'$. Then \cpo{e'}{\sigma.e'} precisely contains two types of \mips: 1) \mips present in \cpo{e}{\sigma} which 
contain edge $e'$, and 2) \mips that start at $e'$ i.e., 
\begin{align}
\text{Given }\onetb& \sigma\in\pwms{e} \land  \text{ $e$ is predecessor edge of $e'$ then }\nonumber \\ 
\twotb\cpo{e'}{\sigma.e'}&=\bset{\symmips\mid (\symmips\in \cpo{e}{\sigma}, e'\in \edges{\symmips}) \lor e'\in 
\mipsstart(\symmips)}\twotb\\
\twotb&=\cpob{\ e'}{\cpo{e}{\sigma}}\onetb\ldots\text{from }~\ref{eq:cpo.set}
\end{align}
\end{observation}

We now formally state the meet over feasible paths (\mofp) solution. Next, we prove that the \fpmfp solution is an 
sound-approximation of \mofp solution.

\subsection{The \mofp solution}
Given a path $\sigt\in\apaths{n_k}$ consisting of nodes $(n_1,n_2,...,n_{k-1},n_k)$, let $f_{\sigt}$ denote the 
composition of functions corresponding to the nodes in \sigt  excluding the last  node $n_k$ i.e., 
$f_{\sigt}= f_{n_{k-1}}\circ f_{n_{k-2}}\circ \ldots\circ f_{n_1}$. 
Note that $f_{\sigt}$ represents the effect of a path that reaches at IN of node $n_k$, hence it excludes $f_{n_k}$. However,  
$f_{\sigt}$ includes the effect of all edges in $\sigt$ including the edge from $n_{k-1}$ to $n_k$. For simplicity, we assume the 
edge flow functions of the input \mfp analysis are identity functions for all edges. Observe that if $\sigt$ contains a single node then 
$f_{\sigt}$ is identity function.

The \mop solution is represented by variables $\In_n$/$\Out_n$ for all nodes $n$ using the following equation. 
\begin{align}
\In_n &=\bigsqcap_{\sigma\in\apaths{n}}\fsigma{\sigma}{BI}\\ 
\Out_n &=\bigsqcap_{\sigma\in\apaths{n}}f_n\circ\fsigma{\sigma}{BI}
\end{align}

Similarly, the \mofp solution is obtained by taking meet over feasible paths that reach $n$ (i.e., \fpaths{n}) as follows.
\begin{align}
\In_n &=\bigsqcap_{\sigma\in\fpaths{n}}\fsigma{\sigma}{BI}\\ 
\Out_n &=\bigsqcap_{\sigma\in\fpaths{n}}f_n\circ\fsigma{\sigma}{BI}
\end{align}

\subsection{Soundness Claim}
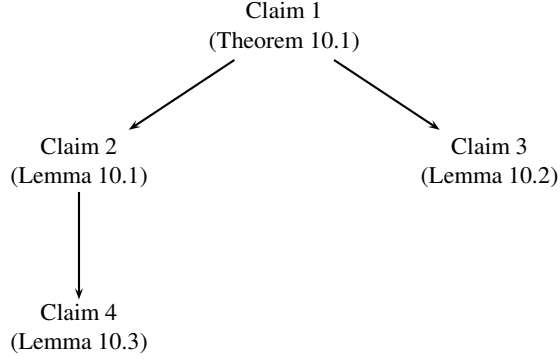
\begin{figure}[t!]
\begin{pspicture}(0,20)(25,75)
\small
\putnode{n00}{origin}{63}{72}{$\begin{array}{c}\text{Claim 1}\\ \text{(Theorem~\ref{theorem.main})}\end{array}$}
\putnode{n0}{n00}{27}{-18}{$\begin{array}{c}\text{Claim 3} \\ \text{(Lemma~\ref{lemma.mpaths.fpaths})}\end{array}$}
\putnode{n1}{n00}{-27}{-18}{$\begin{array}{c}\text{Claim 2} \\ \text{(Lemma~\ref{lemma.in.n})}\end{array}$}
\putnode{n1l}{n1}{0}{-22}{$\begin{array}{c}\text{Claim 4} \\ \text{(Lemma~\ref{lemma.edge.e})}\end{array}$}

\ncline[nodesep=0.5]{->}{n00}{n0}
\ncline[nodesep=0.5]{->}{n00}{n1}
\ncline{->}{n1}{n1l}
\end{pspicture}
\vspace{-2pt}
\caption{Proof Outline. The claim at the root of the tree (claim 1) is the main proof obligation for soundness of \fpmfp. 
A claim at each node is derived using the claims at its child nodes.}\label{fig:proof.outline}
\end{figure}

\begin{mytheorem}\label{theorem.main}
The \fpmfp solution is a sound-approximation of the \mofp solution i.e., if \setN is the set of all nodes in the \cfg of a procedure
then
\begin{align}
\text{Claim 1.}\twotb\ \forall n\in\setN.\onetb\bbIn_n\sqsubseteq\bigsqcap_{\sigma\in\fpaths{n}}\fsigma{\sigma}{BI}
\label{eq:proof.soundnessO.claimO}
\end{align}
\end{mytheorem}
Proof. The proof outline is illustrated in Figure~\ref{fig:proof.outline}. Below, we state Claim 2 and Claim 3 
which together prove Claim 1 i.e.,
\begin{align}
\text{Claim 2} \land \text{Claim 3} \implies \text{Claim 1}
\end{align}

where,
\begin{itemize}
\item Claim 2: at any node $n$, $\bbIn_n$ is a sound-approximation of the meet of data flow values 
reaching along paths in \pwms{n} i.e.,
\begin{align}
\text{Claim 2.}\twotb\ \forall n\in\setN.\onetb\bbIn_n\sqsubseteq\bigsqcap_{\sigma\in\pwms{n}}\fsigma{\sigma}{BI}
\label{eq:proof.soundnessO.claimT}
\end{align}
\item Claim 3: at any node $n$, the meet of data flow values reaching along paths in \pwms{n} is a sound-approximation of
the data flow values reaching along paths in \fpaths{n} i.e.,
\begin{align}
\text{Claim 3.}\twotb\ \forall n\in\setN.\onetb\bbrace{\displaystyle\bigsqcap_{\sigma\in\pwms{n}}\fsigma{\sigma}{BI}}
\onetb\sqsubseteq\onetb \bbrace{\displaystyle\bigsqcap_{\sigma\in\fpaths{n}}\fsigma{\sigma}{BI}}\label{eq:proof.soundnessO.claimTh}
\end{align}
\end{itemize}

Claim 2 and Claim 3 together imply Claim 1 by transitivity of $\sqsubseteq$. We prove the Claim 2 and the Claim 3 
in Lemma~\ref{lemma.in.n} and Lemma~\ref{lemma.mpaths.fpaths} respectively. 

\qed

We first state Claim 4 that will be used to prove Claim 2. 
 
In the \fpmfp computation, the data flow value in each \pairname \pairc{\setm}{d} at an edge 
$e:m\rightarrow n$ is a sound-approximation of the meet of the data flow values reaching along each path $\sigma$ where $\sigma \in 
\pwms{e}$ and 
$\cpo{e}{\sigma}\equ\setm$ i.e., 
\begin{align}
\text{Claim 4}.\twotb\pairc{\setm}{d}\in \bFe_e(\bOut_m) 
\implies d\sqsubseteq\bbrace{\bigsqcap_{\begin{array}{c}\sigma\in\pwms{e},\\ \cpo{e}{\sigma}\equ\setm\end{array}} 
\fsigma{\sigma}{BI}}\label{eq:lemmath}
\end{align}

We prove Claim 4 in Lemma~\ref{lemma.edge.e}. We now show that Claim 2 follows from Claim 4 because for a node $n$ that is 
not the \strt node of the \cfg, \pwms{n} 
is obtained by union of \pwms{e} corresponding to each edge $e$ that is incident on $n$. 

\begin{mylemma}\label{lemma.in.n} At any node $n$ in the CFG of a program the following holds
\end{mylemma}
\begin{align}
\text{Claim 2}.\twotb\forall n\in\setN.\onetb\bbIn_n&\sqsubseteq \bbrace{\bigsqcap_{\sigma\in\pwms{n}}\fsigma{\sigma}{BI}}
\end{align}

Proof. 
We split the proof of Claim 2 in the following two cases which are mutually exclusive and exhaust all possibilities.
\begin{itemize}
\item Case 1: $n$ is the \emph{Start} node of the \cfg. In this case, \pwms{n} contains a single path containing the \strt node only.
 Hence the following holds.
\begin{align}
\bbrace{\bigsqcap_{\sigma\in\pwms{n}}\fsigma{\sigma}{BI}}&=BI&\twotb\label{eq:claimT.caseO.T}\\
\bbIn_n&=BI&\twotb\text{from}~\ref{eq:dfe21}\label{eq:claimT.caseO.O}\\
\bbIn_n&\sqsubseteq \bbrace{\bigsqcap_{\sigma\in\pwms{n}}\fsigma{\sigma}{BI}}&\twotb
\text{from}~\ref{eq:claimT.caseO.O},~\ref{eq:claimT.caseO.T}\label{eq:claimT.caseO}
\end{align}
\item Case 2: $n$ is not the \emph{Start} node of the CFG. In this case, the following holds.
\begin{align}
\bIn_n&=\bigsqcap_{m\in pred(n)}^{\text{\textemdash}} \bFe_{m\rightarrow n}(\bOut_m)\onetb\ldots\text{from}~\ref{eq:dfe21}\\
\pairc{\setmO}{d_1}\in \bIn_n &\implies d_1 \sqsubseteq \bbrace{\bigsqcap_{\begin{array}{c} 
\pairc{\setmO}{d_1'}\in\bFe_{m\rightarrow n}(\bOut_m),\\ m\in pred(n) \end{array}} \!\!\!\!\!\!\!\!d_1'}\onetb\ldots\text{from }~\ref{eq:meet}\\ 
\pairc{\setmO}{d_1}\in \bIn_n &\implies d_1 \sqsubseteq \bbrace{\bigsqcap_{\begin{array}{c}\sigma\in\pwms{n},\\ 
\cpo{e}{\sigma}\equ\setmO \end{array}}\fsigma{\sigma}{BI}}\onetb\ldots\text{from }~\ref{lemma.edge.e}\\
\bbrace{\bigsqcap_{\pairc{\setmO}{d_1}\in\bIn_n}d_1}&\sqsubseteq \bbrace{\bigsqcap_{\sigma\in\pwms{n}}\fsigma{\sigma}{BI}}
\label{eq:lemmaF.bin}\\
\bbIn_n&\sqsubseteq \bbrace{\bigsqcap_{\sigma\in\pwms{n}}\fsigma{\sigma}{BI}}\label{eq:claimT.caseT}
\end{align}
\end{itemize}
Equations~\ref{eq:claimT.caseO} and ~\ref{eq:claimT.caseT} prove the lemma.

\qed

\begin{mylemma}\label{lemma.mpaths.fpaths}
At any node $n$ in the CFG of a program the following holds
\end{mylemma}
\begin{align}
\text{Claim 3}.\twotb\forall n\in\setN.\onetb\bbrace{\bigsqcap_{\sigma\in\pwms{n}}\fsigma{\sigma}{BI}}
\sqsubseteq \bbrace{\bigsqcap_{\sigma\in\fpaths{n}}\fsigma{\sigma}{BI}}
\end{align}

Proof.
The following holds from the definition of \pwms{n} and \fpaths{n}
\begin{align}
\forall n\in \setN.\onetb\fpaths{n}\subseteq\pwms{n} \label{lemma.mpaths.fpaths.one}\twotb\ldots\text{from}~\ref{definition:cfps.that.reach.n}
\end{align}

Claim 3 trivially follows from~\ref{lemma.mpaths.fpaths.one}.

\qed

The following observation states some properties of a path of length 1 starting at \emph{Start} of a \cfg. These properties 
are used for proving claim 4 subsequently.

\begin{observation}\label{observation:path.length.one}
Let $\sigma_1: n_0\xrightarrow{e} n_1$ be a path of length one that begins at \emph{Start} of a procedure, and \setmips be the set of 
all \mips in the procedure, then $\sigma_1$ has the following properties.
\end{observation}
\begin{enumerate}
\item $e$ does not have a predecessor edge. Hence, $e$ cannot be end edge of any \mips i.e.,
\begin{align}
\forall \setm\subseteq\setmips.\ \neg\enof{\setm}{e}& &\label{eq:basis.endof.false}
\end{align}
\item $\cpo{e}{\sigma_1}$ only contains \mips that start at $e$ i.e.,
\begin{align}
\cpo{e}{\sigma_1}&=\set{\symmips\mid\symmips\in\setmips,\ e\in\mipsstart(\symmips)}\onetb\ldots\text{from ~\ref{eq:cpo.path}}
\label{eq:basis.cpo.sigma1}\\
\forall \setm\subseteq\setmips.\ \cpob{e}{\setm}&=\set{\symmips\mid\symmips\in\setmips,\ e\in\mipsstart(\symmips)}
\onetb\ldots\text{from ~\ref{eq:cpo.path}}\\
\forall \setm\subseteq\setmips.\ \cpob{e}{\setm}&=\cpo{e}{\sigma_1}\label{eq:basis.ext.forall}
\end{align}

from~\ref{eq:basis.cpo.sigma1}  and  ~\ref{eq:basis.ext.forall}  we get
\begin{align}
(\setm'\subseteq\setmips) \land (\setm'\not=\cpo{e}{\sigma_1})&\implies (\forall \setm\subseteq\setmips.\ \cpob{e}{\setm}\not=\setm')
\label{eq:basis.not.cpo}
\end{align}
\item There exists only one path of length 1 that begins at \strt
\begin{align}
\pwmsl{1}{e}&=\set{\sigma_1}& \label{eq:basis.one.path}\\
\bbrace{\bigsqcap_{\sigma\in\pwmsl{1}{e}} \fsigma{\sigma}{BI}}&=
\fsigma{\sigma_1}{BI}  \onetb\ldots\text{from}~\ref{eq:basis.one.path}\label{eq:basis.fsigma.n00}\\
&=f_{n0}(BI)  \onetb\ldots f_{\sigma_1}=f_{n0}\label{eq:basis.fsigma.n0}
\end{align}
\item Since $n_0$ is the Start node of the procedure, $\bIn_{n0}$ is the boundary value.
\begin{align}
\bIn_{n0}&=\bBoundary=\bset{\pairc{\nomips}{\Boundary}} \cup 
\bset{\pairc{\setm}{\top}\mid \setm\not=\nomips, \setm\subseteq \setmips}
\onetb\ldots\text{from}~\ref{eq:dfe21}\\
&\text{Applying node flow function }\bF_{n0}\text{ and using ~\ref{eq:nff} we get }\nonumber\\
\bF_{n0}(\bIn_{n0})&= \bset{\pairc{\nomips}{f_{n0}(\Boundary)}} \cup \bset{\pairc{\setm}{f_{n0}(\top)}\mid \setm\not=\nomips, 
\setm\subseteq \setmips}\\
&\text{from }~\ref{eq:dfe22}\text{ we have }\bOut_{n0}=\bF_{n0}(\bIn_{n0}) \text{ hence }\nonumber\\
\bOut_{n0}&= \bset{\pairc{\nomips}{f_{n0}(\Boundary)}} \cup \bset{\pairc{\setm}{f_{n0}(\top)}\mid \setm\not=\nomips, 
\setm\subseteq \setmips}\label{eq:basis.out.no}
\end{align}
\end{enumerate}

\begin{mylemma}\label{lemma.edge.e} In the \fpmfp computation, the data flow value in each \pairname \pairc{\setm}{d} at an edge 
$e:m\rightarrow n$ is a sound-approximation of the meet of data flow values reaching along each path $\sigma$ where $\sigma \in 
\pwms{e}$ and 
$\cpo{e}{\sigma}\equ\setm$ i.e., 
\begin{align}
\text{Claim 4}.\twotb\pairc{\setm}{d}\in \bFe_e(\bOut_m) 
\implies d\sqsubseteq\bbrace{\bigsqcap_{\begin{array}{c}\sigma\in\pwms{e},\\ \cpo{e}{\sigma}\equ\setm\end{array}} 
\fsigma{\sigma}{BI}}\label{eq:lemmath}
\end{align}
\end{mylemma}
Proof.

We prove this by induction on the length of paths reaching from \strt to the edge $e$. Let \pwmsl{l}{e} denote the
 paths of length $l$ from \pwms{e} (edges may be repeated in the path). 

We re-write the proof obligation in the following equivalent form.
\begin{align}
\forall l\ge 1.\ \pairc{\setm}{d}\in \bFe_e(\bOut_m) 
\implies d\sqsubseteq\bbrace{\bigsqcap_{\begin{array}{c}\sigma\in\pwmsl{l}{e},\\ \cpo{e}{\sigma}\equ\setm\end{array}}
 \fsigma{\sigma}{BI}}\label{eq:lemmath.revised}
\end{align}

Basis. $l=1$
\setlength{\abovedisplayskip}{4pt}
\setlength{\belowdisplayskip}{4pt}
There exists only one path of length 1 that begins from \strt node. Let that path be $\sigma_1:n_0\xrightarrow{e}n_1$, where 
$n_0=\strt$. Observe that $e$ cannot be an inner or end edge of any \mips because $e$ does not have a 
predecessor edge. Consequently, $\cpo{e}{\sigma_1}$ only contains \mips that start at $e$. We have stated these properties in
Observation~\ref{observation:path.length.one}. 

Proof obligation for the base case (obtained by substituting $l=1$ and $m=n_0$ in \ref{eq:lemmath.revised}):
\begin{align}
\pairc{\setm'}{d'}\in \bFe_e(\bOut_{n0})
&\implies d'\sqsubseteq \bigsqcap_{\begin{array}{c}\sigma\in\pwmsl{1}{e},\\ \cpo{e}{\sigma}\equ\setm'\end{array}}
 \ntentb\fsigma{\sigma}{BI}&\label{eq:base.case.obligation}
\end{align}

We split the proof of the base case in the following two cases which are mutually exclusive and exhaustive for each \pairname 
$\pairc{\setm'}{d'}$ in $\bFe_e(\bOut_{n0})$. 
\begin{itemize}
\item Case 1:  There exists a path $\sigma\in\pwmsl{1}{e}$ such that $\setm'\equ\cpo{e}{\sigma}$. Since $\pwmsl{1}{e}$ contains a single 
path $\sigma_1$, Case 1 becomes $\setm'\equ\cpo{e}{\sigma_1}$.
\begin{align}
&\bbrace{\pairc{\setm'}{d'}\in \bFe_e(\bOut_{n0})}\land \bbrace{\setm'\equ\cpo{e}{\sigma_1}}&\nonumber \\
&\implies d'=\bigsqcap_{\begin{array}{c}\pairc{\setm}{d}\in \bOut_{n0},\\ \cpob{e}{\setm}\equ\setm'\end{array}} 
\ntentb d &\onetb\text{from}~\ref{eq:eff.eff},~\ref{eq:basis.endof.false}\\
&\implies d'=\bigsqcap_{\begin{array}{c}\pairc{\setm}{d}\in \bOut_{n0},
\\ \cpob{e}{\setm}\equ\cpo{e}{\sigma_1}\end{array}} \ntwentytb d &\onetb\setm'\equ\cpo{e}{\sigma_1}\\
&\implies d'=\bigsqcap_{\pairc{\setm}{d}\in \bOut_{n0}} \ntentb\ \ \  d &
\onetb\text{from}~\ref{eq:basis.ext.forall}\\
&\implies d'=f_{n0}(BI)\sqcap f_{n0}(\top)&\onetb\text{from}~\ref{eq:basis.out.no}\\
&\implies d'=f_{n0}(BI)&\onetb f_{n0}(BI)\sqsubseteq f_{n0}(\top)\\
&\implies d'= \bigsqcap_{\begin{array}{c}\sigma\in\pwmsl{1}{e}\end{array}} \ntentb\fsigma{\sigma}{BI}
&\onetb\text{from}~\ref{eq:basis.fsigma.n00}\text{ and }\ref{eq:basis.fsigma.n0}\\
&\implies d'= \bigsqcap_{\begin{array}{c}\sigma\in\pwmsl{1}{e},\\ \cpo{e}{\sigma}\equ\cpo{e}{\sigma_1}\end{array}} 
\ntentb\fsigma{\sigma}{BI}&
\onetb\text{from}~\ref{eq:basis.one.path}\\
&\implies d'= \bigsqcap_{\begin{array}{c}\sigma\in\pwmsl{1}{e},\\ \cpo{e}{\sigma}\equ\setm'\end{array}} 
\ntentb\fsigma{\sigma}{BI}& \setm'\equ\cpo{e}{\sigma_1}
\end{align}
\item Case 2:  There does not exists a path $\sigma\in\pwmsl{1}{e}$ such that $\setm'\equ\cpo{e}{\sigma}$. Since $\pwmsl{1}{e}$ contains a 
single path $\sigma_1$, Case 2 becomes $\setm'\not\equ\cpo{e}{\sigma_1}$.
\begin{align}
&\bbrace{\pairc{\setm'}{d'}\in \bFe_e(\bOut_{n0})}\land \bbrace{\setm'\not\equ\cpo{e}{\sigma_1}}&\nonumber\\
&\implies d'=\bigsqcap_{\begin{array}{c}\pairc{\setm}{d}\in \bOut_{n0},\\ \cpob{e}{\setm}\equ\setm'\end{array}} 
\ntwentytb d &\onetb\text{from}~\ref{eq:eff.eff},~\ref{eq:basis.endof.false}\\
&\implies d'=\top& \onetb\text{from}~\ref{eq:basis.not.cpo}\\
&\implies d'= \bigsqcap_{\sigma\in\set{}} 
\fsigma{\sigma}{BI}&\onetb \bigsqcap_{\set{}}=\top\\
&\text{Since }\set{\sigma\mid\sigma\in\pwmsl{1}{e}, \cpo{e}{\sigma}\not\equ\cpo{e}{\sigma_1}}\equ\Phi&
\onetb\text{from}~\ref{eq:basis.one.path}\\
&\implies d'= \bigsqcap_{\begin{array}{c}\sigma\in\pwmsl{1}{e},\\ \cpo{e}{\sigma}\equ\setm'\end{array}} 
\ntentb\fsigma{\sigma}{BI}&\onetb\text{from}~\ref{eq:basis.one.path}
\end{align}
\end{itemize}
From Case 1 and 2, we get equation~\ref{eq:base.case.obligation}

Induction Hypothesis. 
Let $n$ be any arbitrary node in the \cfg of a program such that there exist a \cfp of length $l-1$ reaching $n$, and $n$ has atleast one
outgoing edge. We assume that Lemma~\ref{lemma.edge.e}  holds at each edge $e'$ 
that is incident on $n$  (of the form $m\xrightarrow{e'}n$) and is the last edge of the path of length $l-1$ reaching $n$ i.e., 

\begin{align}
\pairc{\setm}{d}\in\bFe_{e'}(\bOut_m)\implies d \sqsubseteq \bigsqcap_{\begin{array}{c}\sigma\in\pwmsl{l-1}{e'},
\\ \cpo{e'}{\sigma}=\setm\end{array}} \ntentb\fsigma{\sigma}{BI}
\label{eq:proof.soundness.hypothesis}
\end{align}

Inductive Step. We prove that Lemma~\ref{lemma.edge.e} holds for any successor edge of $e'$ w.r.t. paths of length $l$ 
i.e., if $m\xrightarrow{e'}n\xrightarrow{e''}o$ then 

\begin{align}
\pairc{\setm''}{d''}\in\bFe_{e''}(\bOut_n)\implies d'' \sqsubseteq \bigsqcap_{\begin{array}{c}\sigma\in\pwmsl{l}{e''},
\\ \cpo{e''}{\sigma}=\setm\end{array}} \ntentb\fsigma{\sigma}{BI}
\label{eq:inductive.proof.obligation}
\end{align}
\emph{Proof of the inductive step.}
We build the proof in two steps.

Let the function \inEdges{n} return the set of incoming edges of a node $n$ in the \cfg, then in step 1, 
we prove that the following holds at $\bOut_n$.
\begin{align}
\pairc{\setm}{d'}\in\bOut_n&\implies d' \sqsubseteq \bigsqcap_{\begin{array}{c}\sigma\in\pwmsl{l-1}{e'},
\\ e' \in \inEdges{n},\\ \cpo{e'}{\sigma}=\setm\end{array}} \ntentb f_n\circ\fsigma{\sigma}{BI}
\end{align}

In step 2, we prove that for any successor edge $e''$ of $e'$, equation~\ref{eq:inductive.proof.obligation} holds.

Step 1:
\begin{align}
\bIn_n&=\mathop{\overline\bigsqcap}_{m\in pred(n)} \bFe_{m\rightarrow n}(\bOut_m)
\onetb\ldots\text{from}~\ref{eq:dfe21}\label{eq:inductive.proof.bin}\\
\pairc{\setm}{d}\in\bIn_n&\implies d \sqsubseteq \bigsqcap_{\begin{array}{c}\pairc{\setm}{d'}\in \bFe_{m\rightarrow n}(\bOut_m),
\\ (m\rightarrow n) \in \inEdges{n}\end{array}} d' \onetb\ldots\text{from }~\ref{eq:inductive.proof.bin},~\ref{eq:meet}\\
\pairc{\setm}{d}\in\bIn_n&\implies d \sqsubseteq \bigsqcap_{\begin{array}{c}\sigma\in\pwmsl{l-1}{e'},\\ e' \in \inEdges{n},
\\ \cpo{e'}{\sigma}=\setm\end{array}} \ntentb\fsigma{\sigma}{BI}\onetb
\ldots\text{from ~\ref{eq:proof.soundness.hypothesis}}\\
&\text{since $f_n$ is monotone}\nonumber\\
\pairc{\setm}{d}\in\bIn_n&\implies f_n(d) \sqsubseteq \bigsqcap_{\begin{array}{c}\sigma\in\pwmsl{l-1}{e'},
\\ e' \in \inEdges{n},\\ \cpo{e'}{\sigma}=\setm\end{array}} \ntentb f_n\circ\fsigma{\sigma}{BI}
\label{eq:inductive.step.one}\\
\pairc{\setm}{d'}\in\bOut_n&\implies \exists \pairc{\setm}{d}\in \bIn_n.\ f_n(d)=d'\onetb\ldots\text{from}~\ref{eq:dfe22} 
\label{eq:inductive.step.two}\\
\pairc{\setm}{d'}\in\bOut_n&\implies d' \sqsubseteq \bigsqcap_{\begin{array}{c}\sigma\in\pwmsl{l-1}{e'},
\\ e' \in \inEdges{n},\\ \cpo{e'}{\sigma}=\setm\end{array}} \ntentb f_n\circ\fsigma{\sigma}{BI}
\onetb\ldots\text{from}~\ref{eq:inductive.step.one},~\ref{eq:inductive.step.two}\label{eq:proof.soundnessO.parto.zero}
\end{align}

Step 2:
We now prove that at any successor edge $e'':n\rightarrow o$ of $e'$ the following holds:

\begin{align}
\pairc{\setm''}{d''}\in \bFe_{e''}(\bOut_n) \implies d'' \sqsubseteq \bbrace{\bigsqcap_{\begin{array}{c}\sigma\in\pwmsl{l}{e''},
\\ \cpo{e''}{\sigma}=\setm'' \end{array}}\fsigma{\sigma}{BI}}\label{eq:proof.soundness.partt}
\end{align}

We consider the following possibilities for any arbitrary pair $\pairc{\setm''}{d''}$ in $\bFe_{e''}(\bOut_n)$. 
These cases are mutually exclusive and exhaustive.

\begin{itemize}
\item Case A: $e''$ is end edge of some \mips \symmips contained in $\setm''$ i.e., 

$\enof{\setm''}{e''}$=\emph{true}. 
\setlength{\abovedisplayskip}{4pt}
\setlength{\belowdisplayskip}{4pt}

In this case, we prove that equation ~\ref{eq:proof.soundness.partt} holds because the following is true.
\begin{align}
d'' =\top \label{eq:proof.soundnessO.caseo.parto.one}\\  
\bbrace{\bigsqcap_{\begin{array}{c}\sigma\in\pwmsl{l}{e''},\\ \cpo{e''}{\sigma}=\setm'' \end{array}}\fsigma{\sigma}{BI}}=
\top\label{eq:proof.soundnessO.caseo.parto.two}
\end{align}

$d''=\top$ follows from the definition of the edge flow function (equation ~\ref{eq:eff.eff}) i.e.,
\begin{align}
\pairc{\setm''}{d''}\in \bFe_{e''}(\bOut_n) \land \enof{\setm''}{e''} \implies 
d''=\top \onetb\ldots \text{from~\ref{eq:eff.eff}}\label{eq:proof.soundnessO.caseo.seven}
\end{align}

Equation~\ref{eq:proof.soundnessO.caseo.parto.two} follows from the fact that there cannot exist a path $\sigma'$ that satisfies 
the constraint $\sigma'\in\pwms{e''} \land \enof{\cpo{e''}{\sigma'}}{e''}$ from the definition of \pwms{e''}.

Hence the following holds,
\begin{align}
\bbrace{\bigsqcap_{\begin{array}{c}\sigma\in\pwmsl{l}{e''},\\ \cpo{e''}{\sigma}=\setm'',\\ \enof{\setm''}{e''} 
\end{array}}\fsigma{\sigma}{BI}}=\top \label{eq:proof.soundnessO.caseo.parto.seven}
\end{align}

From~\ref{eq:proof.soundnessO.caseo.seven} and ~\ref{eq:proof.soundnessO.caseo.parto.seven} we get
\begin{align}
\pairc{\setm''}{d''}\in \bFe_{e''}(\bOut_n) \land \enof{\setm''}{e''} \implies d'' \sqsubseteq \bbrace{\bigsqcap_{\begin{array}{c}
\sigma\in\pwmsl{l}{e''},\\ \cpo{e''}{\sigma}=\setm'' \end{array}}\!\!\!\!\!\!\!\!\!\!\!\!\!\!\!\!\!\!\!\!\fsigma{\sigma}{BI}}
\end{align}	

\item Case B: $e''$ is not end edge of any \mips in $\setm''$ i.e., $\enof{\setm''}{e''}$=\emph{false}.

\begin{align}
&\pairc{\setm''}{d''}\in \bFe_{e''}(\bOut_n) \land \neg\enof{\setm''}{e''} \nonumber\\
&\implies d''=\bigsqcap_{\begin{array}{c}\pairc{\setm}{d}\in\bOut_n,\\ \cpob{e''}{\setm}=\setm''\end{array}} 
d\nonumber\onetb\ldots\text{from}~\ref{eq:eff.eff}\\
&\implies d'' \sqsubseteq \bbrace{\bigsqcap_{\begin{array}{c}\sigma\in\pwmsl{l-1}{e'},\\ e'\in \inEdges{n},\\ 
\cpo{e'}{\sigma}=\setm,\\ \cpob{e''}{\setm}=\setm''\end{array}}f_n\circ\fsigma{\sigma}{BI}}
\nonumber\onetb\ldots\text{from}~\ref{eq:proof.soundnessO.parto.zero}\\
&\implies d'' \sqsubseteq \bbrace{\bigsqcap_{\begin{array}{c}\sigma\in\pwmsl{l-1}{e'},\\ e'\in \inEdges{n},\\ 
\cpo{e''}{\sigma.e''}=\setm''\end{array}}\fsigma{\sigma.e''}{BI}}
\nonumber\onetb\ldots\text{from }~\ref{observation.cpo.path.extend}\\
&\implies d'' \sqsubseteq \bbrace{\bigsqcap_{\begin{array}{c}\sigma\in\pwmsl{l}{e''},\\ \cpo{e''}{\sigma}=\setm'' 
\end{array}}\fsigma{\sigma}{BI}}
\end{align}
\end{itemize}

From case A and B, we get equation~\ref{eq:inductive.proof.obligation}.

\qed





	

\subsection{Soundness of the optimizations of scalability}\label{sec:proof.optimizations}
In this section, we prove that the optimizations proposed in Section~\ref{sec:optimizations} retain the soundness and precision of the 
computed \fpmfp solution. Optimization 1 and Optimization 2 are proved in sections ~\ref{sec:mipsequivalence} and \ref{sec:cso.proof}
respectively.
		
\subsubsection{Proof for Optimization 1}{\label{sec:mipsequivalence}}

	Optimization 1  is applied iff the input \mips satisfy the property \symp (Section \ref{subsection:optimization.one}) which is
	restated as follows.
	\begin{quote}
	\symp: for a \mips 
		$\symmips: e_1\rightarrow e_2\rightarrow\ldots\rightarrow e_n$, if the condition on the end edge $e_n$ is \conden 
		then \conden evaluates to \false at the start edge $e_1$ of \symmips and variables in \conden are not modified along $\symmips$.
	\end{quote}

 	Formally, if $\llbracket p\rrbracket_e$ represents  the evaluation of a condition expression $p$ at an 
		edge $e$, and $\written(p,\symmips)$ is the set of variables in $p$ that are modified at intermediate nodes of a \mips $\symmips$ then 
		\begin{align}
		\text{``\symmips is a \mips''} \iff& \bbrace{(\condenat{e_1}=\false) \land (\written(\conden,\symmips)=\phi)}\label{lemma1}
		\end{align}

		We now prove that if two \mips satisfy \symp
		\footnote{Note that we use algorithms given by Bodik et al for detecting \mips which 
		are presented in Appendix; they allow us to  track if a \mips satisfies the property \symp or not.}, 
		and have the same end edge then they can be considered equivalent. In particular, we prove 
		that if two \mips have the same end edge $e$, then any \cfp 
		that extends from start edge of either \mips till $e$ by traversing edges in either \mips is an infeasible path. Hence, the data
		flow values associated with such \mips can be merged.
		
		 \begin{mytheorem} If $\symmips_1,\symmips_2$ are two \mipss that have the same end edge $e_n$, and they intersect at some 
		node $x_k$, then $\symmips_3,\symmips_4$\textemdash constructed as shown below using split and join of $\symmips_1,\symmips_2$ at 
		the intersecting node\textemdash are also \mipss.	 
		\begin{align}
			\symmips_1: &\uline{x_1\xrightarrow{ex1} x_2\rightarrow\ldots\rightarrow x_k\rightarrow x_{k+1}\rightarrow\ldots\xrightarrow{e_n} n}
			\label{theorem_equivalence}\\ \nonumber
			\symmips_2: &y_1\xrightarrow{ey1} y_2\rightarrow\ldots\rightarrow x_k\rightarrow y_j\rightarrow\ldots\xrightarrow{e_n} n\\ \nonumber
			\\ \nonumber
			\symmips_3: &\uline{x_1\xrightarrow{ex1} x_2\rightarrow\ldots\rightarrow x_k}\rightarrow y_j\rightarrow\ldots\xrightarrow{e_n} n \\ 
			\nonumber
			\symmips_4: &y_1\xrightarrow{ey1} y_2\rightarrow\ldots\rightarrow\uline{x_k\rightarrow x_{k+1}\rightarrow\ldots\xrightarrow{e_n} n} \\ 
			\nonumber
		&\\ \nonumber
		\text{Claim 5.\twotb}&\text{``$\symmips_1$,$\symmips_2$ are \mips''} \implies \text{``$\symmips_3$,$\symmips_4$ are \mips''}
		\end{align}
		\end{mytheorem}
		Proof.
		
		We use the following fact to prove the claim 5: $\symmips_3$ and $\symmips_4$ are constructed using the edges and the nodes from 
		$\symmips_1$ and $\symmips_2$ so the following holds
		\begin{align}
			&(\written(\conden,\symmips_1)=\phi)\ \land \ (\written(\conden,\symmips_2)=\phi)\nonumber \\
			&\implies \nonumber \\
			&(\written(\conden,\symmips_3)=\phi)\ \land\ (\written(\conden,\symmips_4)=\phi) \label{written}
		\end{align}

		\begin{align*}
				LHS:\text{``$\symmips_1$,$\symmips_2$ are \mips''} &\implies (\condenat{ex1}=\condenat{ey1}=\false)\ \land\  \\
				&\threetb (\written(\conden,\symmips_1)=\phi)\ \land\ \\
				&\threetb (\written(\conden,\symmips_2)=\phi)\\
				&\threetb\tentb\threetb\ldots\text{from \ref{lemma1}}\\
				&\implies (\condenat{ex1}=\condenat{ey1}=\false)\ \land\ \\
				&\threetb (\written(\conden,\symmips_3)=\phi)\ \land\ \\
				&\threetb (\written(\conden,\symmips_4)=\phi)\\
				&\threetb\tentb\threetb\ldots\text{from \ref{written}}\\
				&\implies \text{``$\symmips_3$,$\symmips_4$ are \mips''}\ 
		\end{align*}


\subsubsection{Proof of Optimization 2}\label{sec:cso.proof} At an edge, Optimization 2 (Section~\ref{sec:optimization.cso}) merges 
\pairnames that contain the same 
data flow value. In this section, we prove that this optimization retains the precision and the soundness of the \fpmfp solution.
A crucial difference between \fpmfp and \mfp is that in the \fpmfp computation certain data flow values are blocked
at end edges of \mips. We prove that the same set of values are blocked at the same set of end edges with or without 
Optimization 2.

\begin{mytheorem}
If $\bFe_e(\bIn_e)$ contains two \pairnames \pairc{\setmO}{d} and \pairc{\setmT}{d} with the same data flow value $d$ then 
shifting $d$ to the \pairname corresponding to a set of \mips \setmTh defined below does not affect the 
 soundness or precision of the resulting \fpmfp solution.  Specifically, it does not affect the  blocking of $d$ at the end  
edges of \mips in \setmO and \setmT.
\begin{align}
\setmTh=&\set{\symmipsO\mid \symmipsO\in \setmO,\ \exists \symmipsT\in \setmT.\, 
\symmipsT\in\cso{e}{\symmipsO}}\label{eq:definition.mthree}\\
	     &\cup\nonumber\\
			 &\set{\symmipsT\mid \symmipsT\in \setmT,\ \exists \symmipsO\in \setmO.\, \symmipsO\in\cso{e}{\symmipsT} }
			\nonumber
\end{align}
\end{mytheorem}

Proof.
We consider the following two operations each indicating Optimization 2 is performed or not. Next, we prove that the \fpmfp solution is 
identical after both the operations:
\begin{enumerate}
\item Operation 1: \pairnames \pairc{\setmO}{d} and \pairc{\setmT}{d} are kept as it is in $\bFe_e(\bIn_e)$ (i.e., Optimization 2 
is not applied).
\item Operation 2: \pairnames \pairc{\setmO}{d} and \pairc{\setmT}{d} are replaced with \pairc{\setmTh}{d} in $\bFe_e(\bIn_e)$ 
(i.e., Optimization 2 is applied). For simplicity of exposition, we assume there is no value associated with \setmTh initially, 
otherwise we add \pairc{\setmTh}{d'\meet d} where $d'$ is the value associated with \setmTh in $\bFe_e(\bIn_e)$.
\end{enumerate}
The possible cases for \mips in \setmO and \setmT w.r.t. Operation 1 and 2 are shown in Table~\ref{tab:proof.optimizationT}.
Let the data flow value $d$ is not killed along a \mips \symmipsO in \setmO (if it is killed then $d$ does not reach $\mipsend(\symmipsO)$
after either operation is performed). We prove that both the above operations achieve similar effect in the 
following two cases which are mutually exclusive and exhaustive for all \mips in \setmO (similar argument holds for \mips in \setmT):
\begin{enumerate}
\item Case 1: if $\symmipsO\in\setmO$ and $\symmipsO\in\setmTh$, then $d$ is blocked at $\mipsend(\symmipsO)$ after both the operations.
\item Case 2: if $\symmipsO\in\setmO$ and $\symmipsO\not\in\setmTh$, then $d$ reaches $\mipsend(\symmipsO)$  after 
both the operations.
\end{enumerate}
\vspace{15pt}

\textbf{Proof of Case 1.}

If $\symmipsO\in\setmO$ and $\symmipsO\in\setmTh$, then $d$ is blocked at $\mipsend(\symmipsO)$ after both the operations, as explained
 below.
\begin{enumerate}
\item Operation 1: \pairc{\setmO}{d},\pairc{\setmT}{d} are kept as it is in $\bFe_e(\bIn_e)$.

In this case, $d$ is blocked within \pairc{\setmO}{d} at $\mipsend(\symmipsO)$ because $\symmipsO\in\setmO$. Also, $d$ is blocked within
\pairc{\setmT}{d} because there exists a \mips \symmipsT in \setmT  whose suffix is contained in \symmipsO i.e.,
	\begin{align}
	\symmipsO \in \setmO \land \symmips\in\setmTh &\implies \exists \symmipsT\in\setmT. \symmipsT\in\cso{e}{\symmipsO}\\
	\symmipsT\in\setmT \land \symmipsT\in\cso{e}{\symmipsO}&\implies \mipsend(\symmipsT)\in \edges{\symmipsO}\text{ and }\\
	&\twotb \text{$d$ is blocked within \pairc{\setmT}{d} at $\mipsend(\symmipsT)$}
	\end{align}
	Hence, in this scenario, $d$ is blocked within both \pairnames before reaching $\mipsend(\symmipsO)$ or at $\mipsend(\symmipsO)$.
	
\item Operation 2: \pairc{\setmO}{\top},\pairc{\setmT}{\top},\pairc{\setmTh}{d} belong to $\bFe_e(\bIn_e)$ and $\symmipsO\in\setmTh$.  

In this case, $d$ is blocked within \pairc{\setmTh}{d} at $\mipsend(\symmipsO)$ because $\symmipsO\in\setmTh$.  
\end{enumerate}

\begin{table}[t]%
\center
\small
\begin{tabular}{|c|c|c|c|}
\cline{3-4}
	\multicolumn{1}{c}{}
  &\multicolumn{1}{c}{}
	& \multicolumn{2}{|c|}{is d blocked at the edge in $\mipsend(\symmips)$ after}
	\\ \cline{3-4}
	\multicolumn{1}{c}{}
	& \multicolumn{1}{c|}{}
	& Operation 1
	& Operation 2
	\\ \hline
	\multirow{4}{*}{\rotatebox{90}{Cases}}
	& $1.\, \symmips\in\setmO, \symmips\in\setmTh$ 
	& Yes
	& Yes
	\\ \cline{2-4}
	\multicolumn{1}{|c|}{}
	& $2.\, \symmips\in\setmO, \symmips\not\in\setmTh$ 
	& No
	& No
	\\ \cline{2-4}
	\multicolumn{1}{|c|}{}
	& $3.\, \symmips\in\setmT, \symmips\in\setmTh$ 
	& Yes
	& Yes
	\\ \cline{2-4}
	\multicolumn{1}{|c|}{}
	& $4.\, \symmips\in\setmT, \symmips\not\in\setmTh$ 
	& No
	& No
	\\ \hline
\end{tabular}
\caption{Equivalence of Operation 1 and Operation 2 w.r.t. cases in Optimization 2. Optimization 2 involves shifting a data flow value $d$ from 
\pairnames \pairc{\setmO}{d} and \pairc{\setmT}{d} to \pairname \pairc{\setmTh}{d}. 
For simplicity, we assume $d$ is not killed along \symmips.}
\label{tab:proof.optimizationT}
\end{table}

\textbf{Proof of Case 2.}

If $\symmipsO\in\setmO$ and $\symmipsO\not\in\setmTh$, then $d$ reaches $\mipsend(\symmipsO)$ after both the operations.
\begin{enumerate}
\item Operation 1: \pairc{\setmO}{d},\pairc{\setmT}{d} are kept in $\bFe_e(\bIn_e)$ as it is.

In this case, $d$ is blocked within \pairc{\setmO}{d} at $\mipsend(\symmipsO)$. However, $d$ in \pairc{\setmT}{d} is not blocked 
at or before $\mipsend(\symmipsO)$ because \symmipsO does not contain \suffix{\symmipsT}{e} for any \mips \symmipsT in \setmT
(meaning \symmipsT either 1) does not follow the same \cfp as \symmipsO after $e$, or 2) follows the same \cfp as \symmipsO but 
does not end before \symmipsO on the \cfp).
Thus, $d$ in \pairc{\setmT}{d} reaches $\mipsend(\symmipsO)$ along a \cfp that goes through \suffix{\symmipsO}{e}.


\item Operation 2: \pairc{\setmO}{\top},\pairc{\setmT}{\top},\pairc{\setmTh}{d} belong to $\bFe_e(\bIn_e)$. 

In this case,  \symmipsO does not contain 
\suffix{\symmipsTh}{e} for any \mips \symmipsTh in \setmTh i.e.,

$\symmipsO\in\setmO \land \symmipsO\not\in\setmTh \implies \forall\symmipsTh\in\setmTh. \symmipsTh\not\in \cso{e}{\symmipsO}$

This can be proved by proving that following is a contradiction.
\begin{align*}
&\symmipsO\in\setmO \land \symmipsO\not\in\setmTh \land \exists\symmipsTh\in\setmTh.\symmipsTh\in \cso{e}{\symmipsO}&\\
&\implies \symmipsTh\not\in\setmT& \text{from~\ref{eq:definition.mthree}}\\
&\implies \symmipsTh\in\setmO \land \exists\symmipsT\in\setmT.\symmipsT\in \cso{e}{\symmipsTh}& \text{from~\ref{eq:definition.mthree}}\\
&\implies \exists\symmipsT\in\setmT.\symmipsT\in \cso{e}{\symmipsO}& \cso{e}{\symmipsTh}\subseteq\cso{e}{\symmipsO}\\
&\implies \symmipsO\in\setmTh&\text{from~\ref{eq:definition.mthree}}\\
&\implies \emph{false}&
\end{align*}

 Hence $d$ in \pairc{\setmTh}{d} is not blocked at or before $\mipsend(\symmipsO)$. 
Thus, $d$ in \pairc{\setmTh}{d} reaches $\mipsend(\symmipsO)$ along 
a \cfp that goes through \suffix{\symmipsO}{e}.
\end{enumerate}

Similar cases can be proved for \mips in \setmT. Thus, the \fpmfp solutions with or without Optimization 2 are equivalent.
\qed




\section{Appendix: Bodik's Approach for Detecting Minimal Infeasible Path Segments}

In this section, we present the algorithms proposed by Bodik et al.~\cite{bodik1997refining} for detection 
of \mips. We have used these algorithms in the  pre-processing stage of \fpmfp computation to generate a 
set of \mips. 
The section is organized as follows. First, we give an overview of Bodik's approach of 
infeasible path detection (Section~\ref{sec:bodik.overview}).
Next, we describe the algorithms for detection of infeasible paths in detail (Section~\ref{sec:bodik.details}).


\subsection{Overview of Bodik's Approach}\label{sec:bodik.overview}
Bodik et al. observed that many infeasible paths are caused by statically detectable correlations 
between two branch conditions, or between a branch condition and an assignment statement appearing 
on the paths. 
Hence, to find infeasible paths arising from branch correlations, they use the following criteria: 
if the constraint of a conditional edge evaluates to FALSE along any \cfp reaching the edge, then 
the \cfp is infeasible.

\begin{example}~\label{ex:mips.detection.correlation}
In Figure~\ref{fig:abstract.correlation}a, the branch constraint $(b>1=true)$ is 
in conflict with the assignment statement $b=0$, hence the path marked with double 
line arrows is infeasible. Similarly, in Figure~\ref{fig:abstract.correlation}b
the constraint $(b>1=false)$ of the branch at bottom is in conflict with constraint $(b>1=true)$ of the 
branch at top. Hence the path marked with double line arrows is infeasible.
\end{example}

Intuitively, the idea is to enumerate \cfps that reach a conditional edge and identify \cfps 
along which the edge constraint evaluates to FALSE, because such \cfps are infeasible. In particular they proceed as follows. 
They start from a branch node, and backward propagate the corresponding 
 constraint (i.e., branch condition) along the incoming paths that reach the node. 
Here, they evaluate the branch constraint using predefined 
rules at nodes encountered in backward propagation along a path. 
A path is labeled as infeasible if the assertions at a node and the constraint have a conflict that is
detectable using predefined rules. 

\begin{figure}[tp]
\setlength{\tabcolsep}{1pt}
\begin{tabular}{l|l}
\begin{tabular}{c}
\begin{pspicture}(0,0)(25,102)
\putnode{n00}{origin}{30}{85}{\psframebox{\white$b=0$}}
\putnode{n0}{n00}{0}{-20}{\psframebox{$b=0$}}
	\putnode{n2}{n0}{0}{-20}{\psframebox{\white$b=0$}}
		\putnode{n3}{n2}{0}{-20}{\psframebox{$b > 1$}}
				\putnode{n4}{n3}{-10}{-20}{\psframebox{\white$b=0$}}
				\putnode{n5}{n3}{10}{-20}{\psframebox{\white$b=0$}}
\ncline{->}{n00}{n0}
\ncline[doubleline=true]{->}{n0}{n2}
\ncline[doubleline=true]{->}{n2}{n3}
\ncline{->}{n3}{n4}
\nbput{false}
\ncline[doubleline=true]{->}{n3}{n5}
\naput{true}
\end{pspicture}
\end{tabular}
&
\begin{tabular}{c}
\begin{pspicture}(0,0)(25,102)
\putnode{n00}{origin}{30}{85}{\psframebox{\white$b=0$}}
\putnode{n0}{n00}{0}{-20}{\psframebox{$b > 1$}}
	\putnode{n1}{n0}{-10}{-20}{\psframebox{\white$b=0$}}
	\putnode{n2}{n0}{10}{-20}{\psframebox{\white$b=0$}}
		\putnode{n3}{n2}{0}{-20}{\psframebox{$b > 1$}}
				\putnode{n4}{n3}{-10}{-20}{\psframebox{\white$b=0$}}
				\putnode{n5}{n3}{10}{-20}{\psframebox{\white$b=0$}}
\ncline{->}{n00}{n0}
\ncline{->}{n0}{n1}
\nbput{false}
\ncline[doubleline=true]{->}{n0}{n2}
\naput{true}
\ncline[doubleline=true]{->}{n2}{n3}
\ncline[doubleline=true]{->}{n3}{n4}
\nbput{false}
\ncline{->}{n3}{n5}
\naput{true}
\end{pspicture}
\end{tabular}
\\
$\begin{array}{l}
\text{(a) Correlation between assignment}\\
\text{\twotb statement and branch statement. }
\end{array}$ 
& 
$\begin{array}{l} 
\text{\threetb(b) Correlation between two}\\
\text{\threetb\twotb branch statements. }
\end{array}$
\end{tabular}
\caption{Statically detectable correlations between program statements leading to infeasible paths.}
\label{fig:abstract.correlation}
\end{figure}


\begin{figure}[htp]
\psset{unit=2mm}
\begin{pspicture}(-50,0)(-110,-90)
	\putnode{entry}{origin}{-80}{20}{\psframebox{$n_1$:int a=0,b=0}}
	
	\putnode{not1}{entry}{30}{0}{$STEP\ 1\rightarrow$ \psframebox[fillcolor=blue,fillstyle=solid]{}}
	\putnode{not2}{not1}{0}{-5}{$STEP\ 2\rightarrow$ \psframebox[fillcolor=green,fillstyle=solid]{}}
	
	\putnode{c1}{entry}{0}{-17}{\psframebox{$n_2$:$nondet()$}}
	\putnode{c1Ans1}{c1}{13}{12}{\psframebox[linewidth=0.04,linecolor=white]{\blue $A[e_1,q_{e8}]=\{FALSE\}$}}
	\putnode{c1Ans2}{c1}{13}{8}{\psframebox[linewidth=0.04,linecolor=white]{\blue$A[e_1,q_{e9}]=\{TRUE\}$}}
	\putnode{c1top2}{c1}{13}{4}{\psframebox[linewidth=0.04,linecolor=white]{\blue$Q[e_1]=\{q_{e8},q_{e9}\}$}}
	\putnode{c1start1}{c1}{-18}{12}{\psframebox[linewidth=0.04,linecolor=white]{\green Start[$e_{1}$]= \{\sout{$q_{e8}$}\}}}
	\putnode{c1inner1}{c1start1}{0}{-3}{\psframebox[linewidth=0.04,linecolor=white]{\green Inner[$e_{1}$]= \{\}}}
	\putnode{c1end1}{c1start1}{0}{-6}{\psframebox[linewidth=0.04,linecolor=white]{\green End[$e_{1}$]= \{\}}}
	
	\putnode{n1}{c1}{0}{-11}{\psframebox{$n_3$:b=3}}
	\putnode{n1top2}{n1}{8}{5}{\psframebox[linewidth=0.04,linecolor=white]{\blue$Q[e_2]=\{\}$}}
	\putnode{n1start1}{n1}{23}{8}{\psframebox[linewidth=0.04,linecolor=white]{\green Start[$e_{2}$]= \{\}}}
	\putnode{n1inner1}{n1start1}{0}{-3}{\psframebox[linewidth=0.04,linecolor=white]{\green Inner[$e_{2}$]= \{\}}}
	\putnode{n1end1}{n1start1}{0}{-6}{\psframebox[linewidth=0.04,linecolor=white]{\green End[$e_{2}$]= \{\}}}
	
	\putnode{n2}{n1}{0}{-11}{\psframebox{$n_4$:scanf(``\%d",\&a)}}
	\putnode{n2top2}{n2}{8}{4}{\psframebox[linewidth=0.04,linecolor=white]{\blue$Q[e_4]=\{\}$}}
	\putnode{n2start1}{n2}{23}{8}{\psframebox[linewidth=0.04,linecolor=white]{\green Start[$e_{4}$]= \{\}}}
	\putnode{n2inner1}{n2start1}{0}{-3}{\psframebox[linewidth=0.04,linecolor=white]{\green Inner[$e_{4}$]= \{\}}}
	\putnode{n2end1}{n2start1}{0}{-6}{\psframebox[linewidth=0.04,linecolor=white]{\green End[$e_{4}$]= \{\}}}

	\putnode{n3}{n2}{0}{-17}{\psframebox{$n_5$:printf(``\%d",b);}}
	\putnode{n3top2}{n3}{11}{5}{\psframebox[linewidth=0.04,linecolor=white]{\blue$Q[e_5]=\{q_{e8},q_{e9}\}$}}
	\putnode{n3Ans1}{n3}{14}{12}{\psframebox[linewidth=0.04,linecolor=white]{\blue$A[e_5,q_{e8}]=\{UNDEF\}$}}
		\putnode{n3Ans2}{n3}{14}{8}{\psframebox[linewidth=0.04,linecolor=white]{\blue$A[e_5,q_{e9}]=\{UNDEF\}$}}
	\putnode{n3top4}{n3}{-21}{4}{\psframebox[linewidth=0.04,linecolor=white]{\blue$Q[e_3]=\{q_{e8},q_{e9}\}$}}
	\putnode{n3start1}{n3}{30}{8}{\psframebox[linewidth=0.04,linecolor=white]{\green Start[$e_{5}$]= \{\}}}
	\putnode{n3inner1}{n3start1}{0}{-3}{\psframebox[linewidth=0.04,linecolor=white]{\green Inner[$e_{5}$]= \{\}}}
	\putnode{n3end1}{n3start1}{0}{-6}{\psframebox[linewidth=0.04,linecolor=white]{\green End[$e_{5}$]= \{\}}}
	\putnode{n3start2}{n3}{-21}{16}{\psframebox[linewidth=0.04,linecolor=white]{\green Start[$e_{3}$]= \{$q_{e8}$\}}}
	\putnode{n3inner2}{n3start2}{0}{-3}{\psframebox[linewidth=0.04,linecolor=white]{\green Inner[$e_{3}$]= \{$$\}}}
	\putnode{n3end2}{n3start2}{0}{-6}{\psframebox[linewidth=0.04,linecolor=white]{\green End[$e_{3}$]= \{\}}}
	
	\putnode{n4}{n3}{0}{-13}{\psframebox{$n_6$:printf(``\%d'',a);}}
	\putnode{n4top2}{n4}{-17}{5}{\psframebox[linewidth=0pt,linecolor=white]{\blue$Q[e_6]=\{q_{e8},q_{e9}\}$}}
	\putnode{n4start1}{n4}{18}{8}{\psframebox[linewidth=0.04,linecolor=white]{\green Start[$e_{6}$]= \{\}}}
	\putnode{n4inner1}{n4start1}{0}{-3}{\psframebox[linewidth=0.04,linecolor=white]{\green Inner[$e_{6}$]= \{$q_{e8}$\}}}
	\putnode{n4end1}{n4start1}{0}{-6}{\psframebox[linewidth=0.04,linecolor=white]{\green End[$e_{6}$]= \{\}}}
	
	\putnode{c2}{n4}{0}{-13}{\psframebox{$n_7:a>1$}}
	\putnode{c2top2}{c2}{-15}{5}{\psframebox[linewidth=0.04,linecolor=white]{\blue$Q[e_7]=\{q_{e8},q_{e9}\}$}}
	\putnode{c2start1}{c2}{18}{8}{\psframebox[linewidth=0.04,linecolor=white]{\green Start[$e_{7}$]= \{\}}}
	\putnode{c2inner1}{c2start1}{0}{-3}{\psframebox[linewidth=0.04,linecolor=white]{\green Inner[$e_{7}$]= \{$q_{e8}$\}}}
	\putnode{c2end1}{c2start1}{0}{-6}{\psframebox[linewidth=0.04,linecolor=white]{\green End[$e_{7}$]= \{\}}}
	
	\putnode{n5}{c2}{0}{-12}{\psframebox{$n_8$:assert(b!=0)}}
	\putnode{n5top}{n5}{9}{8}{\psframebox[linewidth=0.04,linecolor=white]{\blue$Q[e_8]=\{\}$}}
	\putnode{n5top2}{n5}{13}{5}{\psframebox[linewidth=0.04,linecolor=white]{\blue$q_{e8}$: (($a>1$)==TRUE)}}
	\putnode{n5start1}{n5}{30}{8}{\psframebox[linewidth=0.04,linecolor=white]{\green Start[$e_{8}$]= \{\}}}
	\putnode{n5inner1}{n5start1}{0}{-3}{\psframebox[linewidth=0.04,linecolor=white]{\green Inner[$e_{8}$]= \{\}}}
	\putnode{n5end1}{n5start1}{-2}{-6}{\psframebox[linewidth=0.04,linecolor=white]{\green End[$e_{8}$]= \{$q_{e8}$\}}}
	
	\putnode{exit}{n5}{0}{-12}{\psovalbox{$n_9$:Exit}}
	\putnode{exittop}{exit}{9}{6}{\psframebox[linewidth=0.04,linecolor=white]{\blue$Q[e_{10}]=\{\}$}}
	\putnode{exittop1}{exit}{-15}{8}{\psframebox[linewidth=0.04,linecolor=white]{\blue$Q[e_9]=\{\}$}}
	\putnode{exittop4}{exit}{-18}{5}{\psframebox[linewidth=0.04,linecolor=white]{\blue$q_{e9}$: (($a>1$)==FALSE)}}
	\putnode{exitstart1}{exit}{20}{8}{\psframebox[linewidth=0.04,linecolor=white]{\green Start[$e_{10}$]= \{\}}}
	\putnode{exitinner1}{exitstart1}{0}{-3}{\psframebox[linewidth=0.04,linecolor=white]{\green Inner[$e_{10}$]= \{\}}}
	\putnode{exitend1}{exitstart1}{0}{-6}{\psframebox[linewidth=0.04,linecolor=white]{\green End[$e_{10}$]= \{\}}}
	\putnode{exitstart2}{exit}{-21}{18}{\psframebox[linewidth=0.04,linecolor=white]{\green Start[$e_{9}$]= \{\}}}
	\putnode{exitinner2}{exitstart2}{0}{-3}{\psframebox[linewidth=0.04,linecolor=white]{\green Inner[$e_{9}$]= \{\}}}
	\putnode{exitend2}{exitstart2}{0}{-6}{\psframebox[linewidth=0.04,linecolor=white]{\green End[$e_{9}$]= \{\}}}

	\ncline[linecolor=red]{->}{entry}{c1}
	\nbput{$e_1$}
	\ncline{->}{c1}{n1}
	\nbput{(true) $e_2$}
	\ncangle[armB=5,linecolor=red,angleA=180,angleB=180]{->}{c1}{n3}
	\nbput{(false) $e_3$}
	\ncline{->}{n1}{n2}
	\nbput{$e_4$}
	\ncline{->}{n2}{n3}
	\nbput{$e_5$}
	\ncline[linecolor=red]{->}{n3}{n4}
	\nbput{$e_6$}
	\ncline[linecolor=red]{->}{n4}{c2}
	\nbput{$e_7$}
	\ncline[linecolor=red]{->}{c2}{n5}
	\nbput[npos=0.7]{(true) $e_8$}
	\ncangle[armB=4,angleA=180,angleB=180]{->}{c2}{exit}
	\nbput{(false)$e_9$}
	\ncline{->}{n5}{exit}
	\nbput{$e_{10}$}
	\end{pspicture}
\caption{Example illustrating detection of \mips using Bodik's Approach}
\label{fig:mips.detection}
\end{figure}
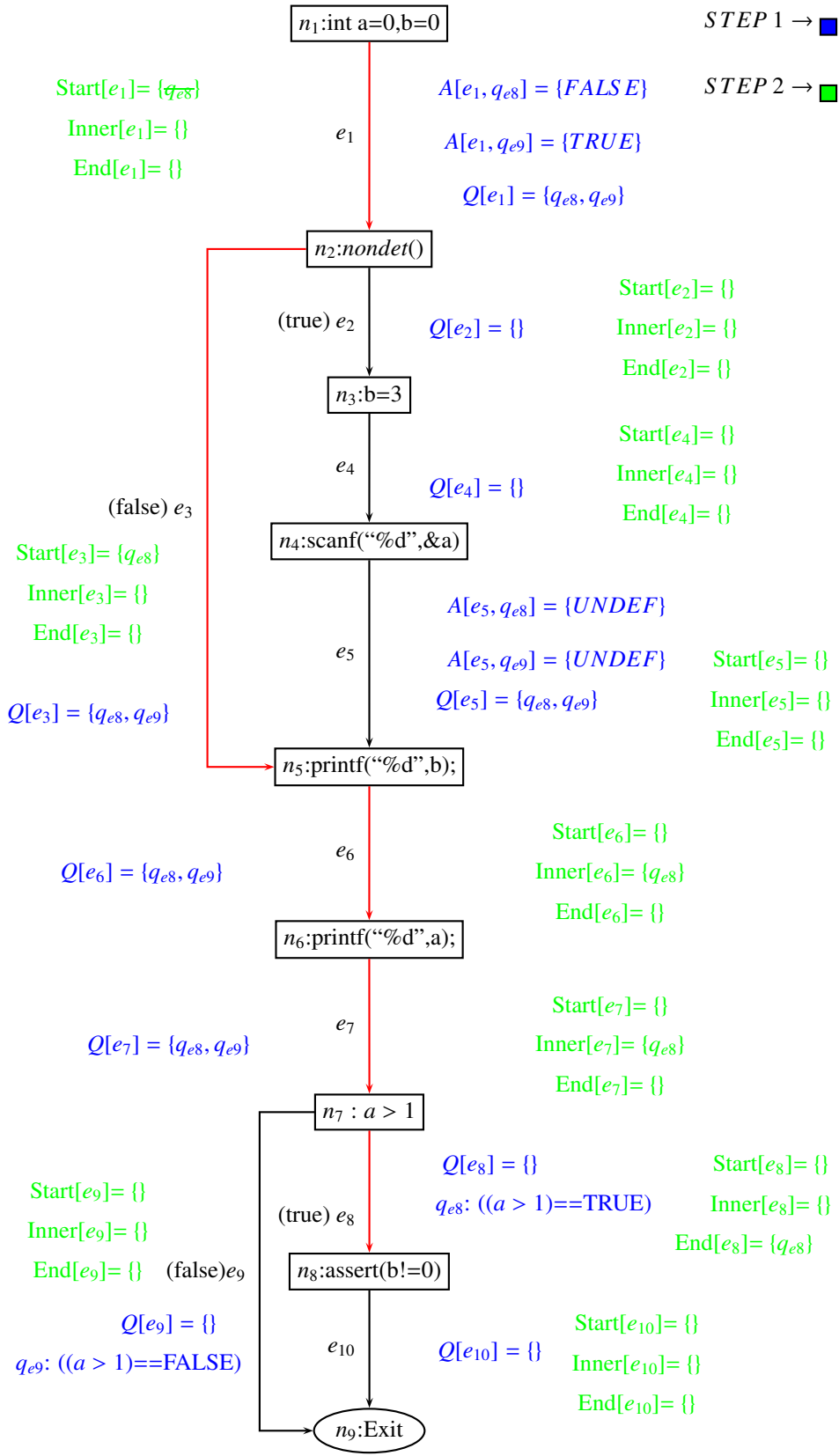

\begin{example}
For example in Figure~\ref{fig:mips.detection}, the branching node $n_7:\ a>1$ is correlated with node $n_1$ in that  
the constraint $a>1$ evaluates to FALSE at $n_1$. Therefore, the following path $\sigma$ that connects
$n_1$ to the TRUE branch of $n_7$ is infeasible,  
$\sigma: n_1\xrightarrow{e_1}n_2\xrightarrow{e_3}n_5\xrightarrow{e_6}n_6\xrightarrow{e_7}n_7\xrightarrow{e_8}n_8$. 
Bodik's approach detects this path by backward propagating the constraint $(a>1)$ from node $n_7$ to $n_1$ along $\sigma$. 
\end{example}

\newcommand{\predEdge}[1]{\text{$\text{\emph{predEdge}($#1$)}$}\xspace}
\newcommand{\raiseQuery}[1]{\text{$\text{\emph{raiseQuery}($#1$)}$}\xspace}
\newcommand{\comnt}[1]{\text{//\emph{#1}}\xspace}
\newcommand{\fivetb}{\fourtb\onetb}

\begin{algorithm}[H]
\caption{Step 1 to detect \mips that end at a conditional edge $e$. 
Let the condition on $e$ be $(v\leq c) = x$, 
where $x$ is either \{TRUE or FALSE\}, and $v\leq c$ is the expression 
at the source node of $e$. The comments are prefixed with //.}
\label{alg:mips.step1}
\begin{algorithmic}[1]
\item[]
\STATE Initialize Q[e] to \{\} at each edge $e$; set \emph{worklist} to \{\}.\\
  \comnt{raise the initial query $q_e:((v\leq c)=x)$, at each immediate predecessor}\\
	\comnt{edge of $e$ these are edges incident on source node of e, and }\\
	\comnt{are represented by $\predEdge{e}$}
\STATE For each $e_m\in \predEdge{e}$ 
\STATE \twotb\raiseQuery{e_m,q_e}
\STATE End For
\STATE While \emph{worklist} not empty
\STATE \twotb remove pair ($e$, $q$) from \emph{worklist}\\
\twotb \comnt{assume unknown outcome of q at the entry edge} \\
\twotb \comnt{(i.e., the outgoing edge of program entry).}\\
\STATE \twotb If $e$ is entry edge 
\STATE \threetb A[e,q]:=UNDEF\\
\STATE \twotb Else
\twotb \comnt{answer q using assertions generated at source
node of e.}\\
\STATE \threetb $answer:=resolve(e,source(e),q)$\\
\STATE \threetb If $answer\in \{TRUE,FALSE,UNDEF\}$ \\
\STATE \fourtb $A[e,q]:={answer}$ 				
\STATE  \threetb Else \\
\STATE  \fourtb For each $e_m\in \predEdge{e}$ 
\STATE \fivetb \raiseQuery{e_m, q}\\
\STATE \fourtb End For\\
\STATE \threetb End If\\
\STATE \twotb End If\\
\STATE End While
\item[]
Procedure \raiseQuery{\text{Edge $e$, Query $q$}}\\
\STATE  \twotb If $q\notin Q[e]$ 
\STATE  \threetb add $q$ to $Q[e]$ \\
\STATE  \threetb add pair $(e,q)$ to \emph{worklist}\\
\STATE  \twotb End If
\end{algorithmic}
\end{algorithm}

\begin{algorithm}[H]
\caption{Step 2: mark the edges with \mips information}
\label{alg:mips.step2}
\begin{algorithmic}[1]
\item[] \comnt{begin analysis from the edges where any query was resolved to FALSE} 
\item[] \comnt{at lines 5 to 19 of Step1}
\STATE  \emph{worklist}:=\{$e$\ $\mid$ \ a query was resolved to FALSE at edge $e$\}
\item[] \comnt{raise the initial query at the analyzed edge, to mark end of \mips.}
\STATE  While \emph{worklist} not empty 
\STATE \onetb remove an edge $e$ from the \emph{worklist}
\item[] \onetb \comnt{if query resolved to false at this edge then mark }
\item[] \onetb \comnt{the edge as start for corresponding \mips}
\STATE  \onetb For each query $q_{ex}$ such that $A[e,q_{ex}]=FALSE$
\STATE  \twotb add $q_{ex}$ to $Start[e]$
\item[] \twotb \comnt{mark e as start and ex as end edge for \mips resulting from} 
\item[] \twotb \comnt{query $q_{ex}$ resolving to FALSE at edge $e$.}
\STATE \twotb add $q_{ex}$ to $End[ex]$
\STATE  \onetb End For
\item[] \onetb \comnt{determine answers for each query that was propagated backward}
\STATE \onetb For each query $q$ from $Q[e]$ s.t. $q$ was not resolved at $e$
\item[] \twotb \comnt{move the \mips started along all predecessor edges of $e$}
\STATE \twotb If (for all $e_m\in \predEdge{e}, q\in Start[e_m]$)
\STATE \threetb add $q$ to $Start[e]$
\STATE \threetb For each $e_m\in \predEdge{e}$
\STATE \fourtb remove $q$ from $Start[e_m]$ 
\STATE \threetb End For
\STATE  \twotb Else \twotb \comnt{make $e$ as the inner edge of started \mips}
\STATE \threetb For each $e_m\in \predEdge{e}$ 
\STATE \fourtb copy $Start[e_m]$ to $Inner[e]$
\STATE \fourtb copy $Inner[e_m]$ to $Inner[e]$
\STATE \threetb End For
\STATE \twotb End If
\STATE \onetb End For
\STATE End While
\end{algorithmic}
\end{algorithm}

\subsection{Detailed Explanation of Bodik's Approach}\label{sec:bodik.details}
The Bodik's approach involves two steps as described below.
\begin{enumerate}
\item In the first step, the constraints from branch nodes are 
propagated backwards (towards \cfg entry) to identify nodes (if any) at which these constraints evaluate to FALSE. 
\item In the second step, a forward traversal of \cfg is done to mark the infeasible paths using data from 
Step 1. 
\end{enumerate}
We now explain each of these steps in detail below.

\subsubsection{Step 1}
The results of Step 1 for the example in Figure~\ref{fig:mips.detection} are marked in blue. 
The details of Step 1 are given in Algorithm~\ref{alg:mips.step1} and are described below. 

At the beginning of Step 1, we raise a query $q_e$ at each conditional edge $e$ such that 
the query $q_e$ 
represents the constraint on $e$. Consequently $e$ is visited in an execution only if $q_e$ is TRUE.

\begin{example}
For example, the query for the edge $e_8$ is $q_{e8}:((a>1)$=TRUE), and for the edge $e_9$ is $q_{e9}:((a>1)$=FALSE). 
This means the edge $e_8$ (resp. $e_9$) will be executed only when 
$q_{e8}$ (resp. $q_{e9}$)  evaluates to TRUE. Similarly, we raise queries $q_{e2}$ and $q_{e3}$ at conditional 
edges $e_{2}$ and $e_{3}$ respectively.
\end{example}

Next, the query raised at an edge $e$ is propagated along each predecessor edge of $e$ say $p_e$. 
Here, we try 
to resolve the query at $p_e$ using the assertions generated at source of $p_e$ if the assertions restrict the 
value of variables in the query, for example, because of assignment of values to the variables, or branching 
out of a conditional expression that tests value of the variables. In the former case, if there is an assignment 
to a variable present in the query then the query resolves to either TRUE, FALSE, or UNDEF (meaning details are not 
sufficient to resolve the query to TRUE or FALSE)
\footnote{In the UNDEF case, Bodik also proposes the idea of \emph{query substitution} which is not described here.}.
 
In the latter case, if a variable present in the query is tested in a conditional at the source of $p_e$ then the 
query may resolve to TRUE or FALSE or remain unresolved. If the query is unresolved at $p_e$ 
then the query is back propagated to predecessors of $p_e$ and so on
\footnote{Note that only intra-procedural predecessor edges are considered. In case, the source of $p_e$ is a call node and 
the variables present in the query are modified inside the callee function, then the query is not propagated to predecessor edges 
of $p_e$.}.
\begin{example}
For example, in Figure~\ref{fig:mips.detection}, 
we propagate queries $q_{e8}$ and $q_{e9}$ along edge $e_7$, since source node of $e_7$ is a \emph{printf} statement, 
none of the queries are resolved, and hence they are back propagated to edge $e_6$ and so on. 
Further, at the edge $e_5$ both the queries $q_{e8}$ and $q_{e9}$ are resolved to UNDEF because of the statement 
\emph{scanf}(``\%d",\&a) at the source of $e_5$. Similarly, at the edge $e_1$ the query $q_{e8}$ 
evaluates to FALSE, and $q_{e9}$ evaluates to TRUE because of assignments at the source 
of $e_1$. 
\end{example}

At the end of Step 1 we have two arrays \emph{Q} and \emph{A}, where \emph{Q} stores the 
queries raised or backward propagated at an edge, and \emph{A} stores queries resolution at an edge.

Recall that we defined the query $q_{ex}$ such that edge $ex$ executes only if $q_{ex}$ is TRUE. 
Consequently, if in Step 1, a query $q_{ex}$ is resolved to FALSE at the edge $e$, then this 
implies there is an infeasible path segment from edge $e$ to the edge $ex$. For example in 
Figure~\ref{fig:mips.detection} at edge $e_1$ the query $q_{e8}:((a>1)$=TRUE) evaluates to FALSE, hence 
there is an infeasible path (marked by red edges) that goes from $e_1$ to $e_8$. 
The infeasible path marking happens in Step 2 (Section~\ref{sec:mips.detection.step2}).

\subsubsection{Step 2} \label{sec:mips.detection.step2}
Step 2 marks the edges in the \cfg with the corresponding infeasible paths that pass through the edges.
In particular, this is achieved by maintaining three sets namely Start, Inner, and End at each edge $e$, 
indicating the \mips that contain $e$ as start, inner, or end edge respectively.
For our running example, the results of Step 2 obtained by using Algorithm~\ref{alg:mips.step2} are marked 
in the green color in Figure~\ref{fig:mips.detection}.

The Step 2 proceeds as follows. For each edge $e$ where any query $q_{ex}$ was resolved to FALSE (in Step 1), 
we mark $e$ as the start edge of an \emph{infeasible path segment} (\ips) that ends at $ex$ 
. We also mark $ex$ (i.e., the  edge at which query $q_{ex}$ was raised in Step 1) as the end 
edge of the \ips. For example, in the Figure~\ref{fig:mips.detection}, edge $e_1$ is marked as start 
edge, and the edge $e_8$ is marked as end edge for \ips that ends at $e_8$.

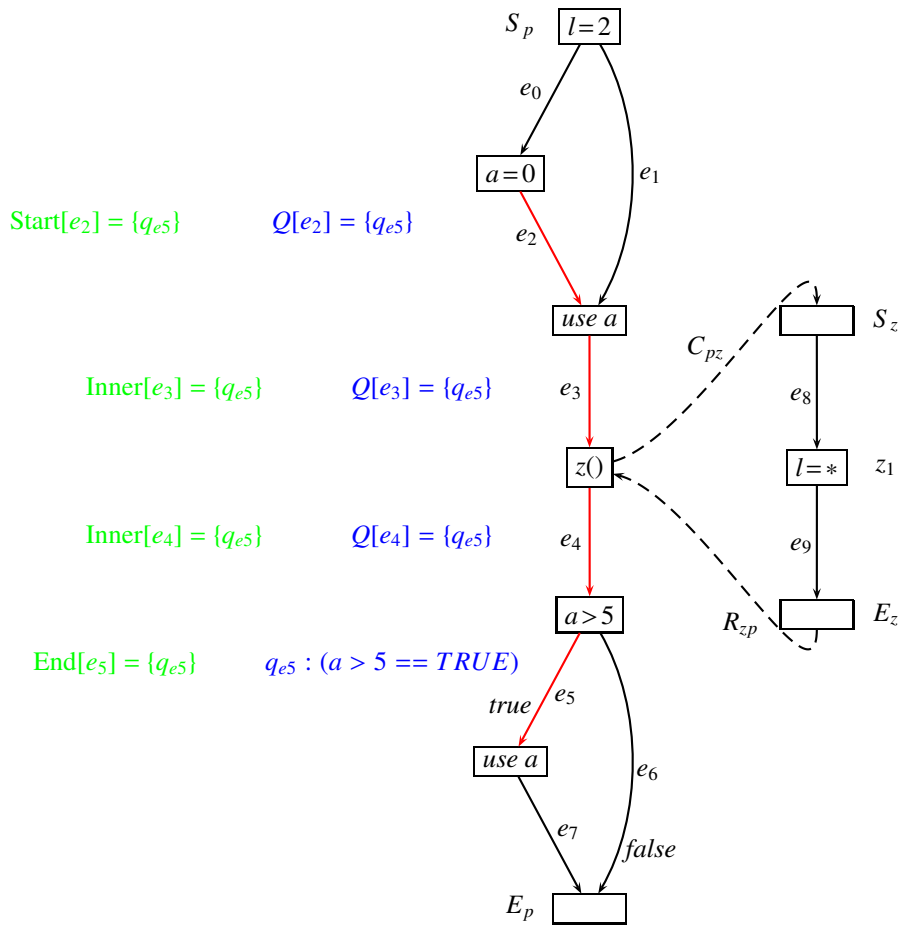
\begin{figure*}[htp]
\psset{unit=1.3mm}
\centering
\begin{pspicture}(-1,0)(55,104)
\putnode{n0}{origin}{42}{94}{\framebox{$l\!=\!2$}}
	\putnode{w}{n0}{-7}{0}{$S_p$}

\putnode{n1}{n0}{-8}{-15}{\framebox{$a\!=\!0$}}
  \putnode{n1top2}{n1}{-17}{-5}{\psframebox[linewidth=0pt,linecolor=white]{\blue$Q[e_2]=\{q_{e5}\}$}}
	\putnode{n1inner1}{n1top2}{-25}{0}{\psframebox[linewidth=0.04,linecolor=white]{\green Start[$e_{2}$] = \{$q_{e5}$\}}}
	
\putnode{n2}{n1}{8}{-15}{\framebox{\em use a}}
  \putnode{n2top2}{n2}{-17}{-7}{\psframebox[linewidth=0pt,linecolor=white]{\blue$Q[e_3]=\{q_{e5}\}$}}
	\putnode{n2inner1}{n2top2}{-25}{0}{\psframebox[linewidth=0.04,linecolor=white]{\green Inner[$e_{3}$] = \{$q_{e5}$\}}}
	
\putnode{n3}{n2}{0}{-15}{\framebox{{\em z}()}}
  \putnode{n3top2}{n3}{-17}{-7}{\psframebox[linewidth=0pt,linecolor=white]{\blue$Q[e_4]=\{q_{e5}\}$}}
	\putnode{n3inner1}{n3top2}{-25}{0}{\psframebox[linewidth=0.04,linecolor=white]{\green Inner[$e_{4}$] = \{$q_{e5}$\}}}

\putnode{n4}{n3}{0}{-15}{\framebox{$a\!>\!5$}}
  \putnode{n4top2}{n4}{-20}{-5}{\psframebox[linewidth=0pt,linecolor=white]{\blue$q_{e5}:(a>5==TRUE)$}}
	\putnode{n4inner1}{n4top2}{-28}{0}{\psframebox[linewidth=0.04,linecolor=white]{\green End[$e_{5}$] = \{$q_{e5}$\}}}

\putnode{n5}{n4}{-8}{-15}{\framebox{\em use a}}

\putnode{n6}{n4}{0}{-30}{\framebox{\white\em use a}}
	\putnode{w}{n6}{-7}{0}{$E_p$}
	
\putnode{z0}{n2}{23}{0}{\framebox{\white\em use a}}
	\putnode{w}{z0}{7}{0}{$S_z$}

\putnode{z1}{z0}{0}{-15}{\framebox{$l\!=\!*$}}
	\putnode{w}{z1}{7}{0}{$z_1$}
	
\putnode{z2}{z1}{0}{-15}{\framebox{\white\em use a}}
	\putnode{w}{z2}{7}{0}{$E_z$}
	
\ncline{->}{n0}{n1}
\nbput[npos=0.5,labelsep=.5]{$e_0$}
\ncline[linecolor=red,doublesep=.2]{->}{n1}{n2}
\nbput[npos=0.35,labelsep=0]{$e_2$}
\nccurve[angleA=300,angleB=60]{->}{n0}{n2}
\naput[npos=0.5,labelsep=.5]{$e_1$}
\ncline[linecolor=red,doublesep=.2]{->}{n2}{n3}
\nbput[npos=0.5,labelsep=.5]{$e_3$}
\ncline[linecolor=red,doublesep=.2]{->}{n3}{n4}
\nbput[npos=0.5,labelsep=.5]{$e_4$}
\ncline[linecolor=red,doublesep=.2]{->}{n4}{n5}
\naput[npos=0.5,labelsep=0.2]{$e_5$}
\nbput[npos=0.7,labelsep=0.2]{\true}
\nccurve[angleA=300,angleB=60]{->}{n4}{n6}
\naput[npos=0.53,labelsep=.2]{$e_6$}
\naput[npos=0.8,labelsep=.2]{\false}
\ncline{->}{n5}{n6}
\naput[npos=0.53,labelsep=.2]{$e_7$}

\nccurve[linestyle=dashed,angleA=15,angleB=90,armB=0.9]{->}{n3}{z0}
\naput[npos=0.43,labelsep=.2]{$C_{pz}$}

\ncline{->}{z0}{z1}
\nbput[npos=0.53,labelsep=.2]{$e_8$}

\ncline{->}{z1}{z2}
\nbput[npos=0.53,labelsep=.2]{$e_9$}
\nccurve[linestyle=dashed,angleA=270,angleB=345,armA=0.5]{->}{z2}{n3}
\naput[npos=0.43,labelsep=.2]{$R_{zp}$}

\end{pspicture}
\caption{Detecting a balanced interprocedural \mips. The variable $a$ is not modified inside the procedure $z$, hence the query
$(q_{e5}:a>5=\text{TRUE})$ is back propagated from $e_4$ to $e_3$.
$S, E$ represent the start and end nodes of a procedure respectively. $C_{pz}$ represents the transfer of 
control from procedure $p$ to procedure $z$ at a call node, and $R_{zp}$ represents the return of control
 from $z$ to $p$. For brevity, cases where $Q$, $Start,Inner$, or $End$ is empty are not shown.}
\label{fig:mips.detection.balanced}
\end{figure*}

Next, if at some edge $e$ query $q_{ex}$ was propagated backwards (in Step 1) and all its predecessor edges are 
marked as start edge for some \ips that ends at $ex$ (in Step 2) then $e$ can also be marked as start edge 
for the \ips. This follows from the fact that the path segment from each predecessor edge $p_e$ to $ex$ is infeasible 
path segment but path segment from $e$ to $ex$ is also \ips  
and is shorter than the \ips from $p_e$ 
to $ex$. Using this justification, we mark edge $e_3$ as the start edge for \ips that ends at $e_8$ in our running example;
observe that the \ips that goes from $e_3$ to $e_8$ is shorter than the one that goes from $e_1$ to $e_8$ 
(hence we un-mark $e_1$ as start edge).

On the other hand, if at some edge $e$ query $q_{ex}$  was propagated backwards (in Step 1) and only some of its 
immediate predecessor edges are marked as start edge  or at least one of the predecessor edge is 
marked as the inner edge for a \ips that ends at $ex$ then we mark $e$ as inner edge for the \ips. 
For instance, we mark edges $e_6,\ \ e_7$ as inner edges for \ips that ends at $e_8$ in our running example.

Finally, at the end of Step 2, each path segment $\sigma: e_1\rightarrow e_2\rightarrow...\rightarrow e_n$ in \cfg 
is a \mips if there exists a query $q$ present in  $Start[e_1]$, $End[e_n]$, and $Inner[e_i], 1<i<n$. 

\begin{example}
Figure~\ref{fig:mips.detection.balanced} shows how balanced inter-procedural \mips (definition~\ref{def:balanced.mips} from 
Chapter~\ref{sec:interprocedural}) are detected using Bodik's approach. In particular, the variables in the query $q_{e5}$
are not modified inside the procedure $z$, hence $q_{e5}$ is propagated from $e_4$ to $e_3$ to $e_2$. Subsequently, in Step 2,
$e_2\rightarrow e_3\rightarrow e_4\rightarrow e_5$ is marked as a \mips.
\end{example}



\end{document}